\documentclass[usenatbib]{mn2e} \voffset=-0.5in

\usepackage{graphicx} 
\usepackage{amsmath} 
\usepackage{amssymb}
\usepackage{exscale, relsize}
\usepackage{amscd}
\usepackage{natbib}
\usepackage{lscape, rotating}      

\usepackage{times}

\newcommand{\secref}[1]{\textsection\ref{#1}}

\newcommand{\eg}{e.g.} 
\newcommand{\ie}{\textit{i.e.}} 
\newcommand{\viz}{\textit{viz.}} 
\newcommand{\cf}{\textit{cf.}}

\newcommand{\Sersic}{S\'ersic}
\newcommand{\model}{\texttt{model}}
\newcommand{\petro}{\texttt{petro}}
\newcommand{\auto}{\texttt{auto}}
\newcommand{\sersictt}{\texttt{sersic}}

\newcommand{\eff}{_\mathrm{e}}

\renewcommand{\sol}{$_{\odot}$}
\newcommand{\age}{t}
\newcommand{\lwage}{\left<t_*\right>}
\newcommand{\dust}{\mathrm{E}_{B\text{-}V}}

\begin{document}



\title[GAMA: Stellar Mass Estimates]
{Galaxy And Mass Assembly: Stellar Mass Estimates}

\author[Edward N. Taylor et al.]
{Edward N.~Taylor,$^{1,2}$\thanks{ent@physics.usyd.edu.au}
Andrew M.~Hopkins,$^{3}$
Ivan K.~Baldry,$^{4}$
Michael J.I.~Brown,$^{5}$ \newauthor
Simon P.~Driver,$^{6}$ 
Lee S.~Kelvin,$^{6}$
David T.~Hill,$^{6}$
Aaron S.G.~Robotham,$^{6}$
Joss \newauthor Bland-Hawthorn,$^{1}$
D.H.~Jones,$^{5}$ 
R.G.~Sharp,$^{17}$ 
Daniel Thomas,$^{14}$ 
Jochen Liske,$^{8}$ \newauthor  
Jon Loveday,$^{9}$ 
Peder Norberg,$^{10}$ 
J.A.~Peacock,$^{10}$  
Steven P.~Bamford,$^{7}$ 
Sarah \newauthor Brough,$^{3}$
Matthew Colless,$^{3}$
Ewan Cameron,$^{11}$ 
Christopher J.~Conselice,$^{7}$
Scott M. \newauthor Croom,$^{1}$
C.S.~Frenk,$^{12}$
Madusha Gunawardhana,$^{1}$
Konrad Kuijken,$^{13}$
R.C.~Nichol,$^{14}$  \newauthor
H.R.~Parkinson,$^{10}$
S.~Phillipps,$^{15}$ 
K.A.~Pimbblet,$^{5}$ 
C.C.~Popescu,$^{16}$  
Matthew \newauthor Prescott,$^{4}$ 
W.J.~Sutherland,$^{18}$ 
R.J.~Tuffs,$^{19}$
Eelco~van~Kampen,$^{8}$ 
D.~Wijesinghe$^{1}$\\
$^{1}$Sydney Institute for Astronomy, School of Physics, University of
Sydney, NSW 2006, Australia \\
$^{2}$School of Physics, the University of Melbourne, Parkville,
  VIC 3010, Australia \\
$^{3}$Australian Astronomical Observatory, PO Box 296, Epping, NSW 1710, Australia \\
$^{4}$Astrophysics Research Institute, Liverpool John Moores University,
Twelve Quays House, Egerton Wharf, Birkenhead, CH41 1LD, UK \\
$^{5}$School of Physics, Monash University, Clayton, Victoria 3800, Australia \\
$^{6}$School of Physics \& Astronomy, University of St Andrews, North
Haugh, St Andrews, KY16 9SS, UK \\
$^{7}$Centre for Astronomy and Particle Theory, University of
Nottingham, University Park, Nottingham NG7 2RD, UK \\
$^{8}$European Southern Observatory, Karl-Schwarzschild-Str.~2, 85748
Garching, Germany \\
$^{9}$Astronomy Centre, University of Sussex, Falmer, Brighton BN1 9QH, UK \\
$^{10}$Institute for Astronomy, University of Edinburgh, Royal
Observatory, Blackford Hill, Edinburgh EH9 3HJ, UK \\
$^{11}$Department of Physics, Swiss Federal Institute of Technology
(ETH-Z{\" u}rich), 8093 Z{\" u}rich, Switzerland \\
$^{12}$Institute for Computational Cosmology, Department of Physics,
Durham University, Durham DH1 3LE, UK \\
$^{13}$Leiden Observatory, Leiden University, P.O.~Box 9500, 2300 RA Leiden, The Netherlands \\
$^{14}$Institute of Cosmology and Gravitation (ICG), University of
Portsmouth, Portsmouth PO1 3FX, UK \\
$^{15}$Astrophysics Group, HH Wills Physics Laboratory,
University of Bristol, Tyndall Avenue, Bristol BS8 1TL \\
$^{16}$Jeremiah Horrocks Institute, University of Central Lancashire,
Preston PR1 2HE, UK \\
$^{17}$Research School of Astronomy \& Astrophysics, Mount Stromlo Observatory, Weston Creek, ACT 2611, Australia \\
$^{18}$Astronomy Unit, Queen Mary University London, Mile End Rd, London
E1 4NS, UK \\
$^{19}$Max Planck Institute for Nuclear Physics (MPIK), Saupfercheckweg
1, 69117 Heidelberg, Germany}

\maketitle \begin{abstract} This paper describes the first catalogue of
photometrically-derived stellar mass estimates for intermediate-redshift ($z <
0.65$; median $z = 0.2$) galaxies in the Galaxy And Mass Assembly (GAMA)
spectroscopic redshift survey. These masses, as well as the full set of
ancillary stellar population parameters, will be made public as part of GAMA
data release 2. Although the GAMA database does include NIR photometry, we
show that the quality of our stellar population synthesis fits is
significantly poorer when these NIR data are included. Further, for a large
fraction of galaxies, the stellar population parameters inferred from the
optical--plus--NIR photometry are formally inconsistent with those inferred
from the optical data alone. This may indicate problems in our stellar
population library, or NIR data issues, or both; hese issues will be addressed
for future versions of the catalogue. For now, we have chosen to base our
stellar mass estimates on optical photometry only. In light of our decision to
ignore the available NIR data, we examine how well stellar mass can be
constrained based on optical data alone. We use generic properties of stellar
population synthesis models to demonstrate that restframe colour alone is in
principle a very good estimator of stellar mass--to--light ratio, $M*/L_i$.
Further, we use the observed relation between restframe $(g-i)$ and $M_*/L_i$
for real GAMA galaxies to argue that, modulo uncertainties in the stellar
evolution models themselves, $(g-i)$ colour can in practice be used to
estimate $M_*/L_i$ to an accuracy of $\lesssim 0.1$ dex (1$\sigma$). This
`empirically calibrated' $(g-i)$--$M_*/L_i$ relation offers a simple and
transparent means for estimating galaxies' stellar masses based on minimal
data, and so provides a solid basis for other surveys to compare their results
to $z \lesssim 0.4$ measurements from GAMA. \end{abstract}



%

\section{Introduction} 

One of the major difficulties in observationally constraining the formation
and evolutionary histories of galaxies is that there is no good observational
tracer of formation time or age. In the simplest possible terms, galaxies grow
through a combination of continuous and/or stochastic star formation and
episodic mergers. Throughout this process---and in contrast to other global
properties like luminosity, star formation rate, restframe colour, or
luminosity-weighted mean stellar age---a galaxy's evolution in stellar mass is
nearly monotonic and relatively slow. Stellar mass thus provides a good,
practical basis for evolutionary studies.

Further, it is now clear that stellar mass plays a central role in
determining---or at least describing---a galaxy's evolutionary state.
Virtually all of the global properties commonly used to describe
galaxies---\eg, luminosity, restframe colour, size, structure, star formation
rate, mean stellar age, metallicity, local density, and velocity dispersion or
rotation velocity---are strongly and tightly correlated \citep[see,
\eg,][]{Minkowski1962, FaberJackson1976, TullyFisher1977,
SandageVisvanathan1978, Dressler1980, FP1, FP2, Strateva2001}. One of most
influential insights to come from the ambitious wide- and deep-field galaxy
censuses of the 2000s has been the idea that most, if not all, of these
correlations can be best understood as being primarily a sequence in stellar
mass \citep[\eg][]{Shen2003, Kauffmann2003b, Kauffmann2004, Tremonti2004,
Blanton2005, Baldry2006, Gallazzi2006}. Given a galaxy's stellar mass, it is
thus possible to predict most other global properties with considerable
accuracy. Presumably, key information about the physical processes that govern
the process of galaxy formation and evolution are encoded in the forms of, and
scatter around, these stellar mass scaling relations. \looseness-1

\subsection{Galaxy And Mass Assembly (GAMA)}

This paper presents the first catalogue of stellar mass estimates for galaxies
in the Galaxy And Mass Assembly (GAMA) survey \citep{Driver2009, Driver2011}.
At its core, GAMA is an optical spectroscopic redshift survey, specifically
designed to have near total spectroscopic completeness over cosmologically
representative volume. In terms of survey area and target surface density,
GAMA is intermediate and complementary to wide-field, low-redshift galaxy
censuses like SDSS \citep{York2000,Strauss2002,sdssdr7}, 2dFGRS
\citep{Colless2001, Colless2003,Cole2005}, 6dFGS \citep{Jones2004,Jones2009},
or the MGC \citep{Liske2003,Driver2005} and deep-field surveys of the high
redshift universe like VVDS \citep{LeFevre2005}, DEEP-2 \citep{Davis2003},
COMBO-17 \citep{Wolf2003, Wolf2004}, COSMOS and zCOSMOS \citep{Scoville2007,
Lilly2007}. The intermediate redshift regime ($z \lesssim 0.3$) that GAMA
probes is thus largely unexplored territory: GAMA provides a unique resource
for studies of the evolving properties of the general galaxy population.

In a broader sense, GAMA aims to unite data from a number of large survey
projects spanning nearly the full range of the electromagnetic spectrum, and
using many of the world's best telescopes. At present, the photometric
backbone of the dataset is optical imaging from SDSS and near infrared (NIR)
imaging taken as part of the Large Area Survey (LAS) component of the UKIRT
Infrared Deep Sky Survey \citep[UKIDSS;][]{Dye2006, Lawrence2007}. GALEX UV
imaging from the Medium Imaging Survey \citep[MIS;][]{Martin2005,
Morrissey2007} is available for the full GAMA survey region. At longer
wavelengths, mid-infrared imaging is available from the WISE all-sky survey
\citep{Wright2010}, far infrared imaging is available from the Herschel-ATLAS
project \citep{Eales2010}, and metre-wavelength radio imaging is being
obtained using the Giant Metre-wave Radio Telescope (GMRT; PI: M.\ Jarvis). In
the near future, the SDSS and UKIDSS imaging will be superseded by
significantly deeper, sub-arcsecond resolution imaging from the VST-KIDS
project (PI: K.\ Kuijken) and from the VISTA-VIKING survey (PI: W.\
Sutherland). Looking slightly further ahead, a subset of the GAMA fields will
also be targeted by the ASKAP-DINGO project (PI: M.\ Meyer), adding 21 cm data
to the mix. By combining these many different datasets into a single and truly
panchromatic database, GAMA aims to construct `the ultimate galaxy catalogue',
offering the first laboratory for simultaneously studying the AGN, stellar,
dust, and gas components of large and representative samples of galaxies at
low-to-intermediate redshifts.

The stellar masses estimates, as well as estimates for ancillary stellar
popualtion parameters like age, metallicity, and restframe colour, form a
crucial part of the GAMA value-added dataset. These values are already in use
within the GAMA team for a number of science applications.. In keeping with
GAMA's commitment to providing these data as a useful and freely available
resource, the stellar mass estimates described in this paper are being made
publicly available as part of the GAMA data release 2, scheduled for mid-2011.
Particularly in concert with other GAMA value-added catalogues, and with
catalogues from other wide- and deep-field galaxy surveys, the GAMA stellar
mass estimates are intended to provide a valuable public resource for studies
of galaxy formation and evolution. A primary goal of this paper is therefore
to provide a standard reference for users of these catalogues.

\subsection{Stellar mass estimation} \label{ch:intro}

Stellar mass estimates are generally derived through stellar population
synthesis (SPS) modelling \citep{TinsleyGunn1976, Tinsley1978, Bruzual1993}.
This technique relies on stellar evolution models \citep[\eg,][]{Starburst99,
pegase, BC03, M05, batsi}. Assuming a stellar initial mass function (IMF),
these models describe the spectral evolution of a single-aged or simple
stellar population (SSP) as a function of its age and metallicity. The idea
behind SPS modelling is to combine the individual SSP models according to some
fiducial star formation history (SFH), and so to construct composite stellar
populations (CSPs) that match the observed properties of real galaxies. The
stellar population (SP) parameters---including stellar mass, star formation
rate, luminosity weighted mean stellar age and metallicity, and dust
obscuration---implied by such a fit can then be ascribed to the galaxy in
question \citep[see, \eg][]{BrinchmannEllis, Cole2001, Bell2003,
Kauffmann2003a, Gallazzi2005}.

SPS fitting is most commonly done using broadband spectral energy
distributions (SEDs) or spectral indices \citep[see the comprehensive review
by][]{Walcher2011}. This presents two interrelated challenges. First is the
question of the accuracy and reliability of the spectral models that make up
the stellar population library (SPL) used as the basis of the fitting,
including both the stellar evolution models that underpin the synthetic
spectra, and the SFHs used to construct the SPL. Second, there is the question
of what SED or spectral features provide the strongest and/or most robust
constraints on a galaxy's SP, taking into account the uncertainties and
assumptions intrinsic to the models. \looseness-1

In principle, the accuracy of SPS-derived parameter estimates is limited by
generic degeneracies between different SP models with the same or similar
observable properties---for example, the well known dust--age--metallicity
degeneracy \citep[see, \eg ,][]{Worthey}. Further, the SPS fitting problem is
typically badly under-constrained, inasmuch as it is extremely difficult to
place meaningful constraints on a given galaxy's particular SFH. This issue
has been recently explored by \citet{GallazziBell}, who tested their ability
to recover the known SP parameters of mock galaxies, in order to determine the
limiting accuracy of stellar mass estimates. In the highly idealised case that
the SPL contains a perfect description of each and every galaxy, and that the
photometry is perfectly calibrated, and that the dust extinction is known
exactly, \citet{GallazziBell} argue that SP model degeneracies mean that both
spectroscopic and photometric stellar mass estimates are generically limited
to an accuracy of $\lesssim 0.2$ dex for galaxies with a strong burst
component, and $\sim 0.10$ dex otherwise.

In practice, the dominant uncertainties in SPS-derived parameter estimates are
likely to come from uncertainties inherent to the SP models themselves.
Despite the considerable progress that has been made, there remain a number of
important `known unknowns'. The form and universality (or otherwise) of the
stellar IMF is a major source of uncertainty \citep{Wilkins2008,
vanDokkum2008, Madusha}. From the stellar evolution side, the treatment of
NIR-luminous thermally pulsating asymptotic giant branch stars
\citep[TP-AGBs;][]{M05, Maraston2006, Kriek2010} is the subject of some
controversy. As a third example, there is the question of how to appropriately
model the effects of dust in the interstellar medium (ISM), including both the
form of the dust obscuration/extinction law, and the precise geometry of the
dust with respect to the stars \citep{Driver2007, Wuyts2009, Dinuka}. Many of
these uncertainties and their propagation through to stellar mass
estimates are throughly explored and quantified in the excellent work of
\citet{Conroy2009, Conroy2010}, who argue that (when fitting to full
UV--to--NIR SEDs) the net uncertainty in any individual $z \sim 0$ stellar
mass determination is on the order of $\lesssim 0.3$ dex.

Differential systematic errors across galaxy populations---that is, biases in
the stellar masses of different galaxies as a function of mass, age, SFH,
etc.---are at least as great a concern as the net uncertainty on any
individual galaxy. The vast majority of stellar mass-based science focusses on
differences in the (average) properties of galaxies as a function of inferred
mass. In such comparative studies, differential biases have the potential to
induce a spurious signal, or, conversely, to mask true signal. In this
context, \citet{Taylor2010b} have used the consistency between stellar and
dynamical mass estimates for SDSS galaxies to argue that any such differential
biases in $M_*/L_i$ (\cf\ $M_*$) as a function of stellar population are
limited to $\lesssim 0.12$ dex (40 \%); \ie, small.

In a similar way, systematic differential biases in the masses and SP
parameters of galaxies at different redshifts are a major concern for
evolutionary studies, inasmuch as any such redshift-dependent biases will
induce a false evolutionary signal. Indeed, for the specific example of
measurement of the evolving comoving number density of massive galaxies at $z
\lesssim 2$, such differential errors are the single largest source of
uncertainty, random or systematic \citep{Taylor2009}. More generally, such
differential biases will be generically important whenever the low redshift
point makes a significant contribution to the evolutionary signal; that is,
whenever the amount of evolution is comparable to the random errors on the
high redshift points. In this context, by probing the intermediate redshift
regime and providing a link between $z \approx 0$ surveys like SDSS and 2dFGRS
and $z \gg 0$ deep surveys like VVDS and DEEP-2, GAMA makes it possible to
identify and correct for any such differential effects. GAMA thus has the
potential to significantly reduce or even eliminate a major source of
uncertainty for a wide variety of lookback survey results. \looseness-1

\subsection{This work}

Before we begin, a few words on the ethos behind our SPS modelling procedure:
we have deliberately set out to do things as simply and as conventionally as
is possible and appropriate. There are two main reasons for this decision.
First, this is only the first generation of stellar mass estimates for GAMA.
We intend to use the results presented here to inform and guide future
improvements and refinements to our SPS fitting algorithm. Second, in the
context of studying galaxy evolution, GAMA's unique contribution is to probe
the intermediate redshift regime; GAMA becomes most powerful when combined
with very wide low redshift galaxy censuses on the one hand, and with very
deep lookback surveys on the other. To maximise GAMA's utility, it is
therefore highly desirable to provide masses that are directly comparable to
estimates used by other survey teams. This includes using techniques that are
practicable for high redshift studies.

With all of the above as background, the programme for this paper is as
follows. After describing the subset of the GAMA database that we will make
use of in \secref{ch:data}, we lay out our SPS modelling procedure in
\secref{ch:estimates}. In particular, in \secref{ch:diagnostics}, we show the
importance of taking a Bayesian approach to SP parameter estimation.

In \secref{ch:nir}, we look at how our results change with the inclusion of
NIR data. Specifically, in \secref{ch:residuals}, we show that our SPL models
do not yield a good description of the GAMA optical--to--NIR SED shapes.
Further, in \secref{ch:nirvalues}, we show that for a large fraction of
galaxies, the SP parameter values derived from the full optical--plus--NIR
SEDs are formally inconsistent with those derived from just the optical data.
Both of these statements are true irrespective of the choice of SSP models
used to construct the SPL (\secref{ch:sspmodels}).

In order to interpret the results presented in \secref{ch:nir}, we have
conducted a set of numerical experiments designed to test our ability to fit
synthetic galaxies photometry, and to recover the `known' SP parameters of
mock galaxies. Based on these tests, which we describe in Appendix
\ref{ch:mocks}, we have no reason to expect the kinds of differences found in
\secref{ch:nir}---we therefore conclude that, at least for the time being, it
is better for us to ignore the available NIR data \secref{ch:nonir}.

In light of our decision not to use the available NIR data, in
\secref{ch:gimoverl}, we investigate how well optical data can be used to
constrain a galaxy's $M_*/L$. Using the SPL models, we show in
\secref{ch:gimodels} that, in principle, $(g-i)$ colour can be used to
estimate $M_*/L_i$ to within a factor of $\lesssim 2$. In
\secref{ch:colourrel}, we use the empirical relation between ($ugriz$-derived)
$M_*/L_i$ and $(g-i)$ colour to show that, in practice, $(g-i)$ can be used to
infer $M_*/L_i$ to an accuracy of $\approx 0.1$ dex. The derived colour--$M/L$
relation presented in this section is provided to enable meaningful comparison
between stellar mass-centric measurements from GAMA and other surveys.

Finally, in \secref{ch:discuss}, we discuss how we might improve on the
current SP parameter estimates for future catalogues. In particular, in
\secref{ch:future}, we examine potential causes and solutions for our current
problems in incorporating the NIR data. In this section, we suggest that we
have reached the practical limit for SP parameter estimation based on
grid-search-like algorithms using a static SPL. In order to improve on the
current estimates, future efforts will require a fundamentally different
conceptual approach. However, as we argue in \secref{ch:nirvalue}, this will
not necessarily lead to significant improvements in the robustness or
reliability of our stellar mass estimates.

Separately, we compare the SDSS and GAMA photometry and stellar mass estimates
in Appendix \ref{ch:cfsdss}. Despite there being large and systematic
differences between the SDSS \model\ and GAMA \auto\ SEDs, we find that the
GAMA- and SDSS-derived $M_*/L$s are in excellent agreement. On the other hand,
we also show that, as a measure of total flux, the SDSS \model\ photometry
suffers from structure-dependent biases; the differential effect is at the
level of a factor of 2. These large and systematic biases in total flux
translate directly to biases in the inferred total mass. For SDSS, this may in
fact be the single largest source of uncertainties in their stellar mass
estimates. In principle, this will have a significant impact on stellar
mass-centric measurements based on SDSS data.

Throughout this work, we adopt the concordance cosmology: $(\Omega_\Lambda,~
\Omega_m,~ h) = (0.7,~ 0.3,~ 0.7)$. Different choices for the value of $h$ can
be accommodated by scaling any and all absolute magnitudes or total stellar
masses by $+5 \log h/0.7$ or $-2 \log h/0.7$, respectively (\ie, a higher
value of the Hubble parameter implies a lower luminosity or total mass). All
other SP parameters, including restframe colours, ages, dust extinctions, and
mass-to-light ratios can be taken to be cosmology-independent, inasmuch as
they pertain to the stellar populations at the time of observation. We assume
a \citet{Chabrier} IMF. Our stellar mass estimates are based on the
\citet{BC03} SP models and the dust obscuration law of \citet{Calzetti}. We
briefly consider the effect of using the \citet{M05} or \citet{CB07} SP models
on the inferred SP parameter estimates in \secref{ch:sspmodels}. In
discussions of stellar mass-to-light ratios, we use $M_*/L_X$ to denote the
ratio between stellar mass and luminosity in the restframe $X$-band; where the
discussion is generic to all (optical and NIR) bands, we will drop the
subscript for convenience. In all cases, the $L_X$ in $M_*/L_X$ should be
understood as referring to the absolute luminosity of the galaxy, \ie, without
correction for internal dust extinction. We thus consider effective, and not
intrinsic stellar mass-to-light ratios. Unless explicitly stated otherwise,
quantitative values of $M_*/L_X$s are given using units of $L_X$ equivalent to
an AB magnitude of 0 (rather than, say, $L_{\odot, X}$). All quoted magnitudes
use the AB system.

\section{Data} \label{ch:data}

\subsection{Spectroscopic redshifts} \label{ch:specdata}

The lynchpin of the GAMA dataset is a galaxy redshift survey targeting three
$4^\circ \times 12^\circ$ equatorial fields centred on $9^\mathrm{h}
00^\mathrm{m} +1^\mathrm{d}$, $12^\mathrm{h} 00^\mathrm{m} +0^\mathrm{d}$, and
$14^\mathrm{h} 30^\mathrm{m} +0^\mathrm{d}$ (dubbed G09, G12, and G15,
respectively), for an effective survey area of 144 $\square^\circ$. Spectra
were taken using the AAOmega spectrograph \citep{SaundersEtAl, SharpEtAl},
which is fed by the 2dF fibre positioning system on the 4m Anglo-Australian
Telescope (AAT). The algorithm for allocating 2dF fibres to survey targets,
described by \citet{Robotham2010} and implemented for the second and third
years of observing, was specifically designed to optimise the spatial
completeness of the final catalogue. Observations were made using AAOmega's
580V and 385R gratings, yielding continuous spectra over the range 3720--8850
\AA\ with an effective resolving power of $R \approx 1300$. Observations for
the first phase of the GAMA project, GAMA I, have recently been completed in a
68 night campaign spanning 2008--2010. GAMA has just been awarded AAT long
term survey status with a view to trebling its survey volume; observations for
GAMA II are underway, and will be completed in 2012.

Target selection for GAMA I has been done on the basis of optical imaging from
SDSS \citep[DR6;][]{sdssdr6} and NIR imaging from UKIRT, taken as part of the
UKIDSS LAS \citep{Dye2006, Lawrence2007}. The target selection is described in
full by \citet{Baldry2010}. In brief, the GAMA spectroscopic sample is
primarily selected on $r$-band magnitude, using the (Galactic/foreground
extinction-corrected) \petro\ magnitudes given in the basic SDSS catalogue.
The main sample is magnitude-limited to $r_\mathtt{petro} < 19.4$ in the
G09/G15 fields, and $r_\mathtt{petro} < 19.8$ in G12. (The definitions of the
SDSS \petro\ and \model\ magnitudes can be found in \secref{ch:sdssphot}.) In
order to increase the stellar mass completeness of the sample, there are two
additional selections: $z_\mathtt{model} < 18.2$ or $K_\mathtt{auto} < 17.6$
(AB). For these two additional selections, in order to ensure both photometric
reliability and a reasonable redshift success rate, it is also required that
the $r_\mathtt{model} < 20.5$. The effect of these additional selections is to
increase the target density marginally by $\sim 7$ \% (1 \%) in the G09/G15
(G12) fields. Star--galaxy separation is done based on the observed shape in a
similar manner as for the SDSS \citep[see][for
details]{Baldry2010,Strauss2002}, with an additional $(J-K)$--$(g-i)$ colour
selection designed to exclude those double/blended stars that still fall on
the stellar locus in colour--colour space.

To these limits, the survey spectroscopic completeness is high \citep[$\gtrsim
98$ \%; see][]{Driver2011, Liske2011}. The issue of photometric incompleteness
in the target selection catalogues has been investigated by
\citet{Loveday2010} using SDSS Stripe 82: the SDSS imaging completeness is $>
99$ (90) \% for $\mu < 22.5$ (23) mag/$\square''$.

The process for the reduction and analysis of the AAOmega spectra is described
in \citet{Driver2011}. All redshifts have been measured by GAMA team members
at the telescope, using the interactive redshifting software \texttt{RUNZ}
(developed by Will Sutherland and now maintained by Scott Croom). For each
reduced and sky-subtracted spectrum, \texttt{RUNZ} presents the user with a
first redshift estimate. Users are then free to change the redshift in the
case that the \texttt{RUNZ}-derived redshift is deemed incorrect, and are
always required to give a subjective figure of merit for the final redshift
determination.

To ensure the uniformity and reliability of both the redshifts and the quality
flags, a large subset (approximately 1/3, including all those with redshifts
deemed `maybe' or `probably' correct) of the GAMA spectra have been
independently `re-redshifted' by multiple team members. The results of the
blind re-redshifting are used to derive a probability for each redshift
determination, $p_z$, which also accounts for the reliability of the
individual who actually determined the redshift \citep{Liske2011}. The final
values of the redshifts and quality flags, \texttt{nQ}, given in the GAMA
catalogues are then based on these `normalised' probabilities. (Note that this
work makes use of `year 3' redshifts, which have yet not undergone the
re-redshifting process.) \citet{Driver2011} suggests that the redshift
`blunder' rate for galaxies with $\mathtt{nQ} = 3$ (corresponding to $0.90 <
p_z < 0.95$) is in the range 5--15 \%, and that for $\mathtt{nQ} = 4$
(corresponding to $p_z > 0.95$) is 3--5 \%. A more complete analysis of the
GAMA redshift reliability will be provided by \citet{Liske2011}.

The redshifts derived from the spectra are, naturally, heliocentric. For the
purposes of calculating luminosity distances (see \secref{ch:sedfit}), we have
computed flow-corrected redshifts using the model of \citet{Tonry2000}. The
details of this conversion will be given by \citet{Baldry2011}.

The GAMA I main galaxy sample ($\mathtt{SURVEY\_CLASS} \ge 4$ in the GAMA
catalogues) comprises 119852 spectroscopic targets, of which 94.5 \%
(113267/119852) now have reliable ($\mathtt{nQ} \ge 3$) spectroscopic
redshifts. Of the reliable redshifts, 83 \% (94448/113267) are measurements
obtained by GAMA. The remainder are taken from previous redshift surveys,
principally SDSS \citep[DR7][13137 redshifts]{sdssdr7}, 2dFGRS \citep[][3622
redshifts]{Colless2003}, and MGCz \citep[][1647 redshifts]{Driver2005}. As a
function of SDSS \texttt{fiber} magnitude (taken as a proxy for the flux seen
by the $2''$ 2dF spectroscopic fibres), the GAMA redshift success rate
($\mathtt{nQ} \ge 3$) is essentially 100 \% for $r_\mathtt{fiber} < 19.5$,
dropping to 98 \% for $r_\mathtt{fiber} = 20$, and then down to $\sim 50$ \%
for $r_\mathtt{fiber} = 22$ \citep{Loveday2010}. For the $r$-selected survey
sample ($\mathtt{SURVEY\_CLASS} \ge 6$), the net redshift success rate is 95.4
\% (109222/114250).

Stellar mass estimates have been derived for all objects with a spectroscopic
redshift $0 < z \le 0.65$. For the purposes of this work, we will restrict
ourselves to considering only those galaxies with $z > 0.002$ (to exclude
stars), and those galaxies with $\mathtt{nQ} \ge 3$ (to exclude potentially
suspect redshift determinations). We quantify the sample completeness in terms
of stellar mass, restframe colour, and redshift in \secref{ch:zmax}.


\subsection{Broadband Spectral Energy Distributions (SEDs)} \label{ch:photdata}

This work is based on version 6 of the GAMA master catalogue (internal
designation \texttt{catgama\_v6}), which contains $ugrizYJHK$ photometry for
galaxies in the GAMA regions. The photometry is based on SDSS (DR7) optical
imaging, and UKIDSS LAS (DR4) NIR imaging. The SDSS data have been taken from
the Data Archive Server (DAS); the UKIDSS data have been taken from the WFCAM
Science Archive \citep[WSA;][]{Hambly2008}. 

In each case, the imaging data are publicly available in a fully reduced and
calibrated form. The SDSS data reduction has been extensively described
\citep[see, \eg][]{Strauss2002,sdssdr7}. The LAS data have been reduced using
the WFCAM-specific pipeline developed and maintained by the Cambridge
Astronomical Survey Unit (CASU).\footnote{Online documentation available via
\texttt{http://casu.ast.cam.ac.uk/surveys-projects/wfcam}.}

The GAMA photometric catalogue is constructed from an independent reanalysis
of these imaging data. The data and the GAMA reanalysis of
them are described fully by \citet{Hill2010} and \citet{Kelvin2010}. We
summarise the most salient aspects of the GAMA photometric pipeline below. As
described in \citet{Hill2010}, the data in each band are normalised and
combined into three astrometrically matched Gigapixel-scale mosaics (one for
each of the G09, G12, and G15 fields), each with a scale of $0\farcs 4$
pix$^{-1}$. In the process of the mosaicking, individual frames are degraded
to a common seeing of 2$''$ FWHM.

Photometry is done on these PSF-matched images using SExtractor
\citep{SExtractor} in dual image mode, using the $r$-band image as the
detection image. For this work, we construct multicolour SEDs using
SExtractor's \auto\ photometry. This is a flexible, elliptical aperture whose
size is determined from the observed light distribution within a
quasi-isophotal region \citep[see][for further explanation]{SExtractor,Kron}
of the $r$-band detection image. This provides seeing- and aperture-matched
photometry in all bands.

In addition to the matched-aperture photometry, the GAMA catalogue also
contains $r$-band \Sersic -fit structural parameters, including total
magnitudes, effective radii, and \Sersic\ indices \citep{Kelvin2010}. These
values have been derived using \texttt{GALFIT3} \citep{galfit} applied to
(undegraded) mosaics constructed in the same manner as those described above.
These fits incorporate a model of the PSF for each image, and so should be
understood to be seeing corrected. In estimating total magnitudes, the
\Sersic\ models have been truncated at 10 $R\eff$; this typically corresponds
to a surface brightness of $\mu_r \sim 30$ mag / $\square''$. \citet{Hill2010}
present a series of detailed comparisons between the different GAMA and
SDSS/UKIDDS photometric measures. Additional comparisons between the GAMA and
SDSS optical photometry are presented in Appendix \ref{ch:cfsdss}. In this
work, we use these $r$-band \sersictt\ magnitudes to estimate galaxies' total
luminosities, since these measurements (attempt to) account for flux missed by
the finite \auto\ apertures.

For each galaxy, we construct multicolour SEDs using the SExtractor \auto\
aperture photometry. Formally, when fitting to these SEDs, we are deriving SP
parameters integrated or averaged over the projected \auto\ aperture. In order
to get an estimate of a galaxy's total stellar mass, it is therefore necessary
to scale the inferred mass up, so as to account for flux/mass lying beyond the
(finite) \auto\ aperture. We do this by simply scaling each of the \auto\
fluxes by the amount required to match the $r$-band \auto\ aperture flux to
the \sersictt\ measure of total flux; \ie, using the scalar aperture
correction factor $f_\mathrm{ap} = 10^{-0.4 (r_\mathtt{auto} -
r_\mathtt{sersic})}$.

Note that we elect not to use the NIR data to derive stellar mass estimates
for the current generation of the GAMA stellar mass catalogue. Our reasons for
this decision are the subject of \secref{ch:nir}.

\section{Stellar Population Synthesis (SPS) modelling and stellar mass
estimation} \label{ch:estimates}

\begin{figure*} \centering \includegraphics[width=16.8cm]{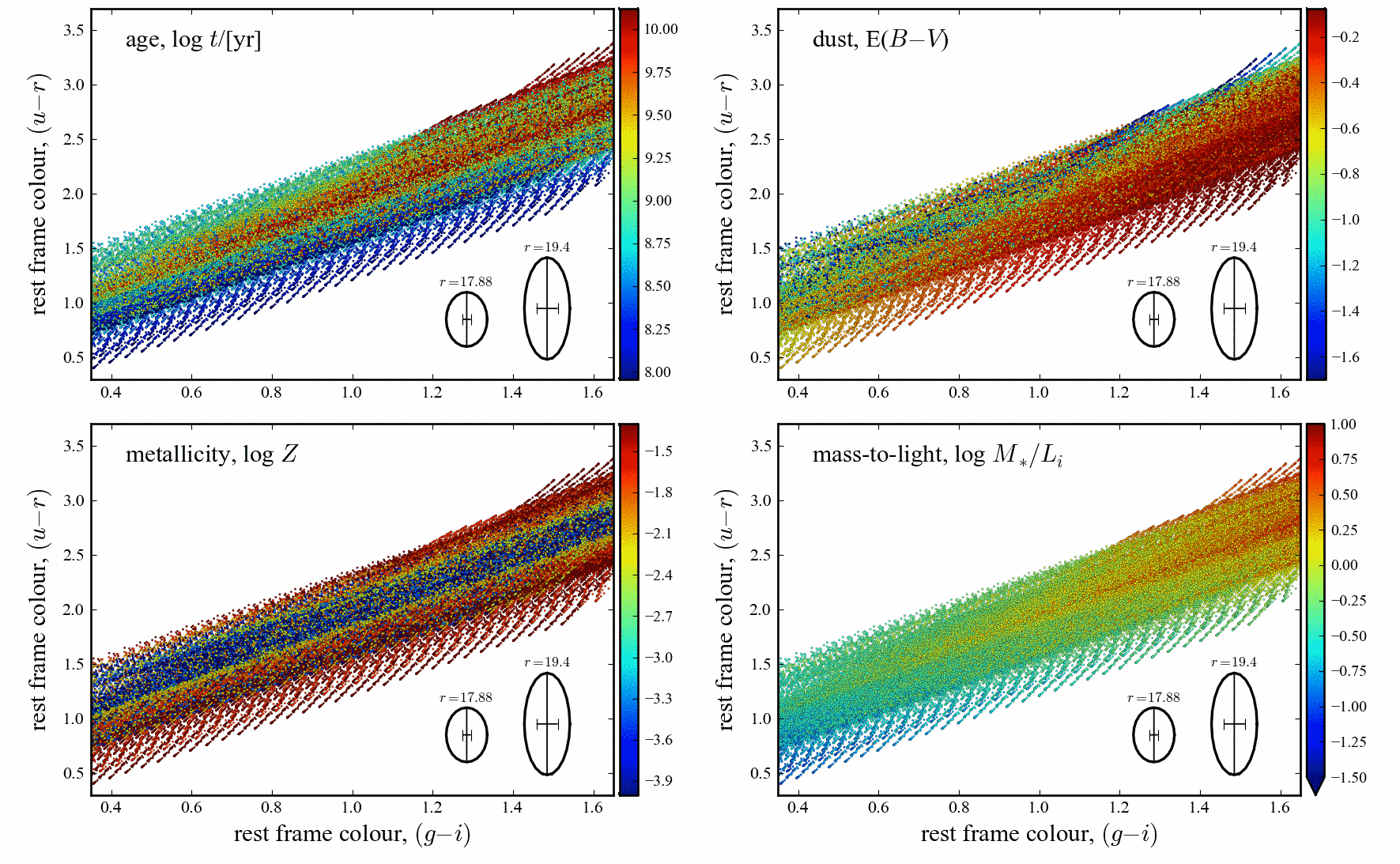}
\caption{Illustrating the basic idea behind stellar population synthesis
modelling.--- In each panel of this Figure, we plot the restframe $(u-r)$ and
$(g-i)$ colours of the models in our library, colour-coded by an important
basic property: CSP age, $t$, dust obscuration, $\dust$, stellar metallicity,
$Z$, and mass-to-light ratio, $M_*/L_i$. In the simplest terms possible, given
a galaxy's restframe $(ugri)$ photometry, the process of stellar population
parameter estimation can be thought of as just `reading off' the value of each
parameter. For comparison, the the median 1$\sigma$ uncertainty in the
observer's frame $ugri$ colours for GAMA galaxies at the SDSS and GAMA
spectroscopic selection limits ($m_r = 17.88$ and $19.4$, respectively) are
shown in each panel. Note that, as is standard practice, we impose an error
floor of 0.05 mag in the photometry in each band; the ellipses show the
effective uncertainties with the inclusion of this error floor. With the
exception of the $u$ band, the catalogued error is almost always less than
0.05 mag; these errors are shown as the error bars within each ellipse. We are
limited not by random noise, but by the systematic errors in the relative or
cross-calibration of the photometry in the different bands. In each panel,
areas where one colour dominates shows where a given parameter can be well
constrained using $ugri$ photometry. Conversely, where models with similar
$ugri$ colours have a wide range of parameter values, this parameter cannot be
well constrained. Thus it can be seen that even though $t$, $\dust$, and
particularly $Z$ are generally not well constrained from an optical SED,
$M_*/L$ can still be relatively robustly estimated.\label{fig:ugri}}
\end{figure*}

The essential idea behind SPS modelling is to determine the characteristics of
the SPs that best reproduce the observed properties (in our case, the
broadband spectral energy distribution; SED) of the galaxy in question. As an
illustrative introduction to the problem, \textbf{Figure \ref{fig:ugri}} shows
the distribution of our SPS model templates in restframe $(u-r)$--$(g-i)$
colour--colour space. In each panel of this Figure, we colour-code each model
according to a different SP parameter.

Imagine for a moment that instead of using the observed $ugriz$ SEDs, we were
to first transform those SEDs into restframe $ugri$ photometry, and then use
this as the basis of the SPS fitting. In the simplest possible terms, the
fitting procedure could then be thought of as `reading off' the parameters of
the model(s) found in the region of the $ugri$ colour--colour space inhabited
by the galaxy. 

In this Figure, regions that are dominated by a single colour show where a
parameter can be tightly constrained on the basis of a (restframe) $ugri$
SED.\footnote{When constructing each panel in Figure \ref{fig:ugri}, we have
deliberately plotted the models in a random order, rather than, say, ranked by
age or metallicity. This ensures that the mix of colour-coded points fairly
represents the mix of model properties in any given region of colour-colour
space.} Conversely, regions where the different colours are well mixed show
where models with a wide range of parameter values provide equally good
descriptions of a given $ugri$ SED shape; that is, where there are strong
degeneracies between model parameters.

In general terms, then, Figure \ref{fig:ugri} demonstrates that it is
difficult to derive strong constraints on $t$ or $Z$; this is the well
known age--metallicity degeneracy.\footnote{In principle, and to foreshadow
the results shown in \secref{ch:moverls}, these degeneracies can be broken by
incorporating additional information. For example, if different models that
have similar $(g-i)$ and $(u-r)$ colours have very different optical-minus-NIR
colours, then the inclusion of NIR data can, at least in principle, lead to
much tighter constraints on the model parameters.} Even where such strong
degeneracies exist, however, note that the value of $M_*/L$ is considerably
better constrained than any of the parameters that are used to define each
model.

\subsection{Synthetic stellar population models} \label{ch:models}

The fiducial GAMA stellar mass estimates are based on the BC03 synthetic SP
model library, which consists of spectra for single-aged or simple stellar
populations (SSPs), parameterised by their age, $\age$, and
metallicity, $Z$; \ie, $f_\mathrm{SSP}(\lambda;~\age,~Z)$. Given these SSP spectra and an assumed star formation history (SFH), $\psi_*(\age)$, spectra for composite stellar populations (CSPs) can be constructed, as a linear combination of different simple SSP spectra;
\ie: \begin{align} 
\label{eq:csp} f_\mathrm{CSP}(\lambda; ~ Z, ~ t, ~ &\dust) = \nonumber \\
k(\lambda;~ &\dust) \int\limits_{t'=0}^{t} \mathrm{d}t' ~ \psi_*(t') \times
f_\mathrm{SSP}(\lambda; ~Z , ~t') ~ . \end{align} 
Here, $k(\lambda;~ \dust)$ is a single-screen dust attenuation law, where the
degree of attenuation is characterised by the selective extinction between the
$B$ and $V$ bands, $\dust$. Note that this formalism works for any quantity
that is additive; \eg\ flux in a given band, stellar mass (including
sub-luminous stars, and accounting for mass loss as a function of SSP age),
the mass contained in stellar remnants (including white dwarfs, black holes),
etc.

When using this Equation to construct the CSP models that comprise our SPL, we
make three simplifying assumptions. We consider only smooth,
exponentially-declining star formation histories, which parameterised by the
$e$-folding timescale, $\tau$; \ie, $\psi_*(t) \propto
e^{-t/\tau}$.\footnote{Whereas the integral in Equation\,(\ref{eq:csp}) is
continuous in time, each set of the SSP libraries that we consider contain
stellar population parameters for a set of discrete ages, $t_i$. In practice,
we compute the integral in Equation (\ref{eq:csp}) numerically, using a
trapezoidal integration scheme to determine the number of stars formed in the
time interval $\Delta t_i$ associated with the time $t_i$. This effectively
assumes that the spectral evolution at fixed $\lambda$ and $Z$ is
approximately linear between values of $t_i$. Note that this is something that
is not optimally implemented in the standard \texttt{galaxev} package
described by \citet{BC03}.} We make the common assumption that each CSP has a
single, uniform stellar metallicity, $Z$. We also make the (equally common)
assumption that a single dust obscuration correction can be used for the
entire CSP.

For our fiducial mass estimates, we use a \citet{Calzetti} dust attenuation
`law'. In this context, we highlight the work of \citet{Dinuka}, who look at
the consistency of different dust obscuration laws in the optical and
ultraviolet. They conclude that the \citet{FischeraDopita} dust curve is best
able to describe the optical--to--ultraviolet SED shapes of GAMA galaxies. In
the optical, the shapes of the \citet{FischeraDopita} and \citet{Calzetti}
curves are quite similar. Using the \citet{FischeraDopita} curve does not
significantly alter our results.



The models in our SPL are thus characterised by four key parameters: age,
$\age$; $e$-folding time, $\tau$; metallicity, $Z$; and dust obscuration,
$\dust$. In an attempt to cover the full range of possible stellar populations
found in real galaxies, we construct a library of CSP model spectra spanning a
semi-uniform grid in each parameter. The age grid spans the range $\log \age /
\left[\mathrm{yr} \right] = 8$--8.9 in steps of 0.1 dex, then from $\log \age
/ \left[\mathrm{yr} \right] = 9$--10.10 in steps of 0.05 dex, and then with a
final value of 10.13 ($\approx 13.4$ Gyr). The grid of $e$-folding times spans
the range $\log \tau / \left[\mathrm{yr} \right] = 7.5$--8.9 in steps of 0.2
dex, and then from $\log \tau / \left[\mathrm{yr} \right] = 9$--10 in steps of
0.1 dex. The dust grid covers the range $\dust = 0.0$--$0.8$ in steps of 0.02
mag. We use the native metallicity grid for the \citet{BC03} models: $Z =$
(0.0001, 0.0004, 0.004, 0.008, 0.02, 0.05). The fiducial model grid thus
includes $34 \times 19 \times 43 \times 6 = 166668$ models for each of 66
redshifts between $z = 0.00$ and $0.65$, for a total of just over 11 million
individual sets of 9 band synthetic photometry.

\subsection{SED fitting} \label{ch:sedfit}

Synthetic broadband photometry is derived using the CSP spectra and a model
for the total instrumental response for each of the $ugriz$- and $YJHK$-bands.
The optical and NIR filter response functions are taken from \citet{Doi2010}
and from \citet{Hewett2006}, respectively. These curves account for
atmospheric transmission (assuming an airmass of 1.3), filter transmission,
mirror reflectance, and detector efficiency, all as a function of wavelength.
For a given a template spectrum $f_\mathrm{CSP}$, placed at redshift $z$, the
template flux in the (observers' frame) $X$-band, $T_X$, is then given by:
\begin{equation} \label{eq:flux} T_X(\mathrm{CSP}; z) = (1+z) ~ \frac{\int
\mathrm{d}{\lambda}~r_X(\lambda)~\lambda ~
f_\mathrm{CSP}(\frac{\lambda}{1+z})} {\int \mathrm{d}{\lambda}
~r_X(\lambda)~\lambda }~ . \end{equation} Here, $r_X$ is filter response
function, and the prefactor of $(1+z)$ accounts for the redshift-stretching of
the bandpass interval. Also note that the factor of $\lambda$ in both
integrals is required to account for the fact that broadband detectors count
{\em photons}, not {\em energy} \citep[see, \eg,][]{Hogg2002, eazy}: $T_X$
thus has units of counts m$^{-2}$ s$^{-1}$.

By construction, each of the template spectra in our library is normalised to
a total, time-integrated SFH (\cf\ instantaneous mass) of 1 M\sol\ observed
from a distance of 10\,pc. A normalisation factor, $A_T$, is thus required to
scale the apparent flux of the base template to match the data, accounting for
both the total stellar mass/luminosity and distance-dependent dimming. It is
thus through determining the value of $A_T$ that we arrive at our estimate for
$M_*$ (for a specific trial template, $T$, and given the observed photometry,
$F$); \viz:
\begin{equation}
	M_*(T;~F) = A_T ~ M_{*,T}(t)
	~ \left(\frac{D_L(z_\mathrm{dist})}{\left[10 ~ \mathrm{pc}\right]}\right)^{2} ~ .
\end{equation}
Here, $M_{*,T}(t)$ is the (age dependent) stellar mass of the template $T$
(including the mass locked up in stellar remnants, but not including gas
recycled back into the ISM), and $D_L(z_\mathrm{dist})$ is the luminosity
distance, computed using the flow-corrected redshift, $z_\mathrm{dist}$.

Given the (heliocentric) redshift of a particular galaxy, we compare the
observed fluxes, $F$, to the synthetic fluxes for the model templates in our
SPL, $T$, placed at the same (heliocentric) redshift. The goodness of fit for
any particular template spectrum is simply given by:
\begin{equation} \label{eq:chi2} 
	\chi_T^2 = \sum_X \left(\frac{A_T~T_X - F_X} 
			{\Delta F_X}\right)^2  ~ , 
\end{equation}
where $\Delta F_X$ is the uncertainty associated with the observed $X$-band
flux, $F_X$.

Following standard practice, we impose an error floor in all bands by adding
0.05 mag in quadrature to the uncertainties found in the photometric
catalogue. This is intended to allow for differential systematic errors in the
photometry between the different bands (for example, photometric calibration,
PSF- and aperture-matching, etc.)\ as well as minor mismatches between the SPs
of real galaxies and those in our SPL. 

It is worth stressing that that in almost all cases, the formal photometric
uncertainties found in the photometric catalogues are considerably less than
0.05 mag (see Figure \ref{fig:ugri}). This implies that, even with the current
SDSS and UKIDSS imaging, we are {\em not} limited by random noise, but by
systematic errors and uncertainties in the relative or-cross calibration of
the different photometric bands. This imposed error floor is thus the single
most significant factor in limiting the formal accuracy of our stellar mass
estimates.

\subsection{Bayesian parameter estimation} \label{ch:bayes}

For a given $F$ and $T$, we fix the value of the normalisation factor $A_T$
that appears in Equation (\ref{eq:chi2}) by minimising $\chi_T^2$. This can be
done analytically. We contrast this approach with, for example, simply scaling
the model SED to match the observed flux in a particular band
\citep[\eg][]{BrinchmannEllis, Kauffmann2003a}. Our approach has the advantage
that the overall normalisation is set with the combined signal--to--noise of
all bands.\footnote{In connection with the results of \secref{ch:nir}, this
approach is also less sensitive to systematic offsets between the observed and
fit photometry, including absolute and relative calibration errors in any
given band, which would produce a bias in the total inferred luminosity in a
given band or bands.}

With the value of $A_T$ fixed, the (minimised) value of $\chi_T^2$ can be used
to associate a probability to every object--model comparison\footnote{This
simply assumes that the measurement uncertainties in the SED $\Delta F_X$ are
all Gaussian and independent. Note that this does not necessarily gel well
with the imposition of an error floor intended to allow for systematics.};
\viz, the probability of measuring the observed fluxes, {\em assuming} that a
given model provides the `true' description of a galaxy's stellar population,
\( \mathrm{Pr}(F|T) \propto e^{-\chi_T^2} \). But this is not (necessarily)
what we are interested in---rather, we want to find the probability that a
particular template provides an accurate description of the galaxy {\em given
the observed SED}; \ie, $\mathrm{Pr}(T|F)$. These two probabilities are
related using Bayes' theorem; \viz\ \( \mathrm{Pr}(T|F) = \mathrm{Pr}(T)
\times \mathrm{Pr}(F|T) \), where $\mathrm{Pr}(T)$ is the {\em a priori}
probability of finding a real galaxy with the same stellar population as the
template $T$.

The Bayesian formulation thus requires us to explicitly specify an {\em a
priori} probability for each CSP. But it is important to realise that {\em all
fitting algorithms include priors}; the difference with Bayesian statistics is
only that this prior is made {\em explicit}. For example, if we were to simply
use the best-fit model from our library, the parameter-space distribution of
SPL templates represents an {\em implicit} prior assumption on the
distribution of SP parameters. In the absence of clearly better alternatives,
we make the simplest possible assumptions: namely, we assume a flat
distribution of models in all of $\age$, $\tau$, $\log Z$, and $\dust$. That
is, we have chosen not to privilege or penalise any particular set of SP
parameter values. The only exception to this rule is that, as is typical, we
exclude solutions with formation times less than 0.5 Gyr after the Big Bang.
\looseness-1

The power of the Bayesian approach is that it provides the means to construct
the posterior probability density function (PDF) for any quantity, $Q$, given
the observations; \ie, $\mathrm{P}(Q = Q_T | F)$, where $Q_T$ is the value of
$Q$ associated with the specific template $T$. The {\em most likely} value of
$Q$ is then given by a probability weighted integral over the full range of
possibilities\footnote{Here, the integral should be understood to be across
the full parameter space spanned by our template library, and the assumption
that our template library covers the full range of possibilities leads to the
integral constraint $\int \mathrm{d}T ~ \mathrm{Pr}(T|F) = 1$.}; \ie:
\begin{eqnarray} \left<Q\right> &=& \int \mathrm{d}T ~ Q(T) ~ \mathrm{Pr}(T|F)
\nonumber \\ &=& \int \mathrm{d}T ~ Q(T) ~ \mathrm{Pr}(T) ~
\exp\left[-\chi_T^2(F)\right] ~ . \label{eq:margin}\end{eqnarray} In the
parlance of Bayesian statistics, this is referred to as `marginalising over
the posterior probability distribution for $Q$'.\footnote{Note that in
practice we do not actually integrate over values of the normalisation
parameter, $A_T$, that appears in Equation (\ref{eq:chi2}). Instead, for a
given $T$ and $F$, we fix the value of $A_T$ via $\chi^2$ minimisation. But
because $\chi^2(A_T)$ is symmetric about the best fit value of $A_T$, this
will only cause problems for galaxies with very low total signal--to--noise
across all bands, where values of $A_T < 0$ may have some formal significance.
Since essentially all the objects in the GAMA catalogue have signal--to--noise
of roughly 30 or more in all of the $gri$ bands, we consider that this is
unlikely to be an important issue.} Similarly, it is possible to quantify the
uncertainty associated with $\left<Q\right>$ as: \begin{equation} \Delta Q =
\sqrt{\langle Q^2\rangle - \left<Q\right>^2} ~ . \end{equation}

\subsection{The importance of being Bayesian} \label{ch:diagnostics}

\begin{figure} \includegraphics[width=7.6cm]{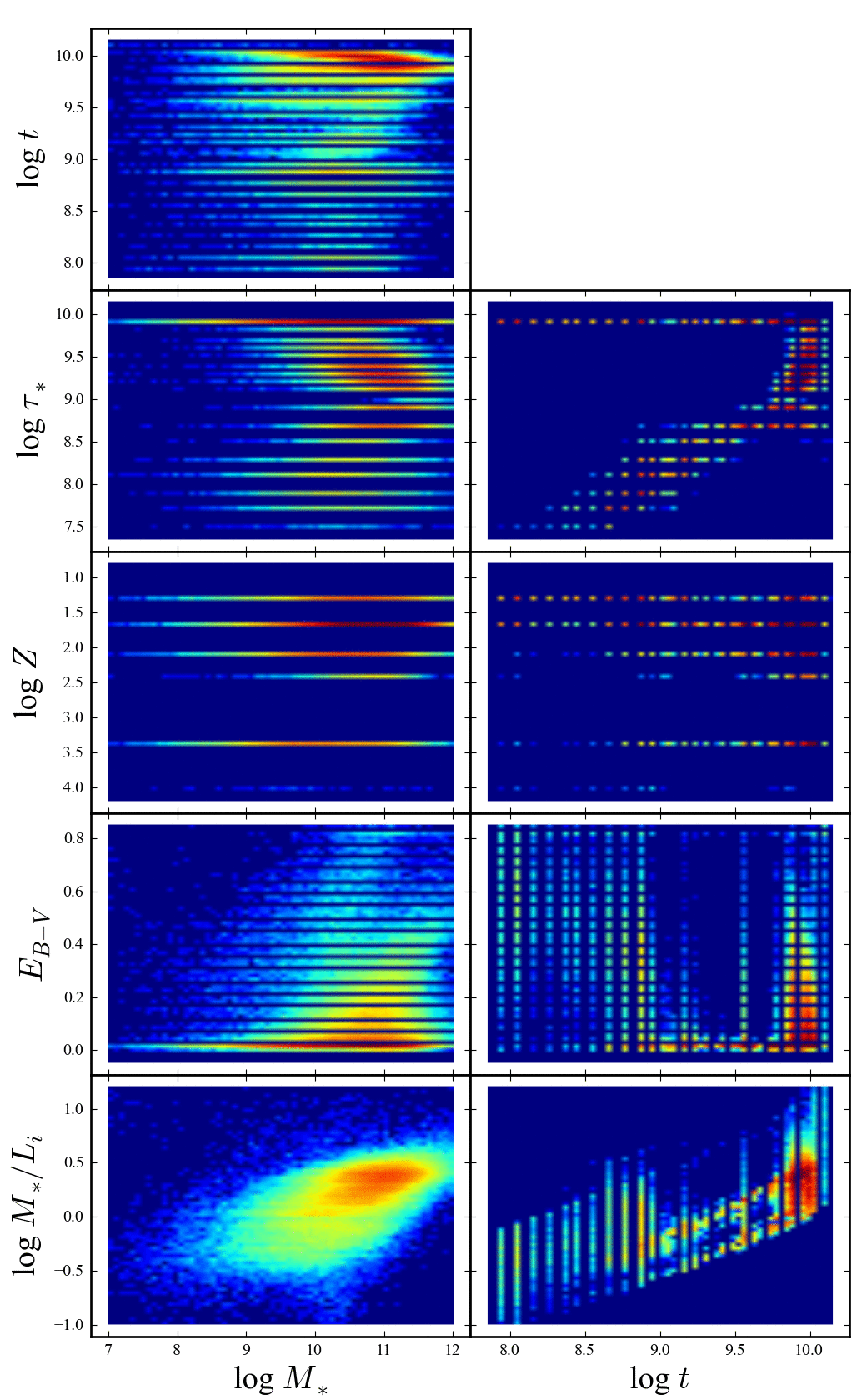} \centering
\caption{Why `best fit' is not the best parameter estimator.--- This Figure
shows the distribution of parameter values corresponding to the single `best
fit' (\ie, maximum likelihood) template. The distributions shown in this
Figure should be compared to the distributions of Bayesian estimators in
Figure \ref{fig:fitdists}. It is immediately obvious how the use of a
semi-regular grid of SP parameter values within our SPL produces strong
discretisation in $t$, $\tau$, $Z$, and $\dust$. In the lower-left panel,
however, it can be seen that there is some quantisation in $M_*/L$, even
though the distribution of $M_*/L$s in the SPL is more nearly continuous. As
described in \secref{ch:diagnostics}, this form of discretisation arises where
there are strong degeneracies in the SPS fit that cannot be properly accounted
for using a frequentist `best fit' approach. \label{fig:bestfit}} \end{figure}

\begin{figure} \includegraphics[width=7.6cm]{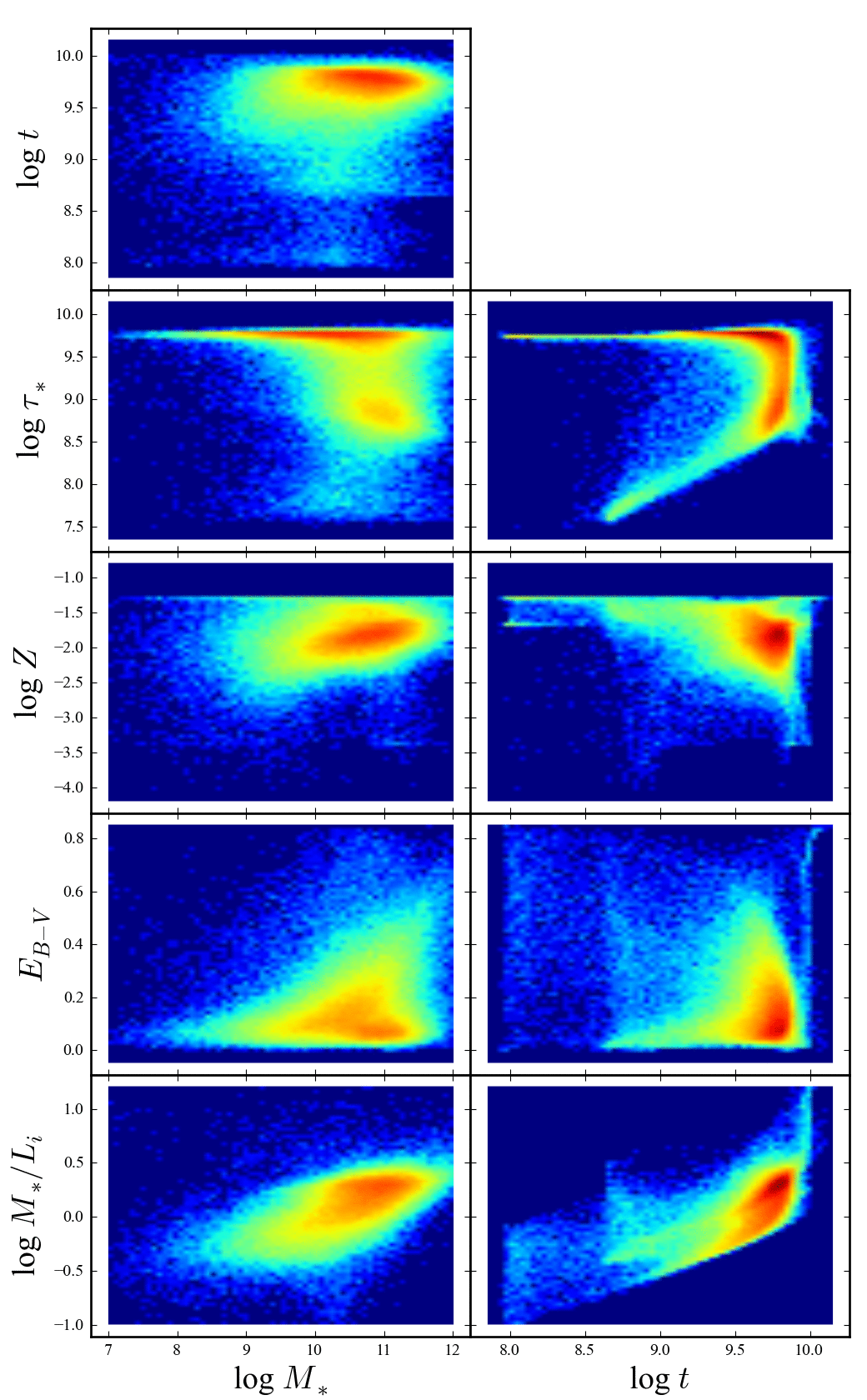} \centering
\caption{Illustrating the inferred distribution of SP parameters for GAMA
galaxies.--- In this Figure, we show the interrelationships between several
important SP parameters for GAMA galaxies. Note that the SP library covers the
full range of $t$, $\tau$, and $\dust$ shown. The observed relations between
these parameters shows that information about the process of galaxy formation
and evolution can be extracted from galaxies' SEDs. From an algorithmic point
of view, the most important point to take from this plot is that by using a
Bayesian approach, we are able for the most part to avoid `discretisation'
errors (\ie, preferred parameter values) associated with the use of a discrete
grid of parameter values (\cf\ Figure \ref{fig:bestfit}). Further, note that
particularly for $t$, $\tau$, and $Z$, these distributions are qualitatively
different to those in Figure \ref{fig:bestfit}. \label{fig:fitdists}}
\end{figure}

\begin{figure}
\includegraphics[width=7.8cm]{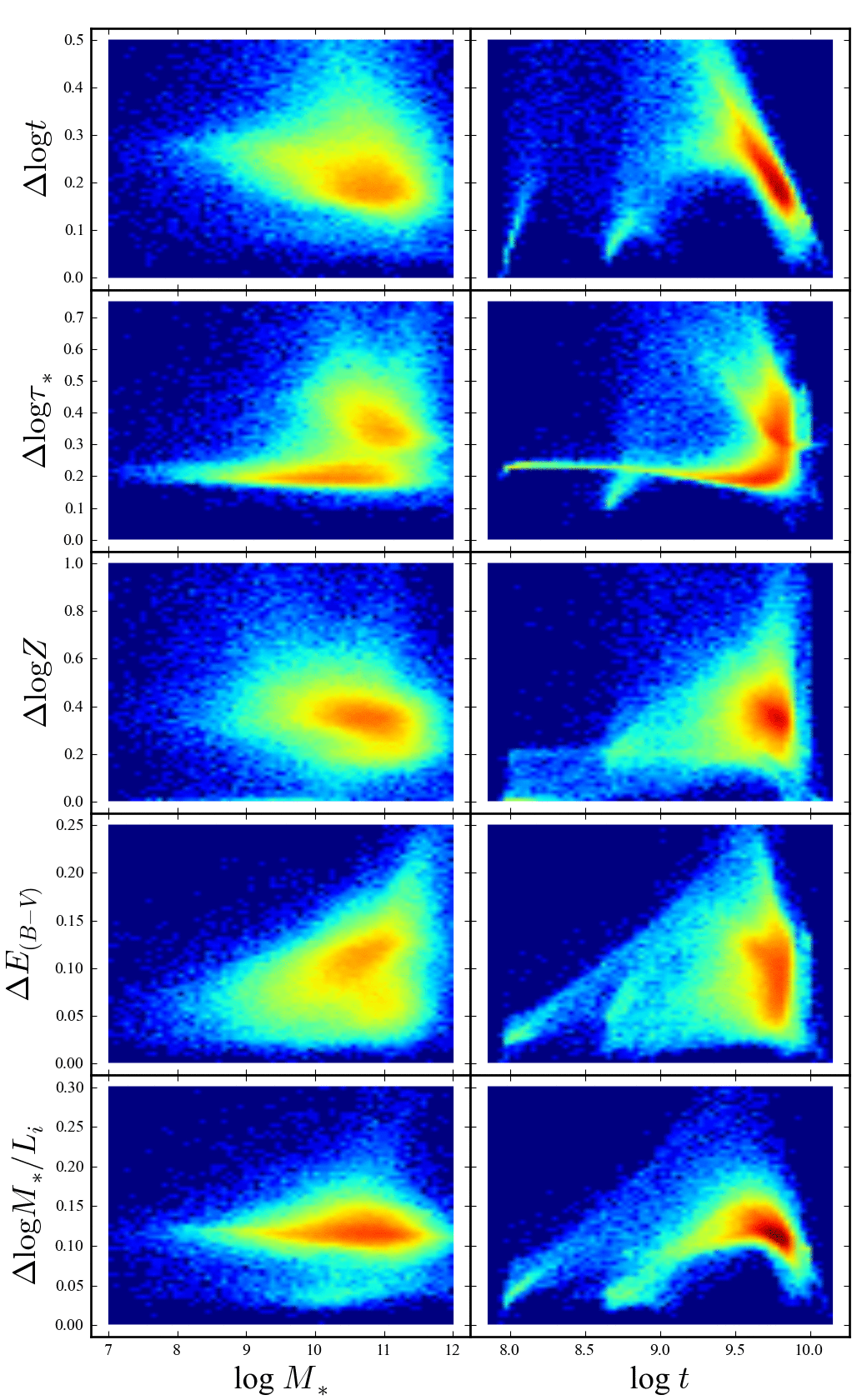}
\caption{Illustrating the distribution of parameter uncertainties.--- In this
Figure, we show the formal uncertainties for several important parameters. The
structure that is apparent in the different panels of this plot shows that our
ability to constrain $t$, $\tau$, and $Z$ is different for different stellar
populations. The crucial point to be made from this Figure, however, is that
the formal uncertainty in $M_*/L_i$ is $\approx 0.1$ dex for the vast majority
of galaxies, with essentially no dependence on the uncertainties in other
parameter values. \label{fig:errors}} \end{figure}

\begin{figure*} \includegraphics[width=16cm]{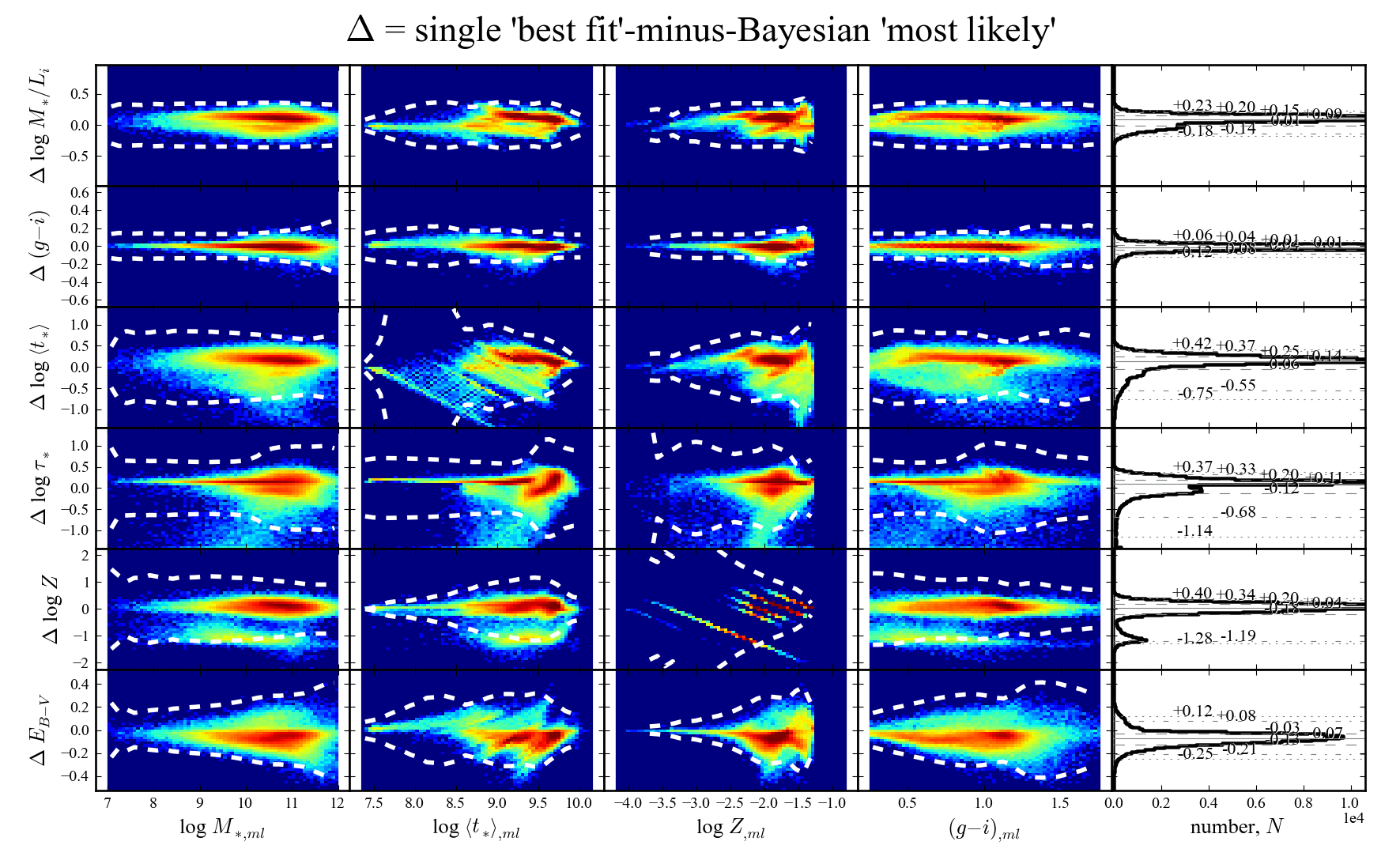} \caption{Using a
frequentist `best fit' estimator leads to significant biases in the inferred
SP parameter values.--- In this Figure, we show the difference between the
`best fit' (\ie, maximum likelihood) value, and the Bayesian `most likely'
value of a number of key parameters. The `$\Delta$' plotted on the $y$ axis
should be understood as being the `best fit'--minus--`most likely' value.
Strong quantisation in the `best fit' values can be clearly seen for $\lwage$
and particularly $Z$. Further, in the right-hand panels of this Figure, we
look at the random and systematic differences between these two parameter
estimates; the numbers in each panel show the equivalent of the $0$, $\pm 1$,
$\pm 2$, and $\pm 3 \sigma$ percentiles for each distribution. In particular,
note that the `best fit' (which again we stress is not the correct estimator
to use) value of $M_*/L$ is systematically higher than the `most likely' value
by 0.09 dex. \label{fig:bfvsml}} \end{figure*}

Before moving on, in this Section we present a selection of diagnostic plots.
Our motivation for presenting these plots is twofold. First, the Figures
presented in this Section illustrate the distribution of derived parameter
values for all $0.02 < z < 0.65$ GAMA galaxies with $\mathtt{nQ} \ge 3$ and
$\mathtt{SURVEY\_CLASS} \ge 4$ (defined in \secref{ch:specdata}). The
different panels in each Figure show the 2D-projected logarithmic data density
in small cells; the same colour-scale is used for all panels in all of Figures
\ref{fig:bestfit}---\ref{fig:bfvsml}. Note that by showing the logarithmic
data density, we are visually emphasising the more sparsely populated regions
of parameter space.

Second, we use these Figures to illustrate the differences between SP
parameter estimates based on Bayesian statistics, and those derived using more
traditional, frequentist statistics. As described above, Bayesian statistics
focuses on the most likely state of affairs given the observation, $P(Q|F)$.
Bayesian estimators can be, both in principle and in practice, significantly
different to frequentist estimators, which set out to identify the set of
model parameters that is most easily able to explain the observations; \ie, to
maximise $P(F|Q)$. To make plain the differences between these two parameter
estimates, we will compare the Bayesian `most likely' estimator as defined by
Equation (\ref{eq:margin}) to a more traditional `best fit' value derived via
maximum likelihood. Note that when deriving the frequentist `best fit' values,
we have applied our priors through weighting of the value of $\chi^2$ for each
template; that is, the `best fit' value is that associated with the template
$T$ which has the highest value of $\log \mathcal{L}(F|T) = \log
\mathrm{Pr}(T) - \chi_T^2(F)$.

The distribution of these `best fit' SP parameters are shown in \textbf{Figure
\ref{fig:bestfit}}, as a function of stellar mass, $M_*$ and SP age, $t$. It
is immediately obvious from this Figure how our use of a semi-regular grid of
SP parameters to construct the SPL leads directly to strong quantisation in
the `best fit' values of $t$, $\tau$, $Z$, and $\dust$. What is more worrying,
however, is that there is also a mild discretisation in the inferred values of
$M_*/L$, seen in the bottom-lefthand panel of Figure \ref{fig:bestfit} as a
subtle striping. This is despite the fact that the SPL samples a much more
nearly continuous range of $M_*/L$s than $t$s, $\tau$s, or $Z$s.

To explain the origin of this effect, let us return to Figure \ref{fig:ugri}.
For a given galaxy, there will be a large number of templates that will be
consistent with the observed $ugriz$ photometry. To the extent that a small
perturbation in the observed photometry can have a large impact on the
inferred SP parameter values, there is a degree of randomness in the selection
of the `best fit' solution from within the error ellipse. This means that
values of $M_*/L$, $(g-i)$, etc.\ that are `over-represented' within the SPL
will be more commonly selected as `best fits'. Note that this problem of
discretisation in $M_*/L_i$ is therefore {\em not} a sign of insufficiently
fine sampling of the SPL parameter space: this problem arises where there very
many, not very few, templates that are consistent with a given galaxy's
observed colours.

Figure \ref{fig:bestfit} should be compared to \textbf{Figure
\ref{fig:fitdists}}, in which we show the distribution of the Bayesian `most
likely' parameter values. Consider again Figure \ref{fig:ugri}: whereas the
`best fit' value is the one nearest the centre of the error ellipse for any
given galaxy, the Bayesian value is found by taking a probability-weighted
mean of all values around the observed data point. The process of Bayesian
marginalisation can thus be thought of as using the SPL templates to
discretely sample a continuous parameter distribution, after effectively
smoothing on a scale commensurate with the observational uncertainties. This
largely mitigates the discretisation in $t$, $\tau$, and $Z$---as well as in
$M_*/L$---that comes from using a fixed grid of parameter values to define the
SPL.

That said, this only works where several different parameter combinations
provide an acceptable description of the data. If one particular template is
strongly preferred---if the observational uncertainties in a galaxy's SED are
comparable to or less than the differences between the SEDs of different
templates---then our approach reverts to a `best fit', and we will again
suffer from artificial quantisation in the fit parameters. For the same
reasons, the formal uncertainty on the SP parameters will be artificially
small in this case. Note that, somewhat perversely, this problem will become
{\em worse} with increasing signal--to--noise. \citep[See also][but note, too,
that the inclusion of a moderate `systematic' uncertainty in the observed SEDs
works to protect against `single template' fits.]{GallazziBell} In this sense,
and in contrast to the quantisation in the `best fit' values discussed above,
quantisation in the Bayesian `most likely' values does indicate inadequately
fine sampling of the SPL parameter space. We have chosen our parameter grids
with this limitation in mind; in particular, we have found that a rather fine
sampling in the $\dust$ dimension is required to avoid strong quantisation.

Although our SPL templates span a semi-regular grid in each of $t$, $\tau$,
$Z$, and $\dust$, the observed distribution in these parameters is anything
but uniform. There is nothing in the calculation to preclude solutions with,
for example, young ages and low metallicities. The fact that these regions of
parameter space are sparsely- or un-populated shows that there are few or no
galaxies with optical SEDs that are consistent with these properties. Figure
\ref{fig:fitdists} thus illustrates the mundane or crucial (depending on one's
perspective) fact that the derived SP parameters do indeed encode information
about the formation and evolution of galaxies. It is particularly striking
that there appears to be a rather tight and `bimodal' relation between $t$ and
$\tau$: there is a population of galaxies that are best fit by very long and
nearly continuous SFHs ($\tau \approx 10$ Gyr), and another with $t/\tau
\approx $ 3---10. There are virtually no galaxies inferred to have $t < \tau$.

The inferred distribution of parameter values is significant in terms of our
assumed priors: it is clear that the derived parameter distributions do not
follow our assumed priors (see also Figure \ref{fig:colourrel}). But this is
not to say that the precise values are not more subtly affected by our
particular choice of priors. In particular, the local slope of the priors on
the scale of the formally derived uncertainties might act to skew the
posterior PDF (see also Appendix \ref{ch:mocks}). In principle, it is possible
to use the observed parameter distributions to derive new, astrophysically
motivated priors. Then, if this were to significantly alter the observed
parameter distribution, the process could be iterated until convergence. Such
an exercise is beyond the scope of this work.

Next, in \textbf{Figure \ref{fig:errors}}, we show the distribution of
inferred uncertainties in each of the parameters shown in Figure
\ref{fig:fitdists}. As in Figure \ref{fig:fitdists}, there is some structure
apparent in these distributions: the uncertainties in some derived properties
are different for galaxies with different kinds of stellar populations. As a
simple example, galaxies with $t \gg \tau$ have considerably larger
uncertainties in $\tau$, as information about the SFH is washed out with the
deaths of shorter-lived stars. In connection to the discretisation problem,
the very young galaxies (seen in Figure \ref{fig:fitdists} to suffer from
discretisation in the values of $Z$) also have low formal values for $\Delta
\log Z$ and/or $\Delta \log t$. But it is worth noting that in comparison to
the uncertainties in other SP parameters, $\Delta \log M_*/L_i$ is more nearly
constant across the population (this is perhaps more clearly apparent in
Figure \ref{fig:bfvsml}, described immediately below).

Our last task for this Section is to directly compare the frequentist `best
fit' and Bayesian `most likely' SP values; this comparison is shown in
\textbf{Figure \ref{fig:bfvsml}} . In each panel of this Figure, the
`$\Delta$' plotted on the $y$ axis should be understood as being the `best
fit'-minus-`most likely' value; these are plotted as a function of the
Bayesian estimator. Within each panel, the dashed white
curves show the median $\pm 3 \sigma$ uncertainty in the $y$-axis quantity,
derived in the Bayesian way, and computed in narrow bins of the $x$-axis
quantity. These curves can thus be taken to indicate the formal consistency
between the best fit and most likely parameter values. 

In practice, there is an appreciable systematic difference between the
frequentist and Bayesian parameter estimates. In general, we find that
traditional, frequentist estimates are slightly older (by $\approx 0.14$ dex),
less dusty (by $\approx 0.07$ mag), and more massive (by $ 0.09$ dex) than the
Bayesian values. In comparison to the formal uncertainties, these systematic
differences are at the 0.5---0.7$\sigma$ level; this is despite the fact that
the `best fit' value is within 1.5$\sigma$ of the `most likely' value for 99
\% of objects. And again, we stress that, formally, the Bayesian estimator is
the correct value to use.

As a final aside for this section, we note that the importance of Bayesian
analysis has been recognised in the context of photometric redshift evaluation
(a problem which is very closely linked to SPS fitting) by a number of
authors, including \citet{BPZ} and \citet{eazy}. While most of the SPS fitting
results for SDSS \citep[\eg][]{Kauffmann2003b, Brinchmann2004, Gallazzi2005}
have been based on a Bayesian approach, it is still common practice to derive
SPS parameter estimates using simple $\chi^2$ minimisation
\citep{Walcher2011}. This is particularly true for high redshift studies
\citep[but see][]{Pozzetti2007, Walcher2008}.

\subsection{Detection/selection limits and $\mathbf{1/V_\mathrm{max}}$ corrections} \label{ch:zmax}

GAMA is a flux-limited survey. For a number of science applications---most
obviously measurement of the mass or luminosity functions---it is important to
know the redshift range over which an individual galaxy would be selected as a
spectroscopic target. To this end, we have used the SP fits described above to
determine the maximum redshift, $z_\mathrm{max}$, at which each galaxy in the
GAMA catalogue would satisfy the main GAMA target selection criterion of
$r_\mathtt{petro} < 19.4$, or, for the G12 field, $r_\mathtt{petro} < 19.8$.
(Recall that the target selection is done on the basis of the SDSS, rather
than the GAMA, $\petro$ magnitude.)

This has been done for each galaxy using the best-fit template
spectrum.\footnote{We have argued in \secref{ch:bayes} that the best fit
template is not appropriate as a basis for deriving SP parameters. For the
same reasons, formally, we should also marginalise over the posterior
probability distribution for $z_\mathrm{max}(T)$. We have checked, and the
value of $z_\mathrm{max}$ derived from the best-fit template typically matches
the Bayesian value to within $\Delta z_\mathrm{max} \sim 0.001$. Given this,
and the fact that using the best-fit value is vastly computationally simpler,
we have opted to use the best-fit template.} Knowing the best-fit template,
including the normalisation factor, $A_T$, we consider how the observers'
frame $r$-band flux of the template declines with redshift. Knowing that
galaxy's observed $r_\mathtt{petro}$, it is then straightforward to determine
the redshift at which the observers' frame $r$-band flux drops to the
appropriate limiting magnitude. The only complication here is accounting for
both the cosmological redshift and the Doppler redshift due to peculiar
velocity in the redshift. This is done by recognising that $(1 +
z_\mathrm{helio}) = (1+z_\mathrm{dist})(1+z_\mathrm{pec})$; the values of
$z_\mathrm{max}$ should be taken as pertaining to $z_\mathrm{dist}$.

\begin{figure*} \centering \includegraphics[height=5.25cm]{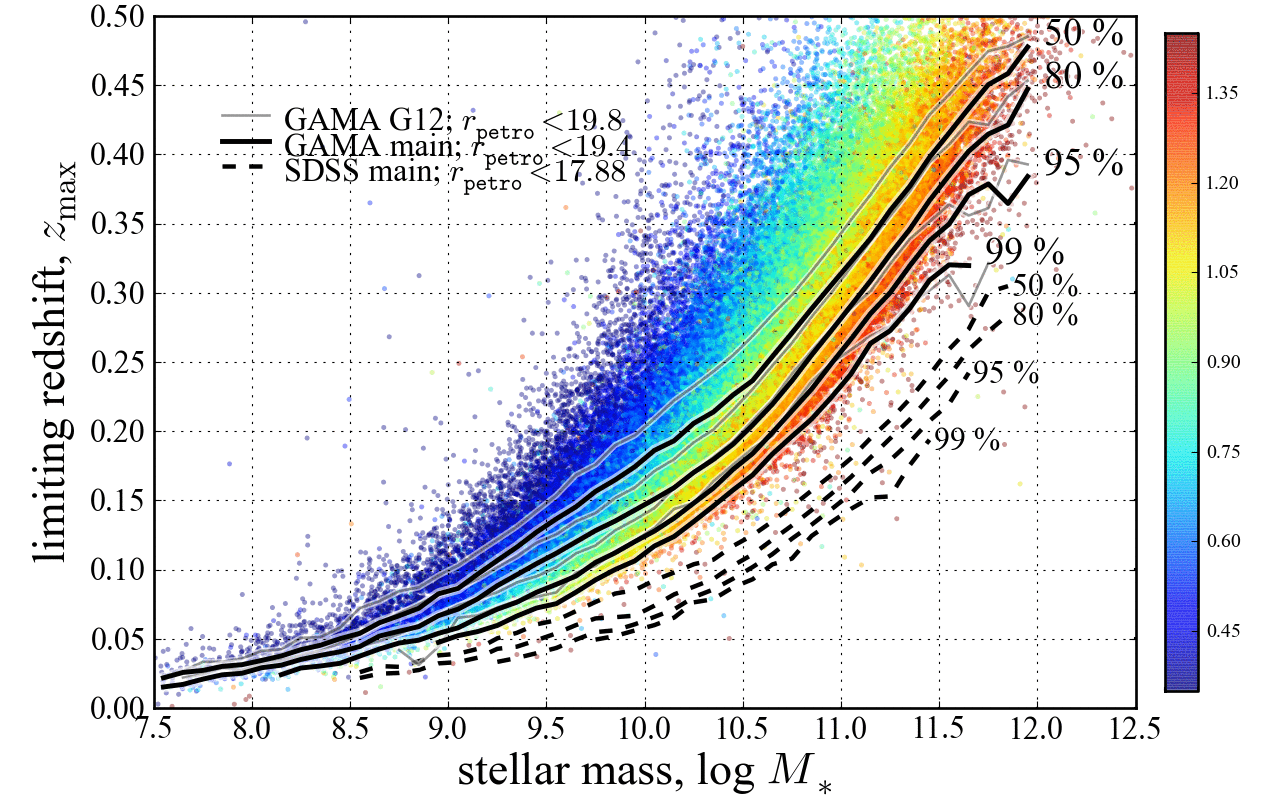}
\includegraphics[height=5.25cm]{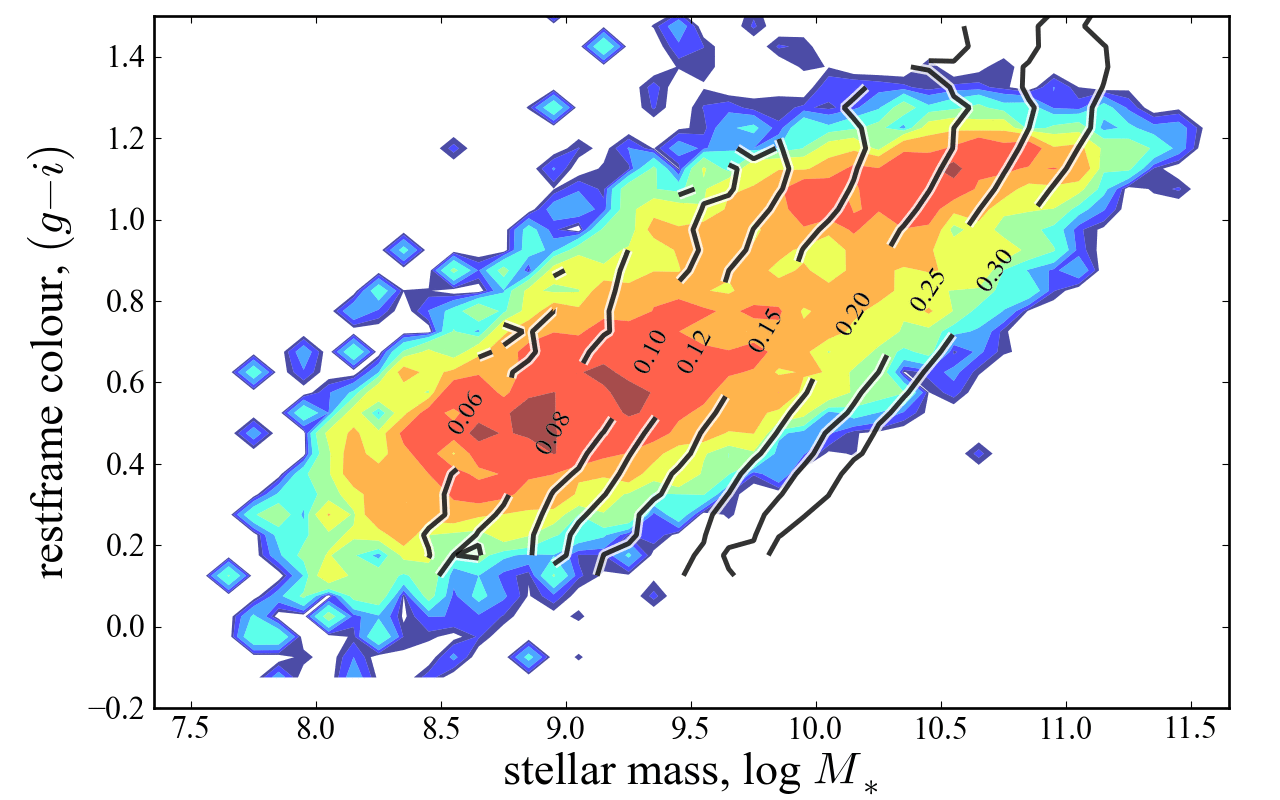} \caption{The GAMA stellar mass
completeness limits as a function of redshift and restframe colour.--- Both
panels of this plot use the derived values of $z_\mathrm{max}$ (that is, the
maximum redshift at which any individual galaxy would satisfy the $r$-band
selection limit) to show the redshift-dependent GAMA completeness limit as a
function of stellar mass and restframe colour. In the left panel, we show the
stellar mass for which GAMA is 50/80/95/99 \% complete, computed in narrow
bins of $z_\mathrm{max}$. The heavier solid lines show the completeness for
the main $r_\mathtt{petro} < 19.4$ GAMA selection limit; the lighter solid
lines show that for the deeper $r_\mathtt{petro} < 19.8$ limit for the G12
field; the dashed lines show the completeness for the SDSS $r_\mathtt{petro} <
17.88$ main galaxy sample selection limit. In this panel, individual plots are
colour-coded according to their restframe $(g-i)$ colour; only galaxies from
the main $r_\mathtt{petro} < 19.4$ sample are shown. In the right panel, the
black contours show the mean value of $z_\mathrm{max}$, again for the main
$r_\mathtt{petro} < 19.4$ sample, projected onto the colour--stellar mass
diagram. For comparison, the filled, coloured contours in this panel show the
incompleteness corrected bivariate colour--stellar mass distribution of $z <
0.12$ galaxies; these contours are logarithmically spaced by factors of 2. In
constructing this plot, individual galaxies have been weighted by $w =
V(z=0.12)/V(z_\mathrm{max})$. We have only counted galaxies with a relative
weighting $w < 30$. In effect, this means that we have not fully corrected for
incompleteness for $z_\mathrm{max} < 0.04$ or $M_* \lesssim 10^{8.5}$ M\sol .
\label{fig:zlim}} \end{figure*}

In \textbf{Figure \ref{fig:zlim}}, we use the value of $z_\mathrm{max}$, so
derived, to show GAMA's stellar mass completeness limit expressed as a
function of redshift and restframe colour. This Figure shows the two-fold
power of GAMA in relation to SDSS. First, for dwarf galaxies, GAMA is $\approx
95$ \% complete for $M_* \approx 10^8$ M\sol\ and $z \approx 0.05$; at these
masses, SDSS completeness is $\lesssim 80$ \% even for $z < 0.02$. GAMA thus
provides the first census of $10^{7.5} < M_* < 10^{8.5}$ M\sol\ galaxies.
Further, for massive galaxies, GAMA probes considerably higher redshifts: for
$M_* \sim 10^{10.5}$ ($10^{11}$) M\sol , where SDSS is limited to $z \lesssim
0.1$ (0.15), GAMA can probe out to $z \approx 0.25$ (0.3). Said another way,
GAMA probes roughly twice the range of lookback times of SDSS. GAMA thus opens
a new window on the recent evolution of the massive galaxy population.

In the right-hand panel of Figure \ref{fig:zlim}, we show these same results
in complementary way. The solid lines in this Figure show the mean value of
$z_\mathrm{max}$ as a function of $M_*$ and restframe $(g-i)$. These values
are for the main $r_\mathtt{petro} < 19.4$ selection only; for the G12 field,
these limits should be shifted down in mass by 0.16 dex.

In this panel, for comparison, we also show the incompleteness-corrected
bivariate colour-mass distribution for $z < 0.12$ galaxies; \ie, individual
galaxies have been weighted by $1/V_\mathrm{max}$, where $V_\mathrm{max}$ is
the survey volume implied by $z_\mathrm{max}$. Note that in the construction
of this plot, we have only included galaxies with a relative weight $< 30$
(\ie, $z_\mathrm{max} > 0.0375$); in effect, this means that we have not fully
accounted for incompleteness for $M_* \lesssim 10^{8}$ M\sol. Again, we see
that GAMA probes the bulk of the massive galaxy population ($M_* \gtrsim
10^{10.5}$ M\sol ) out to $z \approx 0.25$.

Before moving on, we make two further observations.  First, it is clear that the red sequence galaxy population extends well below the `threshold mass' of $10^{10.5}$ M\sol\ suggested by \citet{Kauffmann2003b}. Secondly, it appears that we are seeing the low-mass end of the red sequence population: the apparent dearth of galaxies with $(g-i) \gtrsim 0.8$ and $10^{8.5} \lesssim M_*/$M\sol\ $\lesssim 9.5$ is not a product of incompleteness.  We will investigate these results further in a future work.

\section{How much does NIR data help (or hurt)?} \label{ch:nir}

Conventional wisdom says that that using NIR data leads to a better estimate
of stellar mass. The principal justification for this belief is that, in
comparison to optical wavelengths, and all else being equal, NIR luminosities
1.\ vary less with time, 2.\ depend less on the precise SFH, and 3.\ are less
affected by dust extinction/obscuration. Further, whereas old stellar
populations can have very similar optical SED shapes to younger and dustier
ones (see Figure \ref{fig:ugri}), the optical--NIR SED shapes of these two
populations are rather different. The inclusion of NIR data can thus break the
degeneracy between these two qualitatively different situations, and so
provide tighter constraints on each of age, dust, and metallicity---and hence,
it is argued, a better estimate of $M_*/L$.

There are, however, several reasons to be suspicious of this belief. First,
while stellar evolution models have been well tested in the optical regime,
there is still some controversy over their applicability in the NIR. This has
been most widely studied and discussed recently in connection with TP-AGBs
stars in the models of \citet{BC03} and \citet{M05}
\citep[\eg,][]{Maraston2006, Bruzual2007, Kriek2010}. The different models
have been shown to yield stellar mass estimates that vary by as much as $\sim
0.15$ dex for some individual galaxies \citep[\eg,][]{KannappanGawiser,
Wuyts2009, Muzzin2009}, but only if restframe NIR data are used in the fits.

Separately from the question as to the accuracy of SP models in the NIR, there
are a number of empirical arguments suggesting that optical data alone can be
used to obtain a robust and reliable stellar mass estimate. A number of
authors have found that there is a remarkably tight relation between optical
colour and stellar mass-to-light ratio \citep{BelldeJong, Bell2003,
Zibetti2009, Taylor2010a}. As described in \secref{ch:intro},
\citet{GallazziBell} have argued that a stellar mass estimate based on a
single colour is nearly as reliable and robust as one based on a full
optical--to--NIR SED fit, or even one based on spectral diagnostics. Further,
using their NMF-based \texttt{kcorrect} algorithm that eliminates the need for
assuming parametric SFHs, \citet{kcorrect} have shown that they are able to
use optical SEDs to {\em predict} galaxies' NIR fluxes. Each of these results
implies that the NIR SED does not, in practice, contain qualitatively `new'
information not found in the optical.

With this as background, our goal in this Section is to examine how the
inclusion of NIR data affects our stellar mass estimates. We will take an
empirical approach to the problem, looking at how both the quality of the fits
and the quantitative results themselves depend on the models and photometric
bands used. We will argue that, at least at the present time, the NIR data
cannot be satisfactorily incorporated into our SPS fitting. We explore the
possible causes of our problems in dealing with the NIR SEDs to
\secref{ch:badnir}. In the next Section, we will then look at whether and how
our decision to ignore the NIR data affects the quality of our stellar mass
and SP parameter estimates.

\subsection{How well do the models describe the optical-to-NIR SEDs of GAMA galaxies?} \label{ch:residuals}

\begin{figure*} \centering \includegraphics[width=8cm]{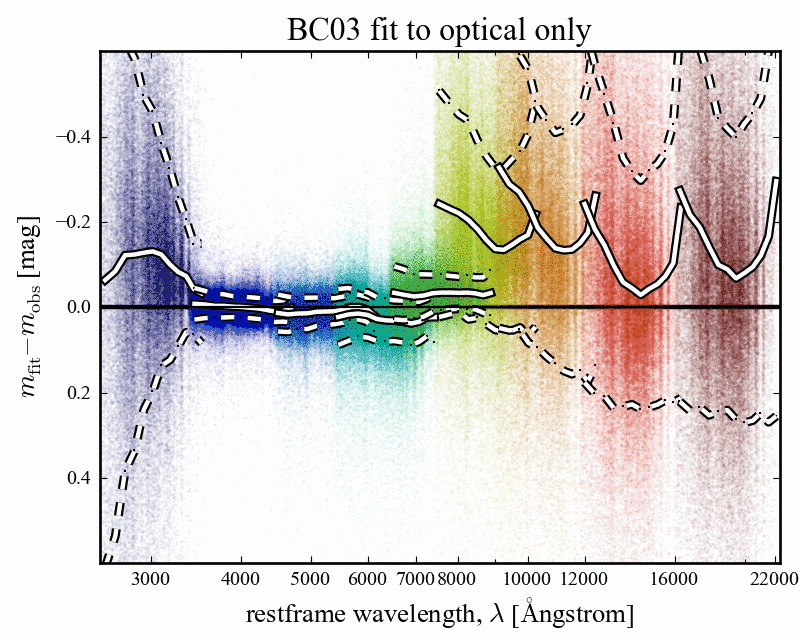}
\includegraphics[width=8cm]{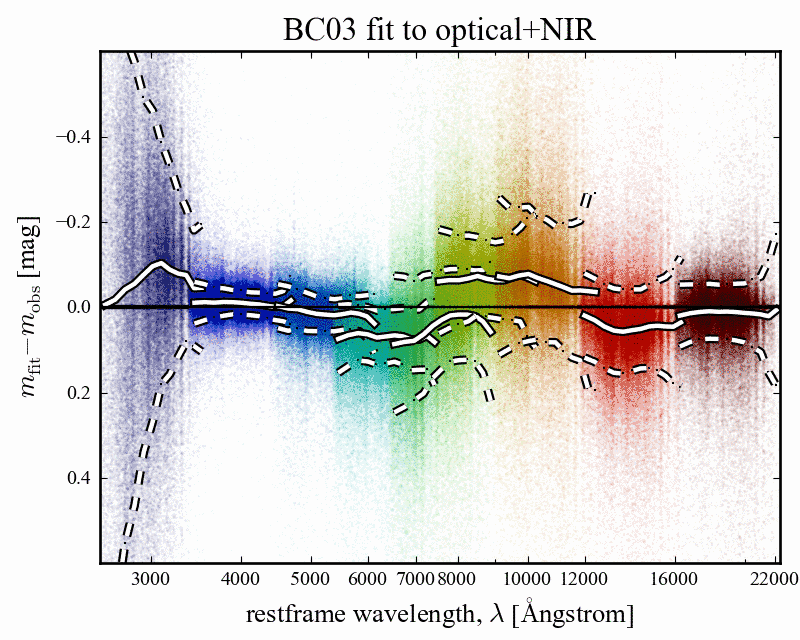} \\ \hspace{0.4cm}
\includegraphics[width=3.9cm]{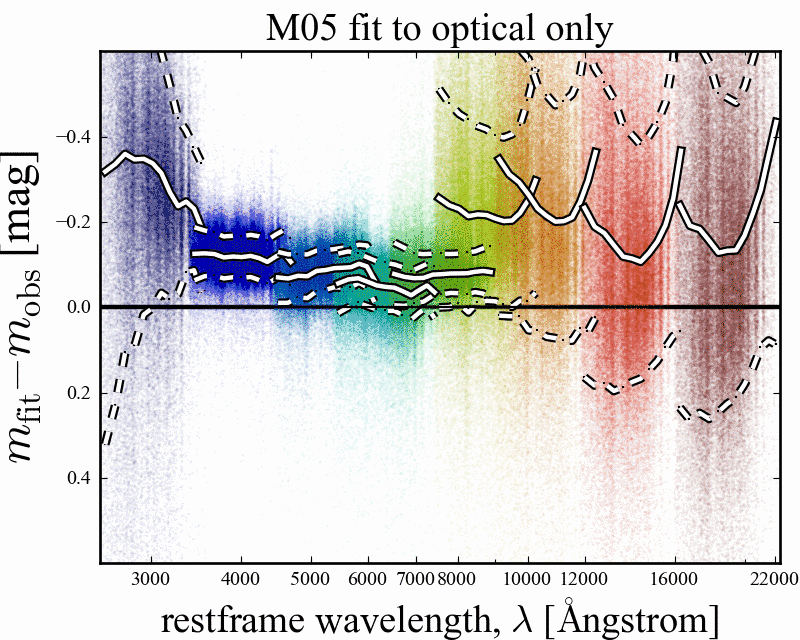}
\includegraphics[width=3.9cm]{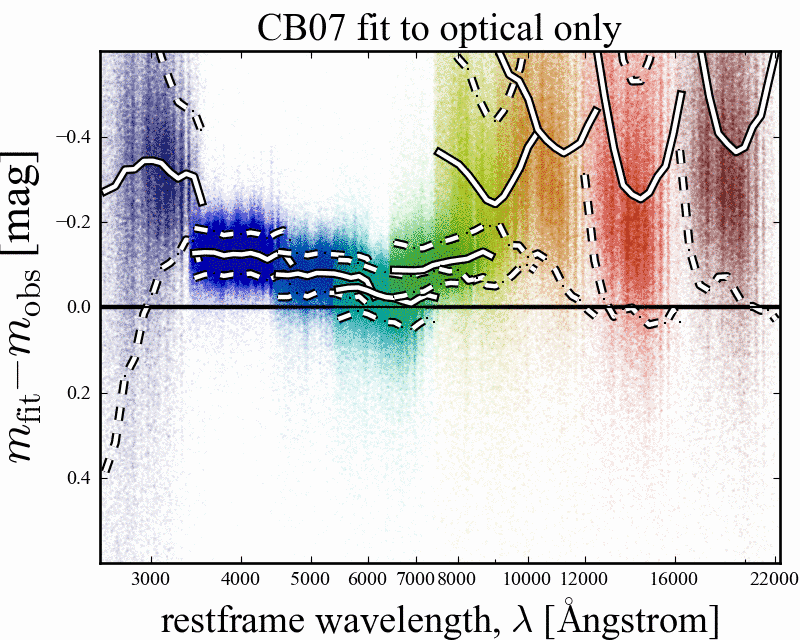}
\includegraphics[width=3.9cm]{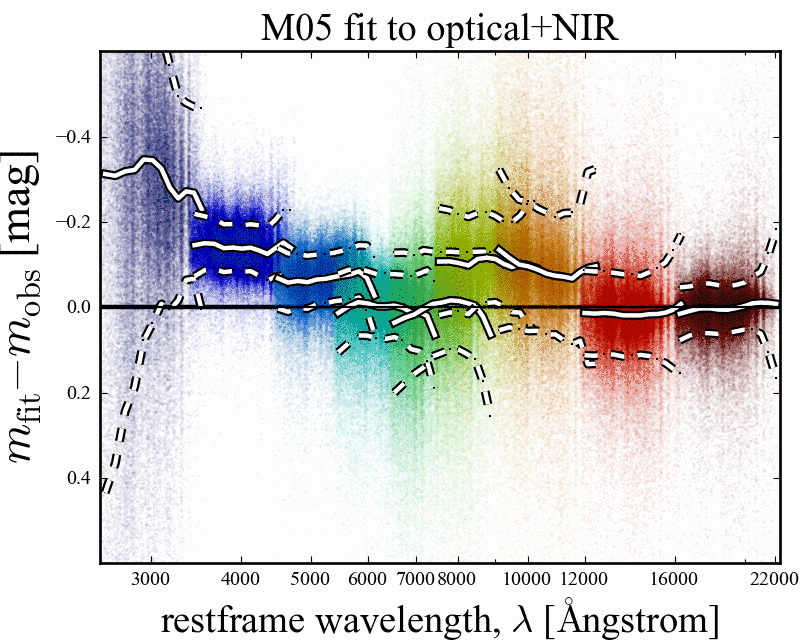}
\includegraphics[width=3.9cm]{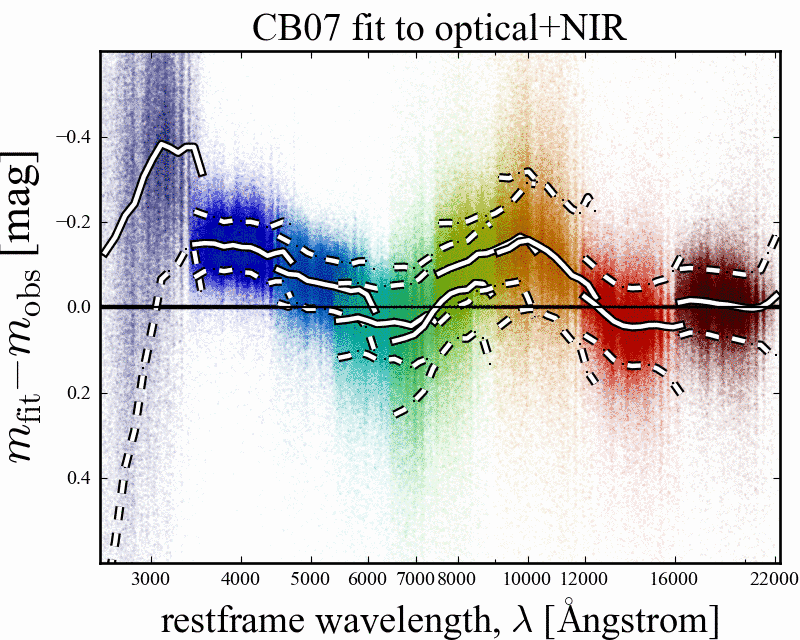} \caption{Illustrating how well
the models describe the observed photometry of real galaxies.--- Each panel of
this Figure shows the residuals when fitting either to the optical $ugriz$
(left panels) or the full $ugrizYJHK$ (right panels) SEDs. The main upper
panels are for fits based on the \citet{BC03} stellar evolution models; the
lower panels show the same using the \citet{M05} or \citet{CB07} models.
Within each panel, the different colours refer to the different observed
bands. For each band the white lines show the median (solid lines) and 16/84
percentile (dashed lines) residual in each band as a function of restframe
wavelength or, equivalently, redshift. This Figure should be compared to
Figure \ref{fig:simresiduals}. From the left-hand panels, it is clear that all
three of these SPS models provide an acceptable description of the optical
SEDs, with relatively small residuals as a function of restframe wavelength.
That said, these optical-only fits overpredict the NIR fluxes by $\sim 0.2$
dex. In the righthand panels, it can be seen that including the NIR data
significantly degrades the quality of the optical fits, particularly in the
$i$- and $z$-bands. Each of the three sets of models show qualitatively and
quantitatively similar residuals: while each set of models yield similarly
good fits to the optical data alone, none of these models provides a good
description of the optical--to--NIR SEDs of GAMA galaxies. Based on our
experiences with mock galaxy photometry in Appendix \ref{ch:mocks}, these
results suggest strong inconsistencies between the optical--minus--NIR colours
of real galaxies and those contained within our SPL. \label{fig:residuals}}
\end{figure*}

In \textbf{Figure \ref{fig:residuals}}, we show the residuals from the SED
fits as a function of restframe wavelength; \ie, $m_{X,\mathrm{fit}} -
m_{X,\mathrm{obs}}$ as a function of $\lambda_X / (1+z)$.\footnote{The values
for the `fit' photometry are obtained in the same way as the other SP
parameters; \viz, via Bayesian marginalisation over the PDF, \'a la Equation
\ref{eq:margin}. They should thus be thought of as estimates of the most
likely value of the `true' observers' frame photometry, given the overall SED
shape.} Figure \ref{fig:residuals} should be compared to Figure
\ref{fig:simresiduals} in Appendix \ref{ch:mocks}. This Appendix describes how
we have applied our SPS fitting algorithm to mock galaxy photometry, which we
have constructed from the fits to the actual $ugriz$ SEDs of GAMA galaxies. In
this way, as in \citet{GallazziBell}, we have tested our ability to fit galaxy
SEDs in the case that the SPL provides perfect descriptions of the
stellar populations of `real' galaxies, and that the data are perfectly
calibrated (\ie, no systematics in the photometric cross-calibration).
Inasmuch as they can inform our expectations for the real data, the results of
these numerical experiments (shown in Figure \ref{fig:simresiduals}) can help
interpret the offsets seen in Figure \ref{fig:residuals}.

In Figure \ref{fig:residuals}, as in Figure \ref{fig:simresiduals}, the
lefthand panels show the residuals when only the optical data is used for the
fit. The NIR points in these panels are thus predictions for the observers'
frame NIR photometry derived from the optical SED. The right-hand panels of
both Figures \ref{fig:residuals} and \ref{fig:simresiduals} show the residuals
for fits to the full 9 band optical--to--NIR SED. In Figure
\ref{fig:residuals}, we show the residuals when using several different sets
of SSP models to construct our SPL. In this Figure, the larger upper panels
are for the fits based on the \citet{BC03} SSP models; the panels below show
the same using the SSP models of \citet{M05} and \citet{CB07} for comparison.

Looking first at the lefthand panels of Figure \ref{fig:residuals}, we see
that our SPS fits provide a reasonably good description of the observed
$ugriz$ SEDs of real GAMA galaxies. The fit residuals are qualitatively and
quantitatively similar when using each of the three different SSP models to
construct the SPL. The median offset in each of the $(ugriz)$-bands is
$\approx$ ($-$0.10, $-$0.00, +0.01, +0.02, $-$0.03) mag. In terms of the
formal uncertainties from the fits, the median offsets are at the level of
$\approx$ ($-$0.3, $-$0.0, +0.2, +0.5, $-$0.5)$\sigma$. The systematic biases
in the fit photometry are thus weakly significant, but, at least for the
$griz$-bands, well within the imposed error floor of 0.05 mag.

How does this compare to what is seen for the mocks in Figures
\ref{fig:simresiduals}? We find qualitatively similar offsets when fitting to
the mock $ugriz$ photometry. More specifically, we see a similar `curvature'
in the residuals, with slight excesses in the fit values for the $u$- and
$z$-band photometry, and the $gri$- band photometry being very slightly too
faint. It is true that, quantitatively, the offsets seen in Figure
\ref{fig:residuals} are about twice as large as we might expect based on our
numerical experiments ($\lesssim 0.5 \sigma$ for the real data, as opposed to
$\lesssim 0.2 \sigma$ for the mocks). But even so, the fact that we see
similar residuals when fitting to the mocks shows that such residuals are to
be expected, even in the ideal case where both the SPL and photometry are
perfect. We do not, therefore, consider the mild systematic offsets between
the fit and observed photometry as evidence for major problems in the $ugriz$
fits.

Unlike \citet{kcorrect}, we seem unable to use the optical SEDs to
satisfactorily predict NIR photometry. The fits to the $ugriz$ data predict
$YJHK$ photometry that is considerably brighter (by up to $\sim 0.2$ mag) than
what is observed. The systematic differences between the predicted and
observed fluxes for the \citet{BC03} models are $-3.3\sigma$ $-2.8\sigma$,
$-1.6\sigma$, and $-2.5\sigma$ in $YJHK$, respectively. For the \citet{M05}
models, the residuals are slightly larger ($-3.3\sigma$, $-3.0\sigma$,
$-2.3\sigma$, and $-2.8\sigma$), and larger again for the \citet{CB07} models
($-4.7\sigma$, $-5.6\sigma$, $-6.0\sigma,$ and $-8.2\sigma$).

\begin{figure*} \includegraphics[width=15cm]{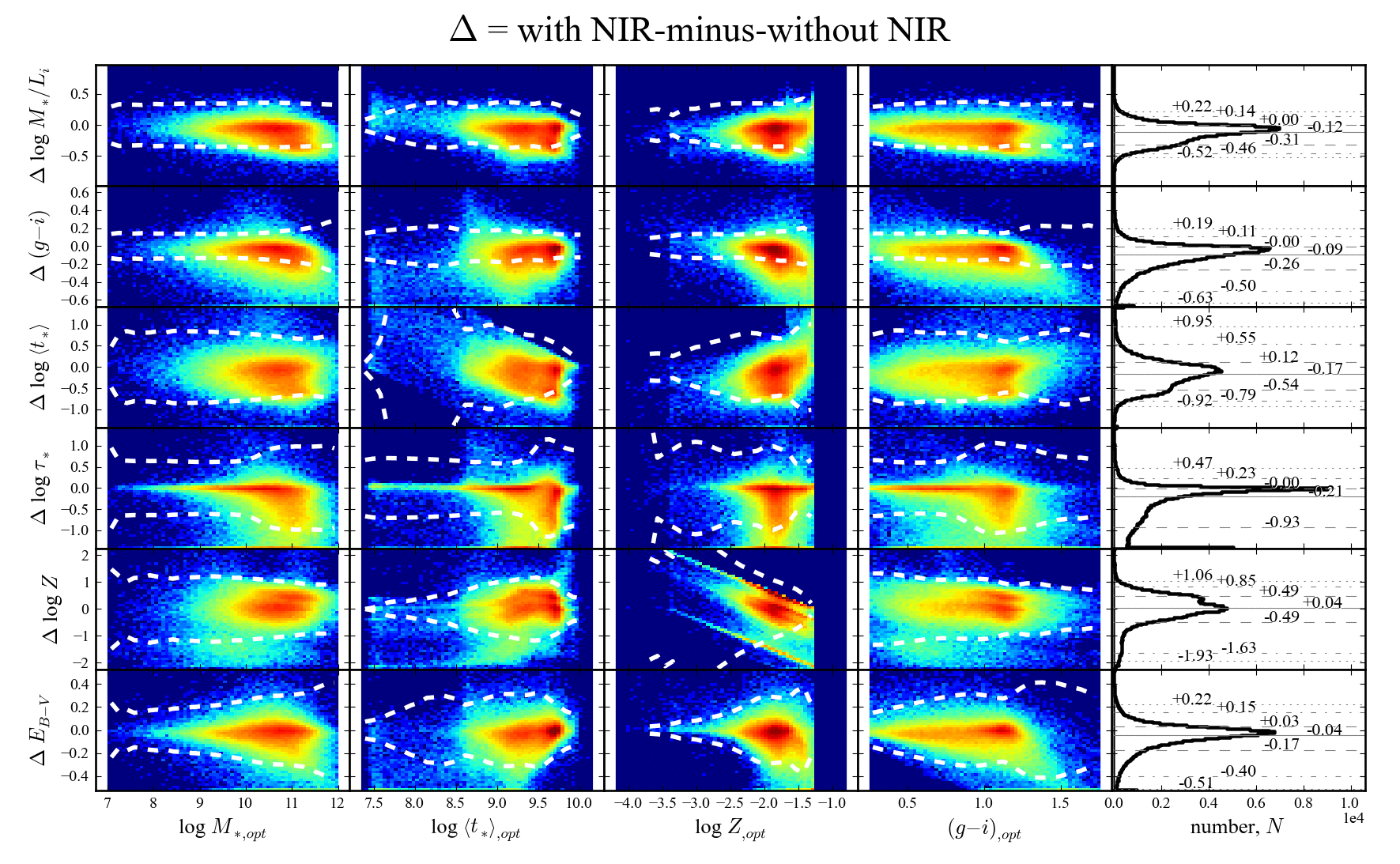} \caption{The effect of
including the NIR data on the inferred stellar population parameters.--- In
each panel of this Figure, we show the difference between the stellar
population parameters derived using only the $ugriz$ photometry, and those
derived using the full $ugrizYJHK$ SEDs. In all cases, the `$\Delta$' should
be understood as `optical-plus-NIR-derived'--minus--`optical-only-derived'. As
in Figure \ref{fig:bfvsml}, the histograms in the right-hand panels show the
distribution of `$\Delta$s' and the 1/2.5/15/50/85/97.5/99 percentiles. In the
other panels, the `$\Delta$' is plotted as a function of the
`optical-only-derived' value; the colourscale shows the logarithmic data
density for the full GAMA sample. In these panels, the dashed lines show the
$\pm 3 \sigma$ uncertainties as derived from the fits to the 5 optical bands.
The median effect of including the NIR data is to systematically reduce the
inferred value of $M_*$ by $\approx 0.1$ dex. It is important to note these
offsets NIR-derived values are formally inconsistent with the
optical-only-derived uncertainties at the $\gtrsim 2.5 \sigma$ level. This is
particularly significant in the case of the restframe $(g-i)$ colour (median
$\Delta(g-i) = 0.1$ mag), which should be independent of the NIR data. Coupled
with the fact that the models do not provide a satisfactory description of the
observed SEDs (see Figure \ref{fig:residuals}), we are thus obliged to
consider the NIR-derived SP parameters as suspect. \label{fig:optnir}}
\end{figure*}

\begin{figure*} \centering \includegraphics[width=15cm]{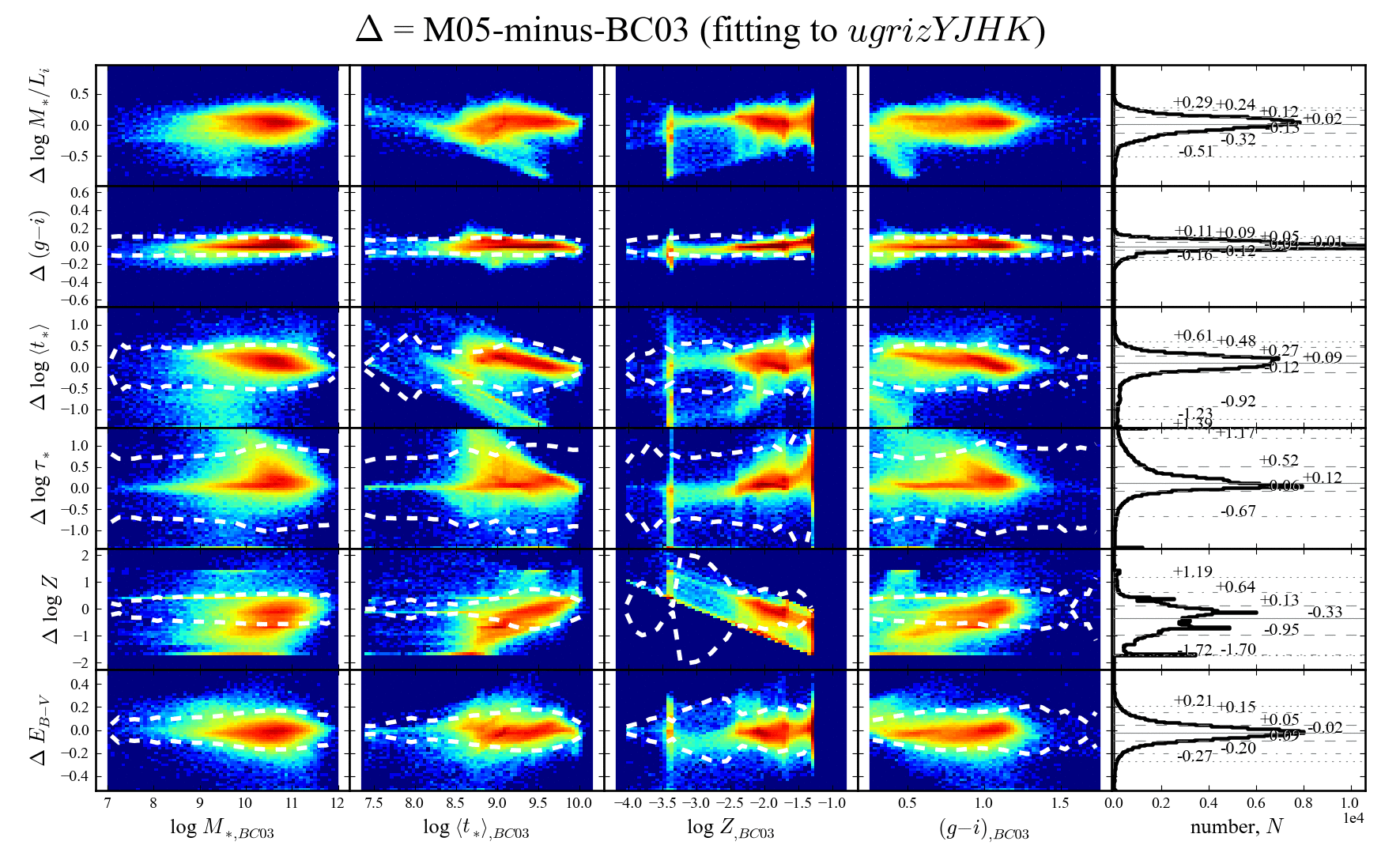}
\caption{Comparison between M05- and BC03-based parameter estimates, derived
from fits to the $ugrizYJHK$ SEDs.--- The `$\Delta$'s in this Figure should be
understood as the M05-based-minus-BC03-based parameter values; all symbols and
their meanings are analogous to Figure \ref{fig:optnir}. While, as expected,
there are some systematics in the inferred values of $M_*$ as a function of
$\lwage$, the global agreement is very good. In comparison to Figure
\ref{fig:optnir}, the systematic differences between parameters derived on the
basis of the different stellar evolution models (but the same $ugrizYJHK$
data) are considerably larger than the differences between the values inferred
with or without the inclusion of the NIR data (but the same stellar evolution
models). This suggests that our apparent inability to adequately fit the
observed optical-to-NIR SED shapes of GAMA galaxies is not a product of errors
in the stellar evolution models. At the same time, however, the `random'
differences between the M05- and BC03-based SP parameter estimates are larger
than the formal uncertainties when the NIR data are included. This is not true
when using only the optical data. That is, the model dependence of the SP
parameter estimates becomes significant if, and only if, the NIR data are
included. \label{fig:bc03m05}} \end{figure*}

The fits to the mock galaxies' optical SEDs also over-predict the `true' NIR
fluxes, but, as can be seen in Figure \ref{fig:simresiduals}, in a
qualitatively different way to what we see for real galaxies. In the case of
the mock galaxies, the offset between the predicted and actual NIR fluxes is a
much smoother function of rest-frame wavelength, as might be expected from
simple extrapolation errors. This is in contrast to the sharp discontinuity in
the residuals seen in Figure \ref{fig:residuals} between the optical and NIR
bands.

Looking now at the righthand panels of Figure \ref{fig:residuals}, we see that
none of the three stellar population libraries are able to satisfactorily
reproduce the optical--NIR SED shapes of GAMA galaxies without significant
systematic biases. Each of the models shows a significant excess of flux for
7000 \AA $\lesssim \lambda \lesssim$ 12000 \AA. The significance of the
offsets in the $i$-, $Y$-, and $J$-bands are $\sim +1.8 \sigma$, $-1.6
\sigma$, and $-1.4 \sigma$, respectively. Based on our numerical experiments,
there is no reason to suspect that we should be unable to reproduce the
observed optical--to--NIR SED shapes of real galaxies. As can be seen in
Figure \ref{fig:simresiduals}, the fits to the mock $ugrizYJHK$ SEDs are near
perfect.

Each of the issues highlighted above point to inconsistencies between the
optical--to--NIR colours of our SPL models on the one hand, and of real
galaxies on the other. Further, the fact that the models fail to
satisfactorily describe the NIR data immediately calls into question the
reliability of parameter estimates derived from fits to the full
optical-to-NIR photometry. The rest of this section is devoted to exploring
the nature of this problem.

\subsection{How including NIR data changes the parameter estimates} \label{ch:nirvalues}

\textbf{Figure \ref{fig:optnir}} shows the difference between stellar
population parameters derived from the $ugriz$ and the $ugrizYJHK$ photometry,
and using the \citet{BC03} models to construct our stellar population library
as per \secref{ch:sedfit}. In this Figure, the `$\Delta$'s plotted on the
$y$-axis should be understood as the 9-band--minus--5-band-derived value;
these offsets are plotted as a function of the 5-band-derived value. 

In the simplest possible terms, the 9-band fits yield systematically lower
values for all of $M_*$, $M_*/L_i$, $\dust$, $\lwage$, and $(g-i)$ than the
5-band fits. Again, based on our experiences with the mock catalogues
described in Appendix \ref{ch:mocks}, we have no reasons to expect these sort
of discrepancies: for the mocks, we are able to recover the input SP
parameters with virtually no systematic bias using either the optical--only or
optical--plus--NIR SEDs (see Figure \ref{fig:reliable}).

In each case, based on the formal uncertainties from the 5-band fits, the
median significance of the offset in the SP parameter estimates is $\gtrsim
1.5 \sigma$. For each of these quantities, the 9-band-derived value is
formally inconsistent with the 5-band-derived value at the $> 3 \sigma$ level
for $\approx$ 25 \% of galaxies. (Using the \citet{M05} models, we find a
similar fraction; using the \citet{CB07} models, this fraction goes up to
30--40 \%.) This shows that the residuals seen in Figure \ref{fig:residuals}
are more than merely a cosmetic problem---they are symptomatic of
inconsistencies between the fits with and without the inclusion of the NIR.

To make plain the importance of these systematic offsets, consider the fact
that there are large and statistically significant differences in the $(g-i)$
colours inferred from the 9-band and 5-band fits. The median values inferred
from the fits with the NIR included are 0.10 mag bluer than those based on the
optical alone. In comparison to the formal uncertainties in the 5-band derived
values of $(g-i)$, this amounts to an inconsistency at the $\sim 2.5 \sigma$
level. And this is despite the fact that the NIR data {\em by definition}
contain no information about $(g-i)$. Looking at Figure \ref{fig:residuals},
it is clear that the 5-band fits are a more reliable means of inferring a
restframe $(g-i)$ colour: for the 9-band fits, the differential offset between
the $g$ and $i$ bands is $\approx 0.10$ mag; for the 5-band fits, the
differential offset is $\lesssim 0.03$ mag.

Said another way, because the 9-band fits have the wrong SED shape, they
cannot be used to infer a restframe colour. But the same is true of any other
derived property---simply put, {\em if the models cannot fit the data, they
cannot be used to interpret them.}


\subsection{The sensitivity of different SSP models to the inclusion/exclusion of NIR data} \label{ch:sspmodels}

One possible explanation for the large residuals seen in Figure
\ref{fig:optnir} is problems with the \citet{BC03} SSP models. In particular,
one might worry that these are related to the NIR contributions of TP-AGB
stars. In this context, let us begin by noting that if this were to be the
source of the problems that we are seeing, then we would expect the
optical--only fits to underpredict the `true' NIR fluxes, particularly for the
\citet{BC03} models. But this is not what we see: the optical--only fits
{\em over}predict galaxies' NIR fluxes using both the \citet{BC03} and the
\citet{M05} models, and by similar amounts in both cases.

In \textbf{Figure \ref{fig:bc03m05}}, we show the comparison between the
\citet{M05}- and \citet{BC03}-derived SP parameter values, based on fits to
the full $ugrizYJHK$ SEDs. It is clear from Figure \ref{fig:bc03m05} that
there are systematic differences between the models, particularly (and as
expected) for $\lwage \sim 10^{8.5}$---$10^{9.5}$ Gyr.

Taking an empirical perspective on the problem, we can consider these
differences as an indication of the degree of uncertainty tied to
uncertainties in the stellar evolution tracks that underpin the SSP spectra.
Using only the optical data, the {\em systematic} differences between any of
the SP parameter values derived using the different models is small: for
$M_*/L_i$, the median offset is 0.01 dex. That is, when using optical data
only, these famously `disagreeing' models yield completely consistent results.
This is in marked contrast to a number of results emphasising the importance
of differences in the modelling of TP-AGB stars in the \citet{BC03} and
\citet{M05} models when NIR data are used \citep[\eg][]{Cimatti2008,
Wuyts2009}. In terms of `random' differences, the inferred values of $M_*$
based on the two sets of models agree to within $\pm 0.3$ dex (a factor of 2)
for 99 \% of galaxies. We can treat the 15/85 percentile points of the
distribution of the `$\Delta$'s as indicative of the 1$\sigma$ random `error'
associated with the choice of SSP model. For $M_*/L$, this `error' is
$\lesssim 0.10$ dex. That is, when using only optical data, the SP parameter
estimates are not significantly model dependent.

When we include the NIR data in the fits, the agreement between the SP values
inferred using the two different sets of SSP models is not as good. The
inferred values of $M_*$ using the \citet{BC03} or \citet{M05} models agree to
within $\pm 0.5$ dex (a factor of 3) for 99 \% of galaxies; the 1$\sigma$
random `error' in $M_*$ is $\approx 0.12$ dex. While the inferred values of
$M_*/L$ agree reasonably well, the differences in the other inferred stellar
population parameters--- $\dust$, $Z$, and especially $\lwage$---are larger.
For $\lwage$, the $1 \sigma$ `error' is $\approx 0.3$ dex; this should be
compared to the formal uncertainty in $\lwage$ of $\sim 0.2$ dex. Thus we see
that the `error' in SP parameter estimates associated with the choice of model
becomes comparable to the formal uncertainties when, and only when, NIR data
is included in the fit.


\subsection{What is the problem with the NIR?} \label{ch:badnir}

What can have possibly gone wrong in the fits to the NIR data? There are (at
least) three potential explanations for our inability to obtain a good
description of the optical--NIR SED shapes of GAMA galaxies using the models
in our SPL. The first is problems in the data. The second is problems in the
stellar evolution models used to derive the SSP spectra that form the basis of
our template library. The third is problems in how we have used these SSP
spectra to construct the CSPs that comprise our SPL.

\subsubsection{Is the problem in the data?  Maybe.}

We cannot unambiguously exclude the possibility of errors in, for example, the
basic photometric calibration of the NIR imaging data. In this context, we
highlight the qualitative difference in our ability to use optical data to
predict NIR fluxes for the real GAMA galaxies on the one hand, and for mock
galaxies on the other. In particular, the sharp discontinuity in the residuals
between the $z$- and $Y$-bands for the real galaxies would seem to suggest a
large inconsistency between these two bands at the level of $\sim 0.1$--$0.2$
mag.

As described in \secref{ch:photdata}, GAMA has received the NIR data fully
reduced and calibrated. In order to ensure that there are not problems in our
NIR photometric methods (which are not different from those in the optical),
we have verified that there are no large systematic offsets between our
photometry and that produced by CASU. This would suggest that any
inconsistencies would really have to be in the imaging data themselves. 

The accuracy of the UKIRT WFCAM data calibration has been investigated by
\citet{Hodgkin2009} through comparison to sources from the 2MASS point source
catalogue \citep{Cutri2003, Skrutskie2006}: they argue that the absolute
calibrations of the $Y$- and $JHK$-band are good to $\sim 2$ and $\sim 1.5$
\%, respectively. \citep[See also XXX][XXX.]{} Taken at face value, this
argues against there being such large inconsistencies in the photometry.

In light of the fact that we have not been directly involved in the reduction
or calibration of these data, and with the anticipated availability of the
considerably deeper VISTA-VIKING NIR imaging in the near future, we will not
investigate this further here.

\subsubsection{Is the problem in the SSP models? Probably not.}

From what we have already seen, we can exclude errors in the SSP models as a
likely candidate. We have shown in Figure \ref{fig:residuals} that {\em none}
of the \citet{BC03}, \citet{M05}, or \citet{CB07} models provides a good
description of the full optical-to-NIR SEDs of real galaxies---these models
all show qualitatively and quantitatively similar fit residuals. Taken
together, the results in Figures \ref{fig:optnir} and \ref{fig:bc03m05} show
that (for the same data) the SP parameters derived using different models show
small systematic differences, while at the same time (for any given set of SSP
models) there is a large systematic difference between the values derived with
or without the NIR data. This is not to say that the models are perfect, but
the offsets seen in Figure \ref{fig:residuals} would appear to be larger than
can be explained by uncertainties inherent in the SSP models
themselves.\looseness-2

\subsubsection{Is the problem in the construction of the SPL?  Probably.}

This leaves the third possibility that the assumptions that we have made in
constructing our SPL are overly simplistic, in the sense that they do not
faithfully describe or encapsulate the true mix of SPs found in real galaxies.
We defer discussion of this possibility to \secref{ch:future}. For now,
however, we stress that the present SPL {\em does} seem to be capable of
describing the optical SED shapes of real galaxies.

\subsection{Summary---why the NIR (currently) does more harm than good}
\label{ch:nonir}

We have now outlined three reasons to suspect that, at least in our case,
SP parameter estimates based only on optical photometry are more robust than if we were to include the NIR data:
\begin{enumerate}
\item Regardless of which set of SSP models we use, we see much larger than
expected residuals in the SED fits when the NIR data are included. If the
models do not provide a good description of the data, then we cannot
confidently use them to infer galaxies' SP properties.
\item The consistency between the SP parameter estimates derived with or
without the inclusion of the NIR data is poor. For a sizeable fraction of GAMA
galaxies ($\gtrsim 25$ \%), the SP parameter values inferred from fits to the
optical--plus--NIR SEDs are statistically inconsistent (at the $3 \sigma$
level) with those based on the optical alone.
\item When using different models to construct the SPL templates, the
agreement between the derived SP parameters is very good when the NIR data are
excluded, but considerably worse when the NIR data are included. That is, the fit results become significantly model-dependent when, and only when, we try to include the NIR data.
\end{enumerate}
For these reasons, and for the time being, we choose not to use the NIR data
when deriving the stellar mass estimates. This begs the question as to how
accurately $M_*/L$ can be constrained based on optical data alone, which is
the subject of the next Section.

\section{The theoretical and empirical relations between $\mathbf{M_*/L}$ and colour} \label{ch:gimoverl}

As we have said at the beginning of \secref{ch:nir}, conventional wisdom says
that NIR data provides a better estimate of stellar mass. Our conclusion in
\secref{ch:nir}, however, is that we are unable to satisfactorily incorporate
the NIR data into the SPS calculation. With this as our motivation, we will
now look at how well $M_*/L$ can be constrained on the basis of optical data
alone. In particular, we want to know whether or to what extent the accuracy
of our stellar mass estimates is compromised by our decision to ignore the NIR
data.

\subsection{Variations in $\mathbf{M_*/L}$ at different wavelengths} \label{ch:moverls}

Part of the rationale behind the idea that the NIR provides a better estimate
of $M_*/L$ is that galaxies show less variation in their NIR $M_*/L$s than
they do in the optical. We address this issue in \textbf{Figure
\ref{fig:moverls}}; this Figure merits some discussion. Each panel of Figure
\ref{fig:moverls} shows a subsample of the models in our SPL. Within each
panel, models are colour-coded according to their metallicity (from the lowest
metallicity in blue to the highest metallicity in red). For each metallicity,
the slightly heavier line shows how the single burst (\ie, $\tau \rightarrow
0$) track evolves with time, $t$; the other single-colour lines then connect
models with the same age (but different $\tau$s) or the same SFH $e$-folding
time (but different ages). Finally, the colour-graded lines connect models
with the same $t$ and $\tau$, but different metallicities. In this way, each
panel shows a 2D projection of the 3D ($t$, $\tau$, $Z$) grid of SPL
templates. Note that we only show zero-dust models in this Figure; the
dust-extinction vector is shown in the lower-right corner of each panel.

Each row of Figure \ref{fig:moverls} shows the mass--to--light ratio in
different bands ($uiJK$, from top to bottom). Let us look first at the first
column, in which we plot each of these $M_*/L$s as a function of time. For
fixed $Z$ and $\tau$, and particularly for $\age \gtrsim 2$ Gyr, it is true
that the NIR $M_*/L$ varies less with $t$ than does the optical $M_*/L$---but
not by all that much. For the SSP models, the total variation in $M_*/L$
between 2 and 10 Gyr is $\lesssim 1.2$ dex in the $u$ band, compared to
$\lesssim 0.7$ dex in the $i$-band, and $\approx 0.6$ dex in the $J$- and
$K$-bands. Similarly, it is also true that at fixed $t$ and $Z$, the spread in
$M_*/L$s for different $\tau$s is slightly smaller for longer wavelengths: the
total variation goes from $\lesssim 1.2$ dex in the $g$-band to $\lesssim 0.4$
dex in the $i$-band, to $\lesssim 0.3$ dex in the $J$- and $K$-bands.
Considering variations in $M_*/L$ with all of $t$, $\tau$, and $Z$, the full
range of $M_*/L$s becomes 2.7, 1.4, 1.2, and 1.3 dex in the $uiJK$-bands,
respectively; these values imply mass accuracy on the order of factors of 22,
5.5, 4.0, and 4.5.

While it is thus true that galaxies tend to show less variation in their
values of $M_*/L$ towards redder wavelengths \citep[see also][]{BelldeJong},
it seems that the most important thing is to use a band that is redder than
the 4000 \AA\ and Balmer breaks---the range in $M_*/L_i$ is not all that much
greater than that in $M_*/L_K$ or $M_*/L_J$.

\subsection{The generic relation between $\mathbf{M_*/L}$ and restframe colour} \label{ch:gimodels}

\begin{figure} \includegraphics[width=7.8cm]{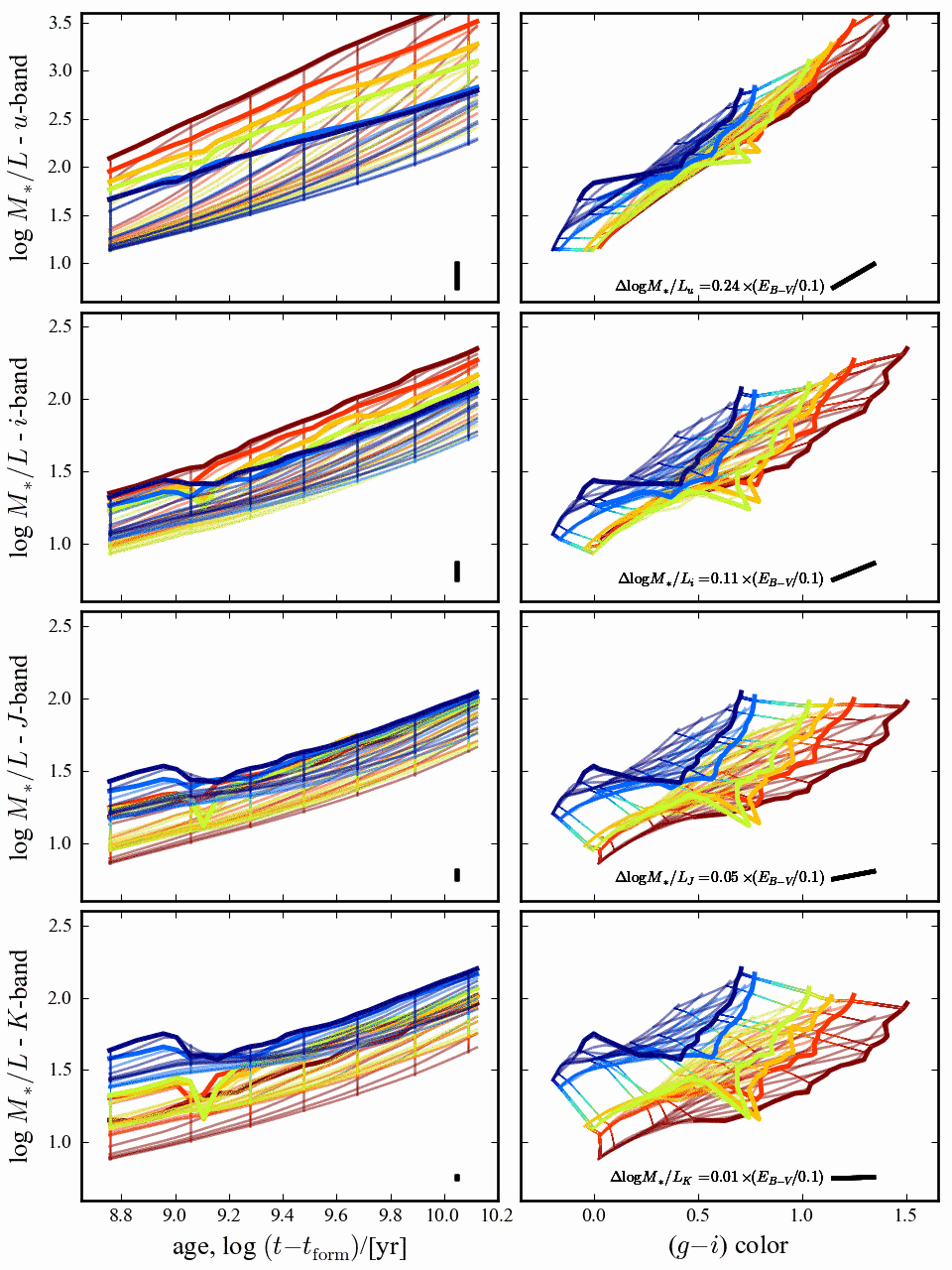} \centering
\caption{Variations in $M_*/L$ as a function of age and colour for models in
our SPL.--- This Figure is discussed at length in
\secref{ch:moverls} and \secref{ch:gimodels}. Each panel of this Figure shows
a projection of our SP model grid. Individual models are colour-coded by their
metallicity. The lines connect models that differ in only one of age,
$e$-folding time, or metallicity. We only show zero-dust models; the dust
vector is given in the lower-right-hand corner of each panel. From top to
bottom, the different rows show the spread of $M_*/L$s in the $u$-, $i-$, $J$
and $K$-bands within the models; these values are plotted as a function of SP
age (left panels), and restframe $(g-i)$ colour (right panels). The results
shown in this panel imply that $u$-, $i$-, $J$-, or $K$-band fluxes on their
own can be used to infer stellar masses to within a factor of 22, 5.5, 4.0,
and 4.5, respectively. That is, on its own, $L_i$ is almost as good an
indicator of $M_*$ as $L_J$ or $L_K$. At the same time, the variation in
$M_*/L_i$ {\em at fixed} $(g-i)$ is considerably smaller: $M_*/L_i$ can in
principle be predicted to within a factor of 2 using $(g-i)$ colour alone.
\label{fig:moverls} } \end{figure}

Let us turn now to the second column of Figure \ref{fig:moverls}, where we
show the relation between $M_*/L$ and restframe $(g-i)$ colour (\cf, \eg,
Figure 2 of Bell \& de Jong 2001; Figure 1 of Zibetti et al.\ 2009). The
principal point to be made here is that, {\em at fixed} ($g-i$), the range
$M_*/L_i$ is $\lesssim 0.5$ dex, whereas, and particularly for blue galaxies,
the spread in the NIR $M_*/L$ is more like 0.65---1.0 dex. That is, by the
same argument we have used above, using only $g$- and $i$-band photometry, it
is possible to derive stellar mass estimates that are accurate to within a
factor of $\lesssim 2$.

\subsubsection{The effects of dust}

In what we have said so far in this Section, we have completely ignored dust.
This may have seemed like a very important oversight, so let us now address
this issue. The dust vector in $(g-i)$--$M_*/L_i$ space is ($\Delta (g-i), ~
\Delta (\log M_*/L_i)$) = (0.19, 0.11) $\times \dust / 0.1$. Compare this to
the empirical $(g-i)$--$M_*/L_i$ relation for GAMA galaxies, which, as we show
in \secref{ch:colourrel} below, has a slope of 0.73. Because these two vectors
are roughly aligned, the first order effect of dust obscuration is merely to
shift galaxies along the $(g-i)$--$M_*/L_i$ relation \citep[see
also][]{BelldeJong, Nicol2010}. This means that the accuracy of
$(g-i)$-derived estimates of $M_*/L_i$ are not sensitive to a galaxy's precise
dust content. Said another way, although there may be large uncertainties in
$\dust$, this does not necessarily imply that there will also be large
uncertainties in $M_*/L_i$.

To see this clearly, imagine that we were only to use zero-dust models in our
SPL, and take the example of a galaxy that in reality has $\dust = 0.1$ mag.
In comparison to the zero-dust SPL model with the same $\age$, $\tau$, and
$Z$, this dusty galaxy's $(g-i)$ colour becomes 0.19 mag redder, and its
absolute luminosity drops by 0.11 dex; the effective $M_*/L_i$ is thus
increased by the same amount. (Recall that $L_i$ denotes the effective
absolute luminosity {\em without} correction for internal dust obscuration,
rather than the intrinsic luminosity produced by all stars.) Now, using the
slope of the $(g-i)$--$M_*/L_i$ relation, the inferred value of $M_*/L_i$ for
the $\dust = 0.1$ mag galaxy will be $0.70 \times 0.19 = 0.13$ dex higher than
it would be for the same galaxy with no dust. That is, in this simple thought
experiment, the error in the value of the effective $M_*/L_i$ implied by
$(g-i)$ would be $0.02 \times (\dust / 0.1)$ dex, even though we would be
using completely the wrong kind of SPS model to `fit' the observed galaxy.

Note that this argument holds only to the extent that dust can be accurately
modelled using a single dust vector; \ie a single screen approximation. Using
the \citet{FischeraDopita} attenuation curve, which is a single screen
approximation to a fractal dust distribution, does not produce a large change
in the derived values of $M_*/L$. Using the model of \citet{Tuffs2004} and
\citet{Popescu2000}, \citet{Driver2007} show how variations in both viewing
angle and bulge--to--disk ratio can produce a spread in the color--$M_*/L$
dust vectors. These results suggest that in some cases, dust geometry may have
a significant effect on the colour-inferred value of $M_*/L$, at the level of
$\Delta M_*/L_V \sim 0.1 \dust$. One avenue for further investigation of the
effects of dust geometry is through detailed radiative transfer modelling for
galaxies whose geometries can be accurately constrained. While this is clearly
impractical for large galaxy samples, an alternative would be to construct
spatially resolved mass maps \citep[see, \eg][]{Conti2003, LanyonFoster2007,
Welikala2008, Zibetti2009}, using a single-screen approximation for each pixel
individually.

\begin{figure*} \centering \includegraphics[width=8.0cm]{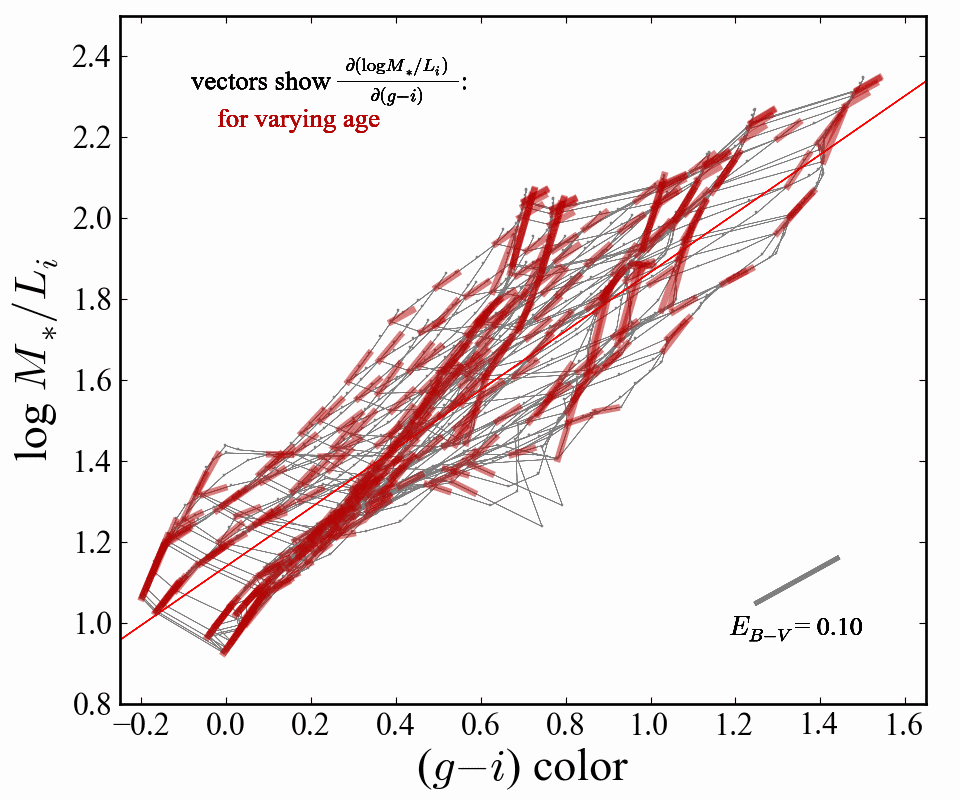}
\includegraphics[width=8.0cm]{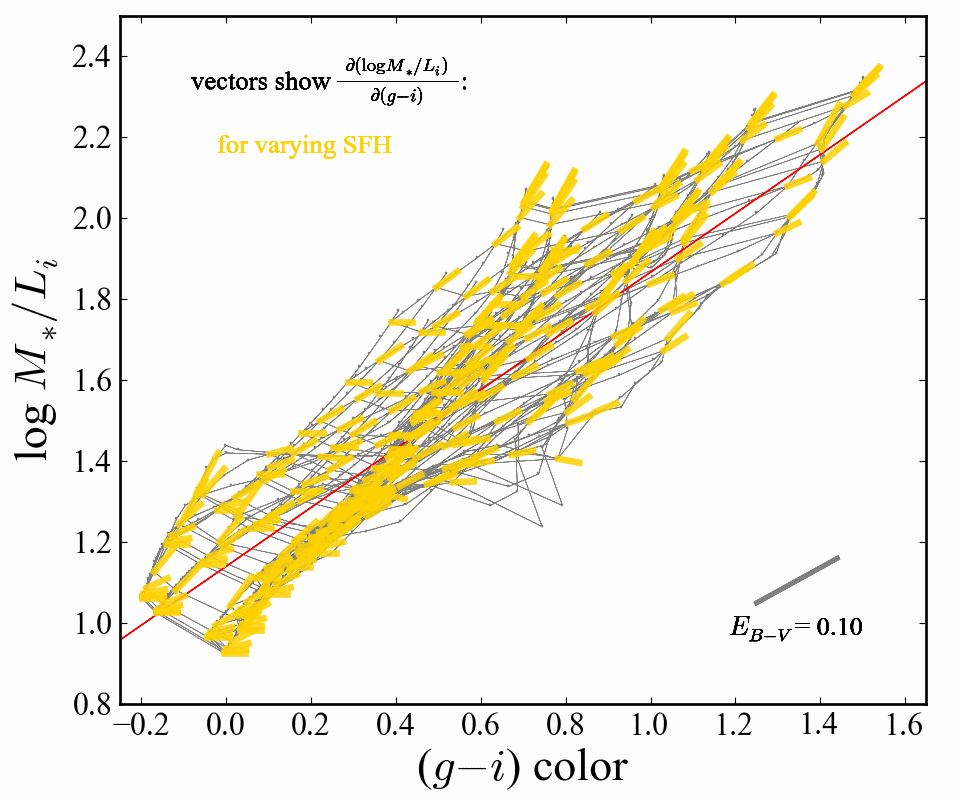}
\includegraphics[width=8.0cm]{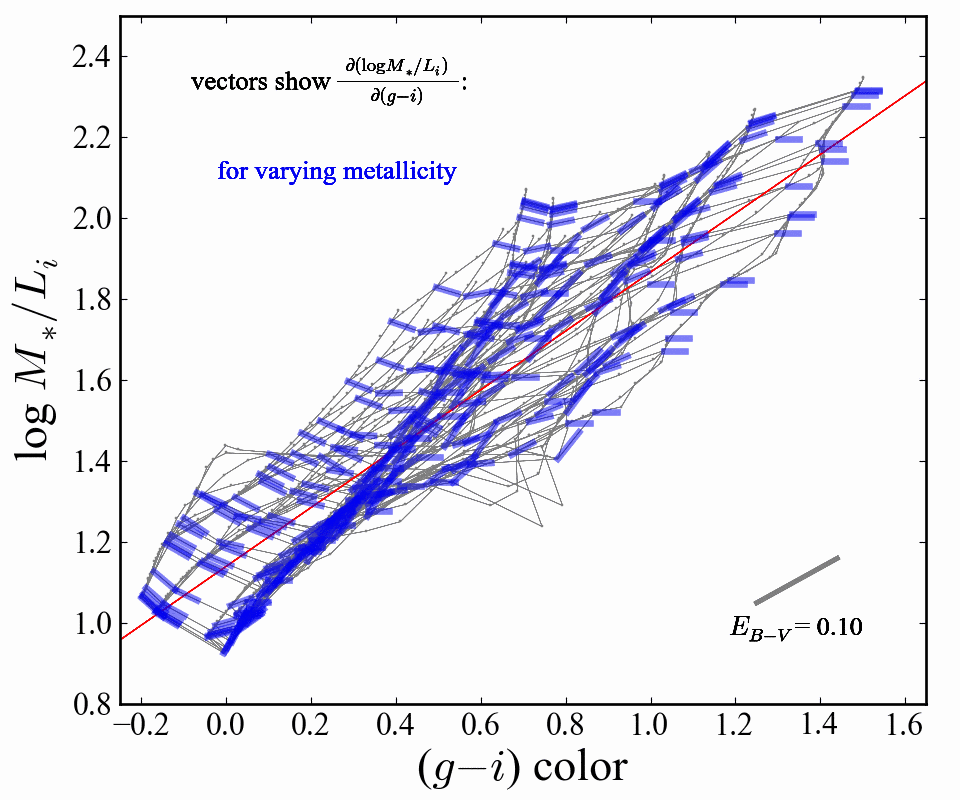}
\includegraphics[width=8.0cm]{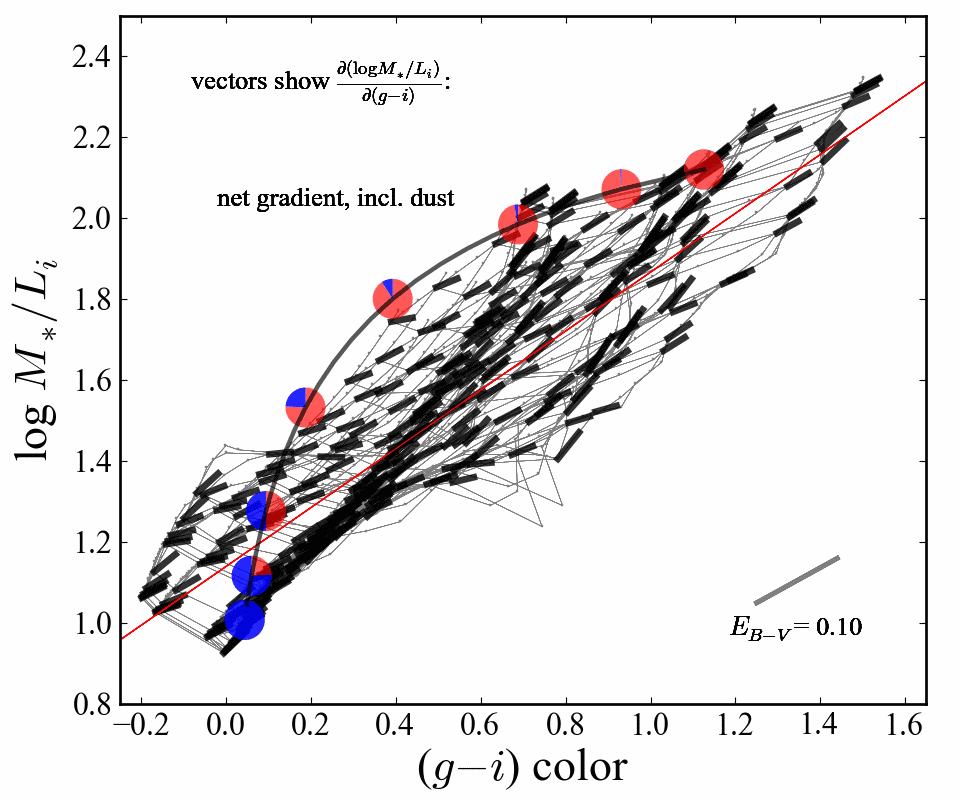} \caption{The generic relation between
restframe $(g-i)$ colour and $M_*/L_i$ for the models in our SPL.--- The first
three panels of this Figure show the movement of the models in our SPL in the
$(g-i)$--$M_*/L_i$ plane with variations age, SFH, and metallicity,
respectively. In each of these panels, the vectors show the age, SFH, or
metallicity analogues of the dust vector for $\Delta(g-i) = 0.05$; that is,
the change in $M_*/L_i$ that is associated with a change in $\age$, $\tau$, or
$Z$ such that the restframe $(g-i)$ colour changes by 0.05 mag. In the fourth
panel, the black vectors show the net variation in $\Delta M_*/L_i$ with a
0.05 mag change in $(g-i)$, obtained by marginalising over the $t$, $\tau$,
and $Z$ priors. The fact that each of these vectors---both individually and
{\em en masse}---are roughly aligned with one another means that variations in
any of these parameters largely preserves both the slope of and scatter in the
relation between $(g-i)$ and $M_*/L_i$ (see \secref{ch:gimodels}). Further, in
the final panel, we provide a qualitative illustration of how multi-component
SPs affect the $(g-i)$--$M_*/L_i$ relation by combining an old and passive SP
with a young, star forming one. These are shown as the large circles connected
by the smooth black curve. The relative young:old mass ratio is indicated by
the relative area of the blue and red regions within each circle for young
mass fractions of approximately (right to left) 0, 1, 3, 9, 24, 50, 76, and
100 \%. Even for this rather extreme example of multi-component stellar
populations, it remains true that $(g-i)$ colour can, in principle, be used to
estimate $M_*/L_i$ to within a factor of $\sim 2$ ($0.3$ dex). \looseness-1
\label{fig:gimodels} } \end{figure*}

\subsubsection{Dependence on $t$, $\tau$, and $Z$}

Just as $(g-i)$ can be used to estimate $M_*/L_i$ without being strongly sensitive to dust, variations in each of age, SFH, and metallicity do not have  a large impact on $(g-i)$-inferred estimates of $M_*/L_i$. This is
demonstrated in \textbf{Figure \ref{fig:gimodels}}, in which we use our SPL
models to show how variations in any one of $\age$, $\tau$, or $Z$ shift
galaxies in the $(g-i)$--$M_*/L_i$ plane. We do this as follows: for an
individual model, we ask how great a change in any one of $\age$, $\tau$, or
$Z$ (while holding the other two parameters fixed) is required to change
$(g-i)$ by 0.05 mag; we then look at the corresponding change in $M_*/L_i$
that comes with this variation. In other words, we are looking at how
uncertainties in each of $\age$, $\tau$, and $Z$ affect the accuracy of
$(g-i)$--inferred estimates of $M_*/L_i$. These are shown as the red, yellow,
and blue vectors, respectively.

Focussing on each set of vectors individually, it can be seen that for the
bulk of the models, the effect of independent variations in any of $\age$,
$\tau$, or $Z$ is to move the model point more or less parallel to the main
cloud. Note, too, that closer to the centre of the main cloud, the three
separate vectors tend to come into closer alignment. 

To first order, then, variations in any one of these quantities simply shift
galaxies along the main $(g-i)$--$M_*/L_i$ relation. By the same argument
presented above in regard to dust, this implies that $(g-i)$ can be used to
infer $M_*/L_i$ to high accuracy, even if the `best fit' model of the same
$(g-i)$ colour has completely the wrong values of $\age$, $\tau$, or $Z$.
Figure \ref{fig:gimodels} thus shows that, uncertainties in any of $\age$,
$\tau$, and $Z$ do not produce large errors in the value of $M_*/L_i$ inferred
from the $(g-i)$ colour.

\subsubsection{The net covariance between $M_*/L_i$ and $(g-i)$}

Considering the combined effect of variations in all three of $\age$, $\tau$,
and $Z$, the robustness of $(g-i)$--derived estimates of $M_*/L_i$ is even
greater. The black vectors in this plot show the net variation in $M_*/L_i$
allowing for the uncertainties in all of $\age$, $\tau$, and $Z$ that come
with an observational uncertainty of $\Delta(g-i) = 0.05$. Notice how closely
aligned these vectors are with the empirical $(g-i)$--$M_*/L_i$ relation for
real galaxies.

This shows that while there may well be a relatively large range of models
with different values of $t$, $\tau$, and/or $Z$ that are consistent with the
observed value of $(g-i)$ for any given galaxy, because all of these models
will follow more or less the same relation between $(g-i)$ and $M_*/L_i$, the
spread of $M_*/L_i$s among these models will still be relatively low. That is,
through a coincidence of dust and SP evolution physics, {\em the
dust--age--metallicity degeneracy actually helps in the estimate of} $M_*/L_i$
\citep[see also, \eg][]{BelldeJong, Nicol2010}. Furthermore, as a corollary to
this statement, because the estimated value of $M_*/L_i$ does not depend
strongly on the accuracy of the inferred values of $\age$, $\tau$, $Z$, or
$\dust$, it is not necessary to model these aspects of the SPL models exactly.

\subsubsection{Multicomponent stellar populations}

The last commonly-cited bugbear of stellar mass estimation is the effect of
`secondary' populations in general, and of bursts in particular, on the
inferred value of $M_*/L$. To explore this issue, consider the case of a
combination of two SPs with ($\age$, $\tau$, $Z$) = (10 Gyr, 0.5 Gyr, $Z$\sol)
and (0.5 Gyr, 30 Myr, $Z$\sol); \ie, an old and passive SP and a very young
and star forming SP. These two individual SPs are highlighted in the fourth
panel of Figure \ref{fig:gimodels}. Now let us combine these two SPs in
varying amounts. The track connecting these two points show where the combined
SP would lie in the $M_*/L_i$--$(g-i)$ plane. The large points highlight the
cases where the mass of the burst component is $10^{-2}$, $10^{-1.5}$, ...,
$10^{0.5}$ times the mass of the old component.

We again see that the $M_*/L_i$--$(g-i)$ relation is largely preserved. In
particular, the effect of small bursts ($f_\mathrm{young} \lesssim 0.05$) on
the $M_*/L_i$--$(g-i)$ colour relation is very slight. At least for the
specific case that we have chosen to illustrate, we see that if a very old
galaxy were to experience a modest burst ($0.05 \lesssim f_\mathrm{young}
\lesssim 25$ \%), then this will shift the galaxy along the edge of the main
cloud. The greatest concern is a largely star-forming galaxy with a sizeable
underlying population of very old stars ($0.25 \lesssim f_\mathrm{old} >
\lesssim 0.75$). In this case, $M_*/L_i$ can shift by up to 0.5 dex with only
a small change in colour; $\Delta (g-i) \lesssim 0.2$. 

It is thus clear from this panel that secondary stellar populations in
general, and those with $0.25 \lesssim f_\mathrm{young} \lesssim 0.75$ in
particular, are better able to shift galaxies away from the main
$M_*/L_i$--$(g-i)$ relation than are variations in any or all of $t$, $\tau$,
$Z$, or $\dust$ within one of our smooth and exponential CSPs. That said, we
stress this specific example represents something of an extreme case: the two
separate SPs we have chosen are at opposite ends of the SP values that we find
for real galaxies, and the problem is only significant where the two
populations are comparable in mass. Furthermore, even for this extreme
example, note that the combined SP still lies within the main cloud of SPs
within our SPL. The claim that $M_*/L_i$ can be estimated to within a factor
of $\lesssim 2$ based on the $(g-i)$ colour alone is true even for galaxies
with relatively complex SFHs.

\subsection{The empirical relation between $\mathbf{(g-i)}$ and $\mathbf{M_*/L_i}$ \label{ch:colourrel}}

\begin{figure*} \centering \includegraphics[width=16cm]{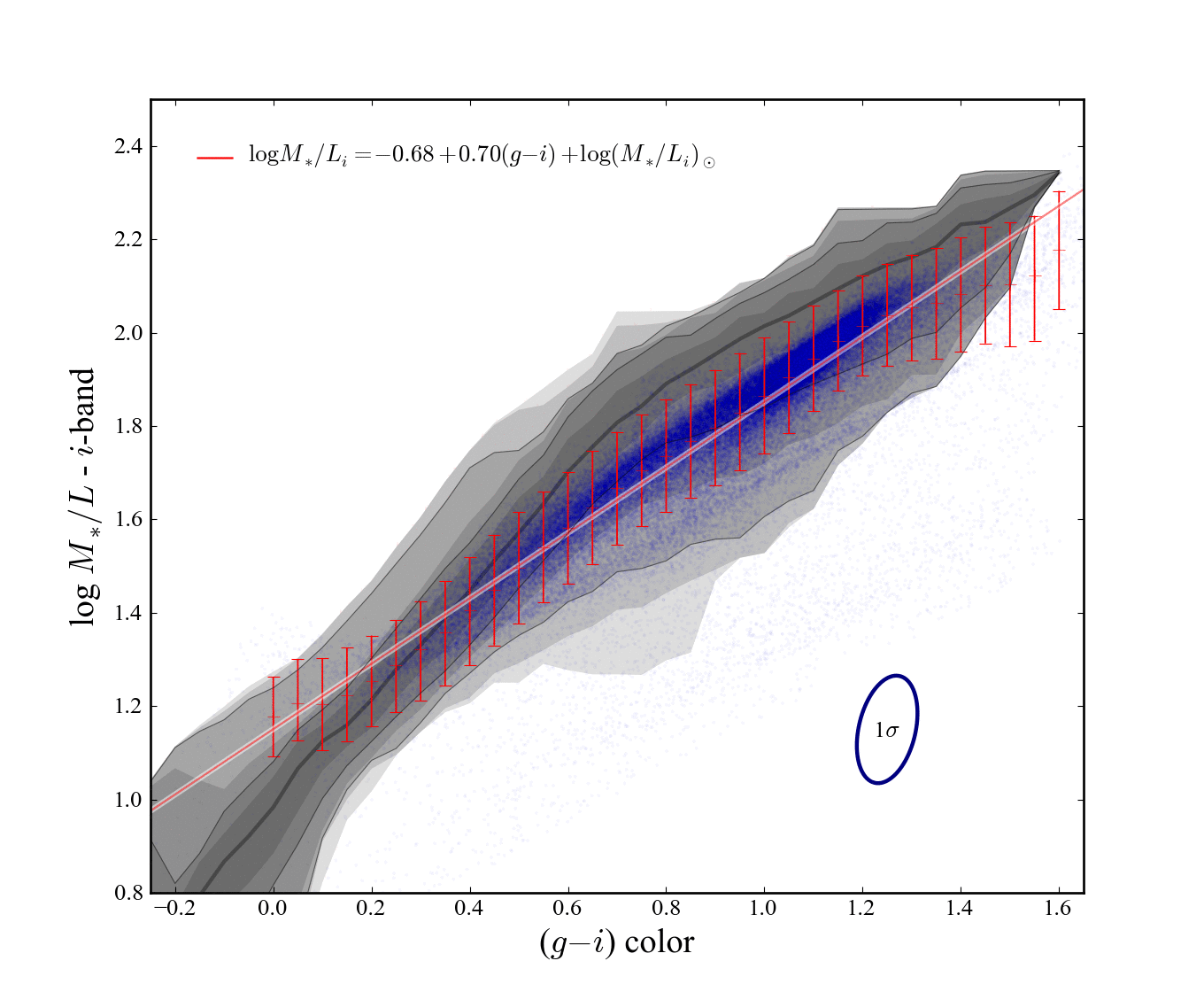} \centering
\caption{The empirical relation between restframe $(g-i)$ colour and
$M_*/L_i$.--- The points show the inferred values of $(g-i)$ and $M_*/L_i$ for
galaxies in the GAMA catalogue. The red points show the median value of
$M_*/L_i$ in narrow bins of $(g-i)$; the error bars on these points reflect
the median $\pm 1 \sigma$ upper and lower limits on $M_*/L_i$ in each bin.
These error bars thus reflect the combination of the (small) intrinsic scatter
in the $(g-i)$--$M_*/L_i$ relation and the (larger) observational
uncertainties in $M_*/L_i$ at fixed $(g-i)$. The solid red line shows the best
fit to the empirical relation between $(g-i)$ and $M_*/L_i$, with the form as
given at top left. In fitting this relation, we have fully accounted for the
covariant errors in $(g-i)$ and $M_*/L_i$; the median error ellipse is shown
at bottom right. For comparison, the underlaid greyscale contours show the
prior-weighted distribution of $M_*/L_i$s for the models in our SPL, computed
in narrow bins of $(g-i)$. The heavy central line shows the prior-weighted
median, and the contours are spaced at the equivalent of the $\pm 0.5$, 1.0,
... 3.0 $\sigma$ percentiles of the prior-weighted distribution. The observed
relation for real galaxies is both more nearly linear and tighter than might
be expected from the models alone. This implies both that the full $ugriz$ SED
contains additional information concerning a galaxy's stellar population not
embodied in the $(g-i)$ colour alone, and that the precise form of and scatter
around the $(g-i)$--$M_*/L_i$ relation is a product of galaxies' formation and
evolutionary histories. In this sense, we have calibrated the
($g-i$)--$M_*/L_i$ relation such that, modulo uncertainties in the stellar
population models used to derive these values, the $(g-i)$ colour can be used
to predict $M_*/L_i$ to a $1 \sigma$ accuracy of $\approx 0.10$ dex.
\label{fig:colourrel} } \end{figure*}

So far in this Section, our discussion has been restricted to the range of
$M_*/L$s and colours spanned by the models in our SPL, and has thus focussed
on the theoretical relation between $(g-i)$ and $M_*/L_i$. In this sense, our
conversation has been completely generic---at least insofar as the SP models
we have used provide an accurate description of the stellar content of real
galaxies. 

Let us now turn to \textbf{Figure \ref{fig:colourrel}}, and consider how well
$(g-i)$ can be used to estimate $M_*/L_i$ in practice. In this Figure, the
grey-scale contours show the prior-weighted distribution on $M_*/L_i$ for the
models, again in bins of $(g-i)$, as the grey-scale contours: the darkest line
shows the median prior-weighted value; the greyscale shows the equivalent of
the $\pm 0.5$, 1.0, ..., 3.0 $\sigma$ percentiles. This Figure also shows the
empirical relation between $(g-i)$ and $M_*/L_i$ that we infer for real GAMA
galaxies, based on our $ugriz$ SPS fits. The data themselves are shown as the
blue points.

Consider for a moment what would happen if we were to have only (restframe)
$g$- and $i$-band photometry for a real galaxy. The inferred value of
$M_*/L_i$ would just be the mean value of all models in our SPL with a similar
colour (weighted both by the consistency between the observed and model
photometry, and the prior probability of that model). Further, for a given
$(g-i)$, the inferred uncertainty in $M_*/L_i$ would simply reflect the
prior-weighted range of values spanned by models of the same colour. That is,
we would expect to `recover' the prior-weighted distribution shown as the grey
contours in Figure \ref{fig:colourrel}.

This is not what we see for the real galaxies: there are clear differences
between the form of and scatter around the relations between $(g-i)$ and
$M_*/L_i$ for the models on the one hand, and for the data on the other.
Particularly for intermediate colours, the median value of $M_*/L_i$ is
considerably lower than the probability-weighted mean of the models. This
demonstrates that the $u$-, $r-$, and $z$-band data do provide additional
information concerning $M_*/L_i$ that cannot be gleaned from $(g-i)$
alone.\footnote{This point is significant in terms of our assumed priors: if
the $urz$-band photometry did not provide additional SP information, then the
observed relation would be strongly dependent on the specific priors used;
particularly the assumed prior on $\tau$.}

Further, the effect of this additional information is to significantly {\em
reduce} the observed scatter in the relation between $(g-i)$ and $M_*/L_i$.
The red points in Figure \ref{fig:colourrel} show the median values of
$M_*/L_i$ for GAMA galaxies in narrow bins of $(g-i)$. The error bars on these
points show the mean values of the formal 1$\sigma$ upper and lower limit on
$M_*/L_i$ in each bin; these error bars thus show the intrinsic, astrophysical
scatter in the relation convolved with the formal, observational
uncertainties. Even considering the formal uncertainties for individual
galaxies, the scatter around the mean colour--$M_*/L_i$ relation is
considerably lower than what would expected from the models alone:
quantitatively, the scatter in $M_*/L_i$ is constrained to being $\lesssim
0.1$ dex for all values of $(g-i)$. This shows that, {\em the apparent
tightness of the relation between $(g-i)$ and $M_*/L_i$ is not a mere
consequence of the central limit theorem}.

In other words, the precise form of the empirical relation between $(g-i)$ and
$M_*/L_i$ encodes information about the distributions of SPs among real
galaxies. Both the linearity and tightness of the observed relation are
therefore fortuitous coincidences of the physics of galaxy formation and
evolution. 

Fitting the empirical relation for GAMA galaxies, we find:
\begin{equation}
		\log M_*/L_i = - 0.68 + 0.70 (g-i) ~ . \label{eq:colourrel}
\end{equation}
(To convert $L_i$ from our preferred AB-centric units to Solar units, note
that the absolute AB magnitude of the sun in the $i$-band is $M_{i,\odot} =
4.58$.) This fit is shown in Figure \ref{fig:colourrel} as the solid red line.
In the fitting, we have fully accounted for the covariant errors in the
derived values of $(g-i)$ and $M_*/L_i$; the mean error ellipse is shown in
the lower-right corner of this panel. Rearranging Equation \ref{eq:colourrel}
to put all observables to one side, we have in effect calibrated the empirical
relation between $(g-i)$ colour, $i$-band luminosity, and stellar mass as:
\begin{equation} \log M_* / [ M_\odot ] = - 0.68 + 0.70 (g-i) - 0.4 M_i
~ , \label{eq:colourrel2} \end{equation} where $M_i$ is the absolute
magnitude in the restframe $i$-band, expressed in the AB system. {\em This
relation can be used to estimate $M_*/L_i$ to a 1$\sigma$ accuracy of $\sim$
0.10 dex using (restframe) $g$- and $i$-band photometry alone.}\looseness-1

\subsection{Comparison with other recent works}

\begin{figure}\includegraphics[width=8.2cm]{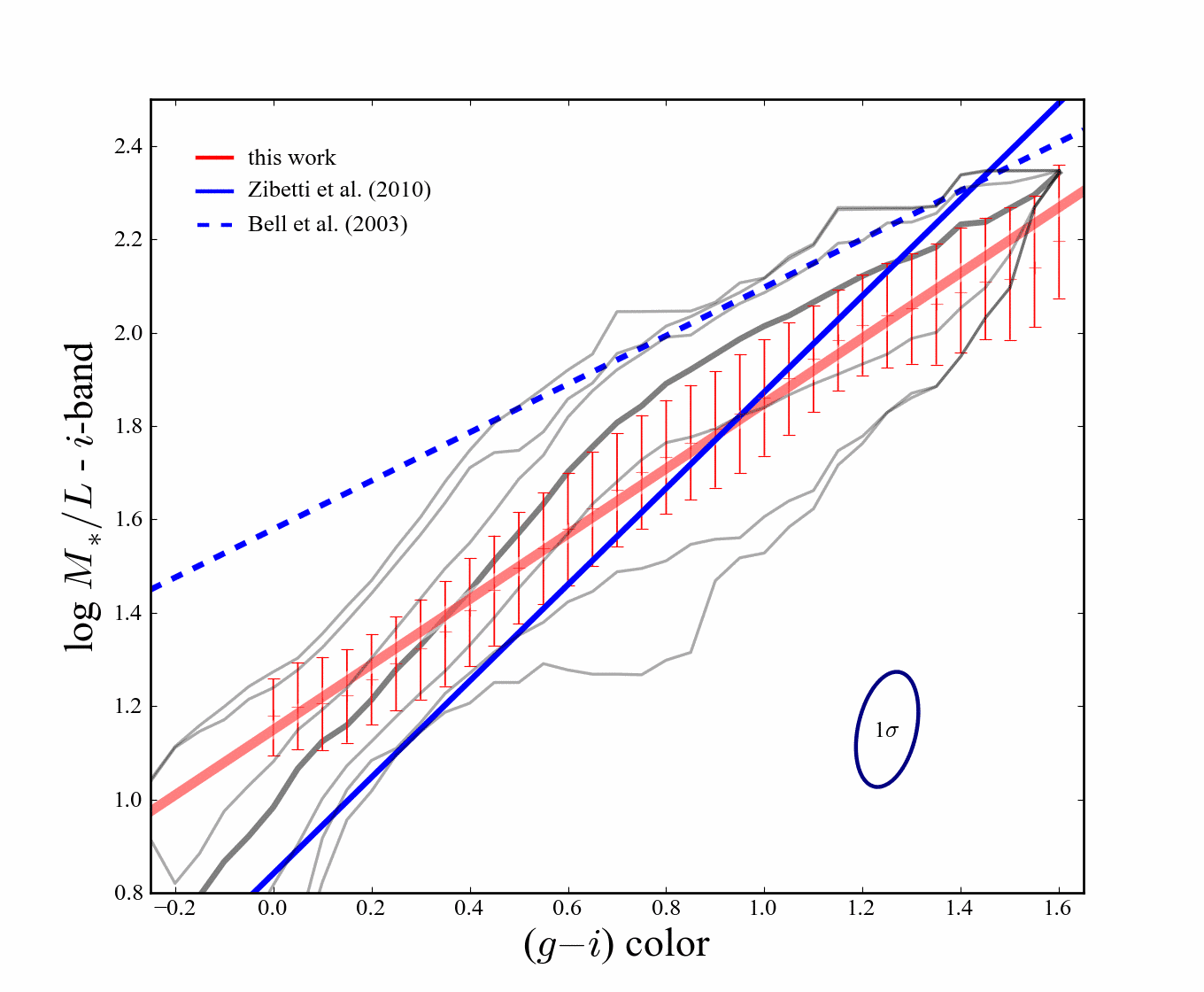} \caption{Comparison
between our empirical $M_*/L_i$--$(g-i)$ relation and other recent works.---
The dashed and solid blue lines in this Figure shows the relations between
$M_*/L_i$ and $(g-i)$ presented by \citet{Bell2003} and \citet{Zibetti2009},
respectively; all other symbols are as in Figure \ref{fig:colourrel}. The
\citet{Bell2003} relation has been derived for galaxies from the SDSS EDR, and should thus be compared to our empirical relation.
Note, however, that the \citet{Bell2003} relation should be understood to
include evolution corrections. The \citet{Zibetti2009} relation has been
obtained by marginalising over the models in their SPL in bins of $(g-i)$.
This relation thus cannot be considered to be `empirical', and should be
compared to the prior-weighted median for our SPL, shown as the heavier grey
line. The significant systematic differences between these relations
underscore the importance of ensuring that any cross-survey comparisons are
based on a comparable mass scale. \label{fig:giothers}} \end{figure}

In \textbf{Figure \ref{fig:giothers}}, we compare our empirically calibrated
$M_*/L_i$--$(g-i)$ relation to two other recent works. From the outset, let us
stress that these relations are {\em not} directly comparable, in the sense
that they have been derived in very different ways, and therefore should be
interpreted as having rather different meanings.

First, the dashed blue line shows the relation given by \citet{Bell2003},
which has been derived from least-squares SED fitting to $ugriz$ SEDs for
galaxies from the SDSS Early Data Release \citep[EDR][]{Stoughton2002} coupled
with $K$-band photometry from 2MASS \citep{Cutri2003, Skrutskie2006}, and
using Pegase \citep{pegase} SSP models as the basis of the SPL. We have scaled
the \citet{Bell2003} relation down by 0.093 dex to account for their use of a
`diet Salpeter' rather than \citet{Chabrier} IMF. Most importantly,
\citet{Bell2003} explicitly attempt to account for evolution between the epoch
of observation and the present day by running forward the implied SFH to
$z=0$. With that caveat, the \citet{Bell2003} relation is derived from fits to
the observed relation between $M_*/L$ and colour for real galaxies in a
similar way as in this work, and so can be compared to our relation, shown in
Figure \ref{fig:giothers} as the solid red line.

Second, the solid blue line shows the relation derived by \citet{Zibetti2009},
which is based on a Monte Carlo realised SPL modelled after \citet
{Kauffmann2003a} (\ie, including secondary SF bursts), with a sophisticated
treatment of dust extinction using the formalism of \citet{CharlotFall}. The
relation shown has been derived by marginalising over all SPL models in bins
of $(g-i)$. The \citet{Zibetti2009} relation should therefore be compared to
the heavier gray line in Figure \ref{fig:giothers}, which shows the
prior-weighted median value of $M_*/L_i$, computed in bins of $(g-i)$, for our
SPL. That said, there is still one important caveat: \citet{Zibetti2009} have
marginalised over their dust priors, whereas the gray line in Figure
\ref{fig:giothers} is based only on the zero-dust models in our SPL. Note that
the precise form of the \citet{Zibetti2009} line is determined almost entirely
by their assumed priors (\ie, the relative weighting given to the different
models in their SPL); no observational data has been used to derive this
relation.

The reader may be forgiven for being startled by the apparently poor agreement
between these relations in the first instance, and then equally by the
subtleties in their meanings. The point to take from this comparison is simply
that there are important systematic differences between each of these mass
determinations. (Although, again, we stress that we have shown our $M_*/L_i$
estimates to be in excellent agreement with the well-tested and widely used
SDSS values.) It is clear from Figure \ref{fig:giothers} that comparing
results based on different mass determination methods would not be fair, or at
best, would be misleading. The utility of these relations is therefore
primarily that they provide a means for simply and transparently reproducing
the results of more sophisticated calculations, `warts and all'; \ie,
including any and all systematics. The derived relation between $(g-i)$ and
$M_*/L_i$ thus provides a solid basis for comparison between results from GAMA
(and, by extension, SDSS), and from other survey projects.

\section{Discussion---Where to from here?} \label{ch:discuss}

In this penultimate Section, we look at how our SP parameter estimates might
be improved for future GAMA catalogues. First, in \secref{ch:nirvalue}, we
look at what might be gained by successfully integrating NIR data into our SPS
calculations. Then, in light of our present difficulties in incorporating the
available NIR data, in \secref{ch:future}, we explore potential avenues for
improving on the present SP parameter estimates. In particular, we argue that
any future improvements in our SPS calculations will require a new conceptual
framework.

\subsection{The value of NIR data} \label{ch:nirvalue}

\begin{figure} \includegraphics[width=7.8cm]{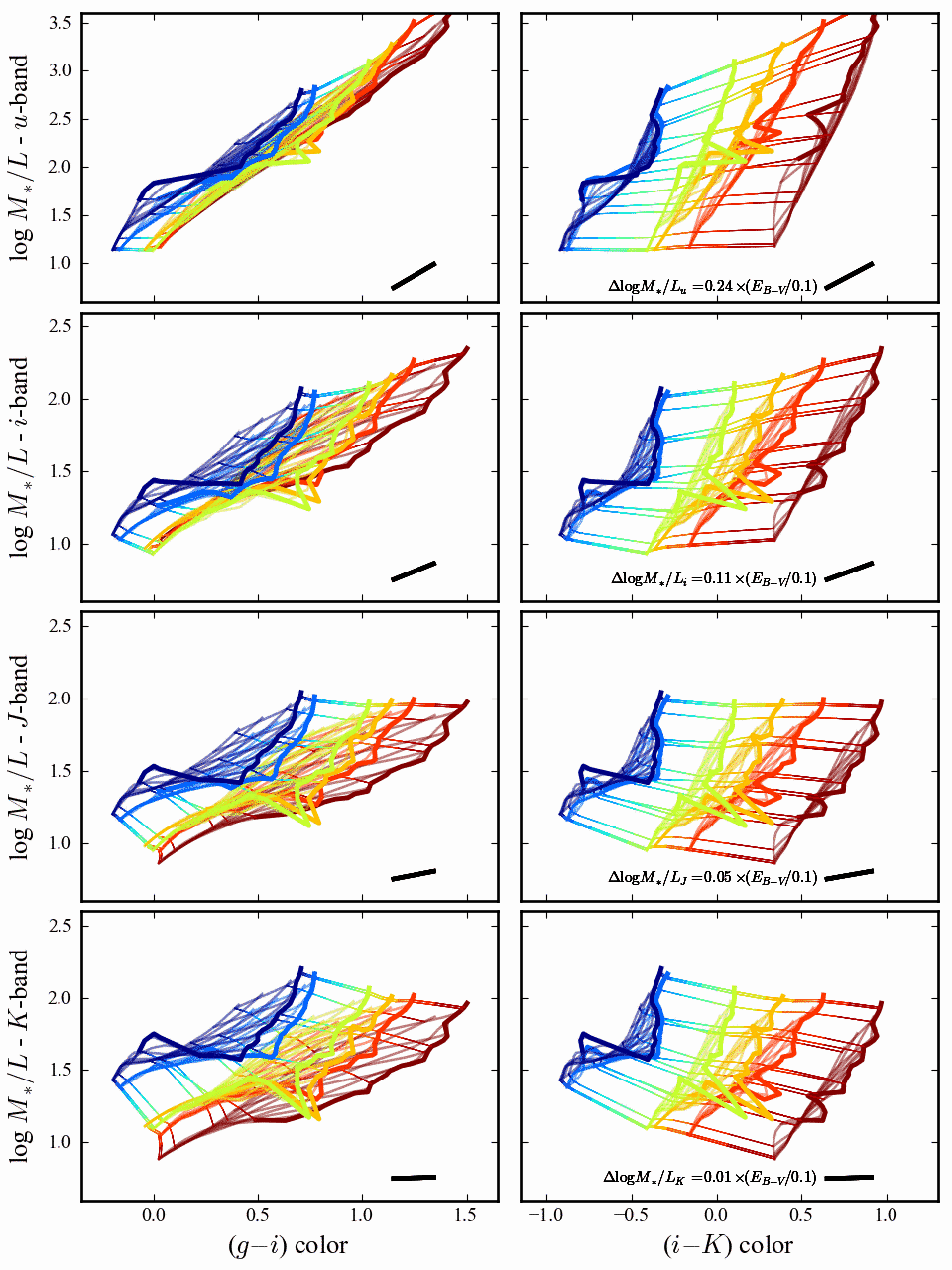} \centering
\caption{Variations in $M_*/L$ as a function of restframe $(g-i)$ and $(i-K)$
colour for models in our SPL.--- All symbols and their meanings are as in the
directly comparable Figure \ref{fig:moverls}. In contrast to $(g-i)$,
optical--minus--NIR colours contain virtually no information directly
pertaining to $M/L$, or to age. Instead, the optical--to--NIR colour is
sensitive primarily to variations in dust and metallicity. While the inclusion
of NIR data into the SPS fitting calculation may in principle lead to tighter
constraints on $t$, $\tau$, $Z$, and $\dust$, it will have little to no impact
on the accuracy of our $M_*/L$ estimates. \label{fig:moverls2} } \end{figure}

Let us now consider what additional information may be provided by the
inclusion of NIR data, or, conversely, what we have sacrificed by excluding
the available NIR data for the present catalogue of stellar masses and SP
parameters. Our discussion of this question is based on \textbf{Figure
\ref{fig:moverls2}}. As in Figure \ref{fig:moverls}, these panels show the
variation in $M_*/L$ at different wavelengths for our SPL. The lefthand panels
show $M_*/L$ as a function of $(g-i)$ colour; the righthand panels show the
same as a function of $(i-K)$ colour. Using this Figure, then, we can compare
the information encoded in optical--minus--optical and optical--minus--NIR
colours.

Figure \ref{fig:moverls2} shows that most of the additional information
encoded within optical--to--NIR colours is concerning metallicity: the fact
that each of the single-metallicity surfaces are well separated in the
righthand panels shows that the different metallicity models can be easily
distinguished by their $(i-K)$ colours. While the single-age metallicity
surfaces are well-separated, however, the fact that each of these surfaces
spans a narrower range of $(i-K)$ colours than $(g-i)$ colours shows that both
$\age$ and $\tau$ are better constrained by $(g-i)$. That is, the inclusion of
NIR data will not necessarily lead to tighter constraints on galaxies'
individual SFHs. Further, by the same argument that we have used in
\secref{ch:moverls} and \secref{ch:gimodels}, it is immediately clear from
Figure \ref{fig:moverls2} that the $(i-K)$ colour encodes virtually {\em no}
information directly pertaining to $M_*/L$: the range of $M_*/L$ within our
SPL is nearly constant as a function of $(i-K)$. To be sure, optical--to--NIR
SED shape is a powerful means of breaking degeneracies associated with
metallicity, but {\em this has very little bearing on the inferred value of
$M_*/L_i$}.

In this way, Figure \ref{fig:moverls2} offers a means of understanding the
results of the numerical experiments presented in Appendix \ref{ch:mocks}. In
this Appendix, we find that the principal gain that comes with the inclusion
of the NIR is in our ability to recover the known values of $Z$ for the mock
galaxies. Although the inclusion of NIR data has little to no effect on our
ability to recover $t$ or $\tau$ individually, our ability to recover $\lwage$
is improved (from $\sim 80$ \% to $\sim 55$ \%) with the inclusion of the NIR.
That is, while NIR data do help to break the degeneracy between metallicity
and luminosity-weighted mean stellar age, it does not help to constrain
galaxies' precise SFHs. We also find that including the NIR data has little
effect on our ability to recover the known values of $M_*/L$ for synthetic
galaxies: we are able to recover the known values of $M_*/L$ to within
$\approx 0.05$ and $\approx 0.06$ dex with and without the inclusion of NIR
data, respectively.

We therefore conclude that the robustness and reliability of our stellar mass
estimates will not necessarily be improved simply by folding the NIR data into
the SPS calculations---or, said another way, our decision to exclude the NIR
data for the current catalogue does not necessarily have a large adverse
effect on the quality of our stellar mass estimates. 

\subsection{Building a better synthetic stellar population library} \label{ch:future}

In \secref{ch:badnir}, we have suggested that our problems in satisfactorily
incorporating the available NIR data into the SPS calculation may reflect that
our present SPL is inadequate to the task of describing galaxies' full
optical--to--NIR SED shapes. In this Section, with an eye towards the
availability of the much deeper VST-KIDS and VISTA-VIKING optical and NIR
imaging in the near future, we discuss possible avenues for deriving improved
SP parameter constraints. In particular, we are interested in the first
instance in what kinds of expansions of our SPL are likely to have the
greatest impact on our SPS calculation; secondarily to this, we want to know
whether and what modifications to our SPS algorithm will be required to fully
exploit these high quality data.

\begin{figure} \includegraphics[width=7.8cm]{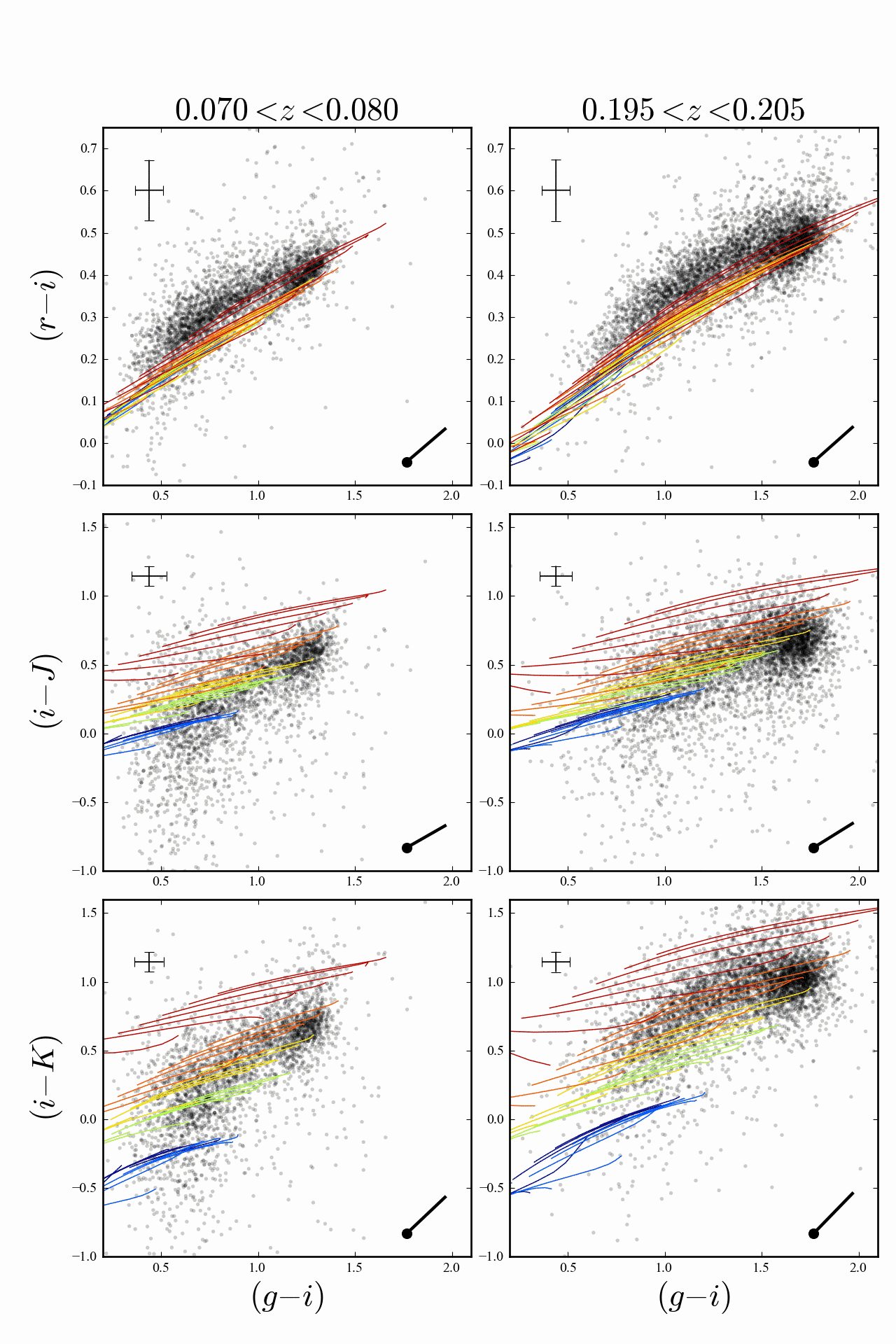} \caption{Comparing the
model and observed optical and NIR colours for galaxies in two narrow redshift
intervals.--- Each panel of this Figure shows a (observers' frame)
colour-colour diagram for these two redshift intervals: from top to bottom, we
plot $(r-i)$, $(i-J)$, and $(i-K)$ against $(g-i)$. The grey points show the
observed colours of GAMA galaxies. We have deliberated selected two
well-populated spikes in the GAMA redshift distribution. The coloured tracks
show models from our SPL, colour-coded by their metallicity. The $\dust = 0.1$
dust obscuration vector is shown in the lower right corner of each panel. The
point to be made from this Figure is that the single metallicity stellar
populations in our library only sparsely cover the observed optical--NIR
colour--colour space of real galaxies. A finer metallicity grid in the stellar
evolution models is required for adequate synthetic stellar population
modelling including NIR data. \label{fig:badnir}} \end{figure}

Our discussion is based on \textbf{Figure \ref{fig:badnir}}, in which we show
colour-colour diagrams for two heavily populated redshift intervals in the
GAMA sample. The coloured lines in Figure \ref{fig:badnir} show the
evolutionary tracks for models in our SPL with different values of $\tau$ and
$Z$. Each track is colour-coded according to its metallicity. For clarity, we
only show the models with zero dust; the $\dust = 0.1$ dust vector is shown at
the bottom right of each panel. These tracks should be compared to the actual
observations, which are shown as the black points. Based on this Figure, let
us now consider how our ability to fit the optical--to--NIR SEDs of real
galaxies might change with an expanded SPL template set.

\subsubsection{Expanding the metallicity grid}

The upper panels of Figure \ref{fig:badnir} show $(g-i)$ versus $(r-i)$. Note
that in the optical, the different uniform metallicity models in our SPL
almost completely overlap. But as you go further towards the NIR (lower
panels), the distance between the different metallicity tracks steadily grows.
Looking at the bottom panels, it is clear that the relatively coarse grid of
$Z$ values used for our present SPL only sparsely samples the $giK$ color
space of real galaxies. Particularly for the gap between $\log Z = -3.4$ and
$-2.4$, the distance between the distinct metallicity tracks in $(i-K)$
becomes comparable to the imposed error floor of 0.05 mag (see
\secref{ch:sedfit}). This explains the origin of the most striking feature of
Figure \ref{fig:bc03m05}: the rather strong quantisation in $Z$. Galaxies with
colours that lie between the distinct metallicity tracks can only be fit by
adopting the too-blue, lower metallicity model, with the addition of some dust
to compensate.

The implication, then, is that a finer grid of metallicities is required when
working with NIR data than when working with optical data alone. (This is a
direct corollary to the fact that optical--minus--NIR colours are sensitive to
metatllicity in a way that optical colours are not.) The problem here is
twofold. The first problem is a mundane, practical one: the size of our
current SPL is already about as big as we can deal with. With the current
architecture of our code, we cannot easily expand the grid in any one
dimension without reducing its size in some other dimension to compensate.

The bigger problem is that the BC03 (like the M05 and CB07) SSP models cover
only a relatively coarse grid in stellar metallicities. In principle, it is
trivial to generate models with arbitrary metallicity by interpolating between
the SSP models provided by \citet{BC03}, \citet{M05}, or whomever. However, at
fixed age, the $Z$-dependence of both flux and SP properties is complex. For
this reason, we consider it unwise to blindly interpolate between models of
different metallicities.\footnote{While some authors have used interpolation
to generate models of arbitrary metallicity, our suggestion would be that it
might be more appropriate to interpret these models as being linear
combinations of the different metallicities; that is, mixed metallicity SPs,
rather than intermediate metallicity SPs. This issue will be addressed in more
detail by Robotham et al.\ (in prep.).}

To ensure that our coarse metallicity grid is not responsible for the problems
we are seeing with the NIR data, however, we have tried re-fitting the
galaxies in our main sample with $z < 0.12$ galaxies (\ie, we have reduced the
size of our redshift grid by a factor of $\sim 5$) using a finer $Z$ grid for
the SPL. This grid, which we have generating by interpolating between the
different metallicity SSP models at fixed age and wavelength, spans the same
range as the native \citet{BC03} SSP grid in 24 logarithmically spaced steps.

Using a finer metallicity grid makes no appreciable difference to the quality
of the fits to the optical--plus--NIR data. The biggest difference between the
two fits is that, with the finer $Z$ grid, the inferred values of $Z$ for
galaxies with $-3.5 \lesssim \log Z \lesssim -2.5$ are systematically higher
by $\approx 0.3$ dex. But even so, compared to the optical--plus--NIR fits
using the native \citet{BC03} metallicity grid, the change in the inferred
value of $M_*/L_i$ is less than 0.06 dex for 99 \% of galaxies; the median
change is $< 0.01$ dex. We therefore conclude that simply expanding our
metallicity grid does not fix our current problems with incorporating the NIR
data into the SPS fits, nor does it significantly improve the accuracy of our
stellar mass estimates.

In principle, it is easy to accommodate more sophisticated treatments of mixed
metallicities by generalising Equation (\ref{eq:csp}) so that the SFR is an
explicit function of $Z$ as well as $t$. In practice, however, the principal
disadvantage to doing so is that we would want to specify or parameterise the
relations between $\psi_*(t)$, and $Z(t)$, whether explicitly, or in terms of
appropriate priors. (Further, this does not address the issue of whether and
how one can safely generate SSP models of arbitrary metallicity.) One simple
way to accomplish this would be to assume an exponentially-declining gas
accretion rate with a characteristic timescale $\tau_\mathrm{gas}$, coupled
with assumptions about stellar gas recycling back into the ISM \citep[as done
by, \eg, P\'egase;][]{pegase}. Again, the apparent insensitivity of our $M_*/L$ estimates argues against this having a large impact on our stellar mass estimates; it may, however, lead to improvements in our estimates of both $Z$ and $\lwage$.

\subsubsection{Allowing for secondary stellar population components}

For the present work, we have limited ourselves to considering smooth,
exponentially declining SFHs. A number of authors have attempted to
incorporate or allow for more complicated SFHs in their SPS calculations. One
approach has been to increase the dimensionality of the SPL parameter space by
introducing additional SP components as short bursts
\citep[\eg][]{Kauffmann2003a, Brinchmann2004, Gallazzi2005}. 

We can also use Figure \ref{fig:badnir} to explore what impact the inclusion
of more complicated SFHs in our SPL might have. Consider what happens to any
of the models in Figure \ref{fig:badnir} with the addition of a secondary
burst of star formation. If any two of the models shown in Figure
\ref{fig:badnir} are combined in any proportion, the evolutionary track of the
resultant SP must necessarily lie between the individual tracks of the two
distinct SP components. If you were to combine any two SPs with the same dust
and metallicity, but different SFHs, the result will necessarily still lie
trapped within the region of colour space spanned by the smooth models. That
is, so long as any secondary stellar population has the same dust and
metallicity, it will not be easily distinguishable from any of our existing
smooth models.

This insight is significant in terms of the results of \citet{GallazziBell}.
These authors find that SFH-related degeneracies mean that the inclusion of
bursty SFHs among the SPL model templates does little to reduce this bias for
bursty galaxies. Further, there is the potential that the inclusion of too
many bursty SPL models can lead to biases in {\em non}-bursting galaxies. In
other words, because these scenarios cannot be distinguished on the basis of
their SEDs, a bias in $M_*/L$ is inevitable, whether that be a small bias for
the many `smooth SFH' galaxies, or a large bias for the fewer bursty galaxies.
Note, too, that in this picture, the degree of the bias is strongly dependent
on the assumed priors. 

The implication from the above, then, is that the inclusion of models with
mixed metallicities and/or multi-component SFHs will improve our SP parameter
estimates only to the extent that they expand the high-dimensional colour
space spanned by the full ensemble of SPL models. And then, multi-component
SFHs will only expand the SPL colour space if and only if the different
components are allowed to have different amounts of dust and/or metals. 

In order to meaningfully incorporate dual-component SPs thus requires the
addition of at least five parameters to describe the secondary SP: the
equivalent of a `formation time'; some characteristic timescale for the
secondary SFH (\ie, an $e$-folding time, or some equivalent); its mass
relative to the primary; and then both its metallicity, and its dust content.
With the current architecture of our SPS code, such an expansion of parameter
space is completely impractical. 

\subsubsection{The need for a new conceptual framework}

We would therefore appear to have reached the practical limits of complexity
that can be covered by discrete grid-search-like fitting algorithms using a
static SPL. Independently of the question of NIR data issues, any future
expansion of the model parameter space will have to be accompanied by a change
in the conceptual framework that underpins our SPS modelling procedure.

Alternative approaches apply standard dimensionality-reducing techniques,
developed in the context of data compression, to the problem. One example is
the \texttt{MOPED} algorithm \citep{Heavens2000, Panter2003}, which uses a
variant of principal component analysis (PCA) to efficiently perform a
23-component SPS fit to full SDSS spectra, including a generalised 10-bin SFH
\citep[see also \texttt{VESPA};][]{Tojeiro2007}. Another example is
\texttt{kcorrect} \citep{kcorrect}, which uses the technique of non-negative
matrix factorisation (NMF). The idea here is to determine the basis set of
template spectra that optimally describes the observed SEDs of real galaxies.
The NMF basis set is constructed as a combination of SP template spectra; this
means that the basis templates constructed using the NMF algorithm can be
considered as SPS template spectra with realistic, multi-component,
non-parametric SFHs. The principal motivation for and advantage of these
approaches is that they can eliminate entirely the need to assume parametric
forms for the SFH. In the context of the above discussion, the operational
advantage of such approaches can be thought of as shifting from sampling a
static and semi-regular grid of parameter values to a dynamic sampling of an
expanded but continuous parameter space.

The main point to take from the above discussion is that proper modelling of
the optical-to-NIR SED shapes of galaxies is considerably more challenging
than modelling just the optical SED. Part of the reason for this is that, for
a fixed optical colour, a galaxy's optical--NIR colour is sensitive to both
$Z$ and to $\lwage$. Said another way it is precisely because a galaxy's
optical--NIR SED can break metallicity-related degeneracies that it becomes
necessary to model each of these quantities in more detail---indeed, in more
detail than is practical within the present architecture of our code. On the
other hand, the relatively strong degeneracies between a galaxy's SP
properties and its optical SED shape means that SPS fitting of optical SEDs
can be done using a relatively crude SPL.

\section{Summary} \label{ch:nirsummary}

The primary purpose of this work has been to present and describe the `first
generation' estimates of stellar mass and other ancillary stellar population
parameters for galaxies in the GAMA survey. We have deliberately set out to
use widely used and accepted techniques to derive these values, partially in
order to allow for the fairest comparison between results from GAMA and other
high and low redshift galaxy surveys. Our stellar mass estimates are based on
the synthetic stellar population models of \citet{BC03}, assuming a
\citet{Chabrier} IMF and a \cite{Calzetti} dust law (\secref{ch:models}). In
constructing the stellar population library that forms the backbone of the
calculation, we have used the standard assumptions of a single metallicity and
a continuous, exponentially-declining star formation history for all stellar
populations, with dust modelled as a single, uniform screen
(\secref{ch:sedfit}).

The most significant `non-standard' element of the calculation is that we use
a Bayesian approach when determining the fiducial values of all parameters and
their associated uncertainties (\secref{ch:bayes}). As we show in Figure
\ref{fig:bfvsml}, this decision has an important {\em systematic} effect on
the parameter estimates: averaged over the full GAMA sample, the {\em most
likely} (in a Bayesian sense) values of $M_*/L_i$ are $\sim 0.10$ dex higher
than those taken from the single {\em best fit} (\ie, maximum likelihood) SP
template. While the MPA-JHU mass estimates for SDSS used a Bayesian approach,
this is not (yet) generally done in high redshift studies.

\subsection*{Comparisons between GAMA and SDSS}

Through comparison between the GAMA- and SDSS-derived values of $M_*/L$ and
$M_*$ (Appendix \ref{ch:cfsdss}), we highlight two important issues with the
SDSS \model\ photometry. First, we show that as a measure of total flux, the
SDSS \texttt{model} photometry has serious systematics as a function of (true)
\Sersic -index (Figure \ref{fig:mags}). For galaxies best-fit by an
exponential \texttt{model} profile, the differential bias between $n \sim 0.5$
and $n \sim 1.5$ is $\gtrsim 0.2$ mag ($\sim 20$ \%); for those best-fit by a
de Vaucouleurs \texttt{model} profile, the differential bias between $n \sim
2$ and $n \sim 8$ is $\approx 0.7$ mag (a factor of 2). These systematic
biases in total luminosity translate directly to biases in total stellar mass:
this may be the single largest source of error in the SDSS mass estimates
based on \model\ photometry.

Second, if we apply our algorithm to the SDSS \texttt{model} SEDs, we see very
large differences between our derived values and those from the MPA-JHU
catalogue (Figure \ref{fig:cfsdss}). These differences are directly tied to
strong systematic differences between the SDSS \texttt{model} and GAMA
\texttt{AUTO} colours (\secref{ch:modelseds}), such that the net systematic
offset in $(u-z)$ is as large as 0.2 mag (Figure \ref{fig:seds}). We therefore
suggest that it may be more appropriate to use \texttt{petro} SEDs when
analysing data from the SDSS photometry catalogues.

Despite these differences in the SDSS and GAMA photometry, the fiducial GAMA
values of $M_*/L$ are in excellent agreement with those found in the latest
generation MPA-JHU catalogue for SDSS DR7 (Section \ref{ch:cfsdss}), which
have been shown to be well consistent with dynamical mass estimates
\citep{Taylor2010b}. (We investigate the consistency between GAMA-derived
stellar and dynamical mass estimates in a companion paper.) As we argue in
\secref{ch:modelmasses}, the inclusion of a dust prior in the MPA-JHU stellar
mass estimation algorithm may have effectively circumvented the potential bias
in stellar population parameters based on the SDSS \texttt{model} photometry;
using the GAMA \texttt{AUTO} photometry, we find no need for such a prior.

\subsection*{NIR data (currently) does more harm than good}

For the present generation of stellar mass estimates, we have elected {\em
not} to include the available $YJHK$ NIR photometry in the SED-fitting; the
stellar population parameters presented here are based on fits to the optical
$ugriz$ SEDs only.

As summarised in \secref{ch:badnir}, there are three reasons for this
decision. First, none of the commonly used stellar population models
\citep{BC03, M05, CB07} provide good fits to the full optical--to--NIR SEDs
(Figure \ref{fig:residuals}). Second, while the inclusion of the NIR data does
have an impact on the derived values---the median value of $M_*$ goes down by
0.15 dex when the NIR data are included---the values derived with the NIR are
formally inconsistent with those derived from the just the optical data for a
large fraction of galaxies. Both of these points suggest inconsistencies
between the optical--to--NIR SED shapes of real galaxies and those of the
models in our SPL. The third reason is that we find that the `random'
differences in inferred SP parameters---particularly $\dust$, $Z$, and
$\lwage$---using different SSP models are larger than than the formal
uncertainties once the NIR data are included. That is, our SP parameter
estimates become significantly model dependent when, and only when, the NIR
data are used. 

That said, the {\em systematic} differences in the inferred SP parameters
based on different SSP models are small: for $M_*/L$, the median difference in
the value of $M_*/L$ using the BC03 and M05 SSP models is just 0.02 dex. We
therefore consider it unlikely that the failure of the models to adequately
accommodate the NIR data is due to differences between or uncertainties in the
stellar evolution models themselves. 

This leaves two possibilities: that there may be problems in the NIR data,
and/or that the stellar population library that we have used is insufficient
to describe the full range of stellar populations that exist in the local
universe. These issues will have to be addressed---both through additional
data validation and verification and through expansion of our stellar
population library to include additional metallicities and possibly more
sophisticated star formation histories---in the construction of future
generations of the GAMA stellar mass catalogues. We have discussed possible
avenues for expanding our SPL in \secref{ch:future}, and conclude that any
future expansion to our SPL parameter space will have to be accompanied by a
change in the conceptual framework of our SPS fitting procedure. We would
appear to have reached the practical limit of complexity for a simple
grid-search-like approach using a static SPL.

Note that future catalogues will make use of considerably deeper VST optical
imaging from KIDS and VISTA NIR imaging from the VIKING survey. In this
context, it is highly significant that, even with the present photometry, the
accuracy of our SP parameter estimates are not currently limited by
signal--to--noise, but by systematics. With the exception of the $u$-band, the
photometric errors are smaller than the error floor of 0.05 mag that we have
imposed for the fits (see \secref{ch:sedfit}). The extent to which the deeper
data will improve on the present SPS fitting results will depend crucially on
how well the data can be self-consistency cross-calibrated, including biases
due to PSF- and aperture-matching, colour gradients, and background
subtraction, as well as the basic photometric calibration. As a corollary to
this, it will be incumbent upon us to ensure that the model photometry in any
future SPL can be considered accurate to the same level as the real data;
$\ie, \lesssim 0.05$ mag.

\subsection*{The robustness and reliability of our optical-derived stellar mass estimates}

In light of our decision to ignore the presently available NIR data, we have
reexamined the commonly-held belief that NIR data are crucial to deriving a
robust and reliable estimate of stellar mass (\secref{ch:gimoverl}). We use
generic properties of the stellar population models to demonstrate that on its
own, the $i$-band flux is nearly as good a representation of the total stellar
population as is the NIR flux (\secref{ch:moverls}). More quantitatively,
assuming a constant $M_*/L_i$ or $M_*/L_K$, it is possible to use $L_i$ or
$L_K$ to estimate $M_*$ to within a factor of 5.5 or 4.5, respectively.

Using a similar argument, we show that the variation in $M_*/L_i$ {\em at
fixed} $(g-i)$ is $\lesssim 0.5$ dex (Figure \ref{fig:moverls}). The effect of
dust is largely to shift galaxies along the $(g-i)$--$M_*/L_i$ relation. Dust
thus does not significantly affect one's ability to estimate $M_*/L_i$ using
$(g-i)$. Similarly, we show that, for a given model, variations in age, SFH,
and metallicity act to largely preserve the relation between $M_*/L_i$ and
$(g-i)$ colour (Figure \ref{fig:gimodels}). Further, we show that
multi-component stellar populations (\eg, a burst superposed over an old and
passive stellar population) fall within the same region of $(g-i)$--$M_*/L_i$
space as the exponentially-declining star formation histories that comprise
our stellar population library. In this way, based on generic properties of
stellar evolution models, we show that $(g-i)$ colour can be used to estimate
$M_*$ to within a factor of $\lesssim 2$, even considering the well known
dust--age--metallicity degeneracy, and even for multi-component stellar
populations (\secref{ch:gimodels}).

Finally, we consider the empirical relation between $M_*/L_i$ and $(g-i)$ for
GAMA galaxies. It is significant that the observed relation between $M_*/L_i$
and $(g-i)$ is both more linear and considerably tighter (at fixed colour, the
scatter in $M_*/L_i$ is $\lesssim 0.1$ dex) than we might expect by simply
taking the prior-weighted average of the models in our stellar population
library (Figure \ref{fig:colourrel}). This implies that the full $ugriz$ SED
shape contains additional information not found in the $(g-i)$ colour, and
that this information is sufficient to exclude a significant range of the
models in our library. The tightness of the $(g-i)$--$M_*/L_i$ relation is not
merely a consequence of the central limit theorem.

In other words, there are two completely separate reasons why $(g-i)$ is an
excellent predictor of $M_*/L_i$, both of which are entirely fortuitous.
First, variations in age, SFH, dust, and metallicity---independently and {\em
en masse}---largely preserve the $(g-i)$--$M_*/L_i$ relation. This is a
coincidence produced by the physics of stellar evolution. Second, the stellar
populations of real galaxies produce a $(g-i)$--$M_*/L_i$ relation that is
both tighter and more nearly linear than might be expected from stellar
population models alone. This is a coincidence produced by the physics of
galaxy formation and evolution.

In this sense, presuming that both our derived values and their associated
uncertainties are reasonable, we have effectively `calibrated' the
$(g-i)$--$M_*/L_i$ relation to a precision of $\lesssim 0.1$ dex ($1\sigma$).
The derived relation offers a reliable and robust means for observers to
derive stellar masses based on minimal information. Similarly, under the
(non-trivial) assumption that the relation does not evolve strongly with
redshift, it offers a simple and transparent basis for fair comparison between
results derived from GAMA and other low- and high-redshift surveys. As an
important caveat on the use of this relation, however, any and all systematic
errors or uncertainties in the SPL itself---including, \eg, the IMF and errors
in the treatment of the {\em optical} stellar evolution tracks---are not
included in the quoted uncertainty of 0.1 dex. On the other hand, the relation
given does offer a solid means for other surveys to compare their stellar
mass-centric measures to those from GAMA under the identical assumptions.

\subsection*{Concluding remarks}

The stellar mass estimates we have described have been or will be used for a
wide variety of recent and ongoing studies by the GAMA collaboration. These
include studies of variability in the stellar IMF \citet{Madusha}, measurement
of the $z \approx 0$ mass function \citep{Baldry2011}, the properties of
galaxies at the lowest end of the H$\alpha$ luminosity function
\citep{Brough2011}, and studies of galaxy demographics in the field (Taylor et
al.\ in prep.), in groups (Prescott et al.\ in prep.), and in filaments
(Pimbblett et al.\ in prep.). Further, they provide an important benchmark for
any and all future GAMA stellar mass and SP parameter estimates.

In line with the legacy goals of the GAMA survey, these stellar mass estimates
are also being made publicly available for use by the wider astronomical
community as part of GAMA DR 2 (scheduled for mid-2011). GAMA's unique
combination of depth and survey area have been deliberately chosen to bridge
the gap between large-scale local universe surveys like 6dFGS, SDSS and
2dFGRS, and deep surveys of the high redshift universe like VVDS, DEEP-2, and
zCOSMOS. Particularly in combination with these other major surveys, the GAMA
catalogues are intended to provide a valuable laboratory for studies of galaxy
formation and evolution.

\setlength{\bibhang}{2.0em}
\setlength\labelwidth{0.0em}

\appendix

\section{Numerical Experiments---\\ SPS fitting of mock galaxy photometry} \label{ch:mocks}

\begin{figure*} \includegraphics[width=8.6cm]{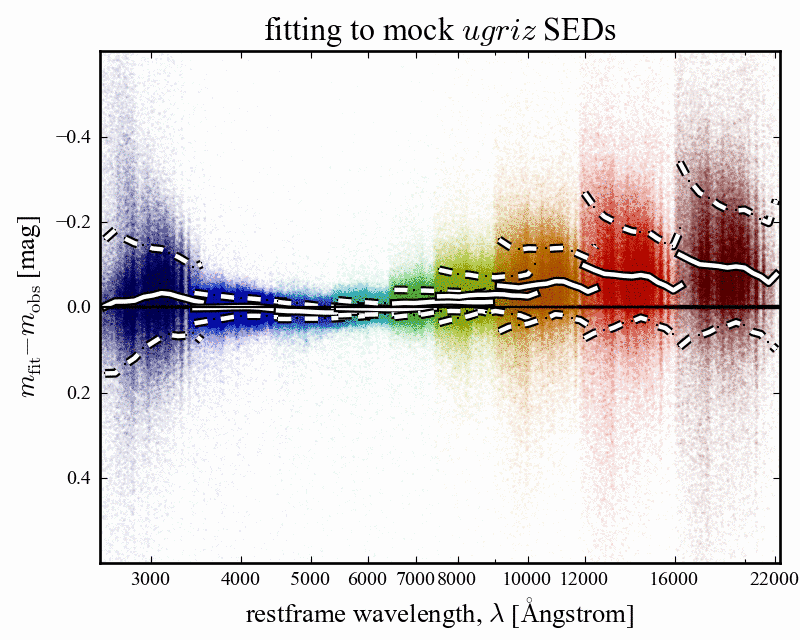}
\includegraphics[width=8.6cm]{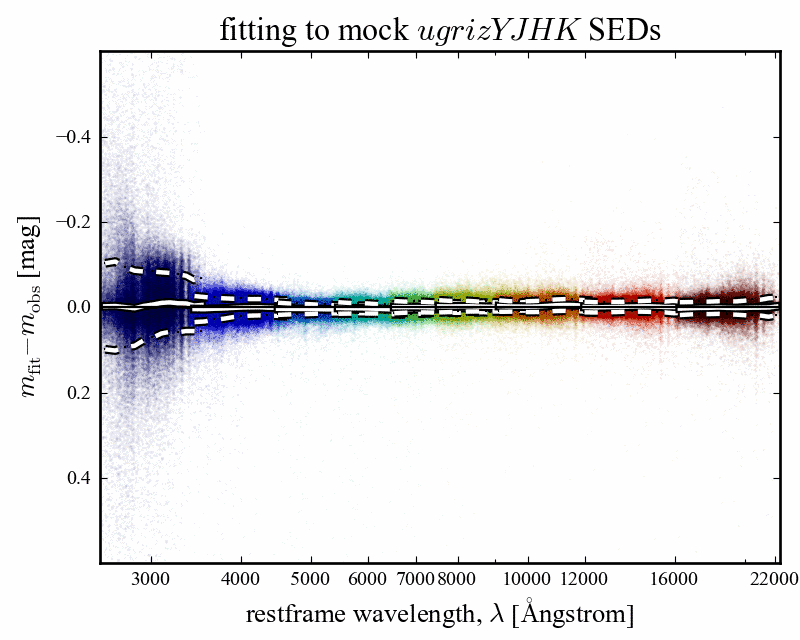} \caption{Quality of the SED
fits for mock galaxy photometry.--- This Figure shows the residuals from the
SED fits to the mock galaxy photometry described in Appendix \ref{ch:mocks}.
The left-hand panel shows the residuals when fitting only the the optical
$ugriz$-bands; the right-hand panel shows those for fits to the full
$ugrizYJHK$ SEDs. All symbols and their meanings are as in Figure
\ref{fig:residuals}, to which these plots should be compared. As argued in
\secref{ch:quality}, the fact that the optical--only fits tend to over-predict
the `true' NIR fluxes is possible due to the mismatch between our assumed
priors and the `true' multi-variate distribution of SP parameters within the
mock catalogue. At least qualitatively, the similar residuals seen in Figure
\ref{fig:residuals} for the real galaxies does not then suggest that the fits
are necessarily `bad'. Further, and in contrast to Figure \ref{fig:residuals},
we are able to reproduce or describe the full optical--to--NIR shapes of the
mock galaxies. This strongly suggests that the large residuals seen in Figure
\ref{fig:residuals} are due to problems in the data, short-comings in the SPL
models, or both. In any case, the large systematics seen in the right-hand
panels of Figure \ref{fig:residuals} do give reason to treat the results of
the optical--to--NIR SED fits with some suspicion. \label{fig:simresiduals}}
\end{figure*}

In this Appendix, we examine our ability to recover the known SP parameters of
a set of mock galaxies with a realistic distribution of SP parameters. To this
end, we have used the results of our `live' SPS fits to the real GAMA $ugriz$
data to construct a catalogue of mock galaxy photometry. Specifically, for
each galaxy, we have take the SPL template SED that most closely matches its
`most likely' SP values, added random perturbations commensurate with the
actual observational uncertainties for the original galaxy, and then fed that
photometry back into the SPS fitting algorithm. Note that the distribution of
SP parameters in this mock catalogue is, by construction, the same as what we
observe for GAMA galaxies; \ie, it is quite different to our assumed priors.

In \secref{ch:residuals}, we have argued that our seeming inability to
satisfactorily fit the optical--to--NIR SEDs of GAMA galaxies calls into
question the validity of SP parameter estimates inferred from such fits. We
have also argued that our inability to predict NIR photometry based on just
the optical SEDs---with the implication that the NIR contains additional
information not found in the optical---does not imply that the optical cannot
be used to reliably infer stellar mass-to-light ratios. In order to provide
context for these arguments, we will first spend some time looking at the
quality of the photometric fits to the mock photometry in \secref{ch:quality}.
We will then go on to look at how accurately and precisely we can recover the
known SP parameters of the mock galaxies in \secref{ch:recovery}.

\subsection{Quality of fits} \label{ch:quality}

\subsubsection{How well can you fit the optical--to--NIR SEDs?}

In the \textbf{Figure \ref{fig:simresiduals}}, we show the analogue of Figure
\ref{fig:residuals} for this mock galaxy catalogue. Let us look first at the
right-hand panel of this Figure, in which we show the difference between the
known, input photometry, and the recovered, output photometry, after fitting
to the mock $ugrizYJHK$ SEDs. Put simply, the quality of the fits is near
perfect. The residuals are at the millimag level for all but the $u$-band; the
median residual in the $u$-band is still just 0.02 mag. None of these
residuals is significant at the 0.1$\sigma$ level.

\subsubsection{How well can you predict NIR photometry based on optical SEDs?}

Now consider the left-hand panel of Figure \ref{fig:simresiduals}, in which we
show the residuals when fitting to the mock $ugriz$ SEDs. Our primary interest
here is in how well we are able to predict NIR photometry for the mock
galaxies using their $ugriz$ SEDs. In comparison to Figure
\ref{fig:residuals}, there are three features of this plot that we find
striking.

First, it is clear that, as in Figure \ref{fig:residuals}, the NIR photometry
predicted from the mock optical SEDs is systematically too bright.
Quantitatively, in comparison to Figure \ref{fig:residuals}, the size of the
discrepancy is considerably smaller. For the mock galaxies, the residuals are
$\lesssim 0.1$ mag; roughly half that seen for real GAMA galaxies. Compared to
the photometric errors, the median significance of these offsets is $0.5$,
$0.7$, $1.3$, and $1.7 \sigma$ in the $YJHK$ bands; again, roughly half that
seen in Figure \ref{fig:residuals}.

Second, the residuals for the mock galaxies show a qualitatively different
dependence on rest-frame wavelength/redshift than is seen in Figure
\ref{fig:residuals}. Whereas for real galaxies, the offsets in each individual
band appear to be greatest at the highest and lowest redshifts, for the mock
galaxies, the offsets grow rather smoothly for longer and longer wavelengths.
That is, for the mock galaxies, our results suggest that one's ability to predict NIR photometry depends primarily on how far one is willing to extrapolate off the red end of the observed optical SED.

Third, we note that, just as in Figure \ref{fig:residuals}, we do see some
residuals in the optical bands. Quantitatively, the median residual in each of
the $ugriz$ bands is $-0.02$, $+0.00$, $+0.01$, $+0.00$, and $-0.01$ mag,
respectively; in all cases, this is insignificant at the level of $\lesssim
0.2\sigma$. In comparison to those seen in Figure \ref{fig:residuals}, these
residuals are again roughly half the size as for real galaxies, but show a
qualitatively similar `curvature' with wavelength.

\begin{figure*} \centering \includegraphics[width=8cm]{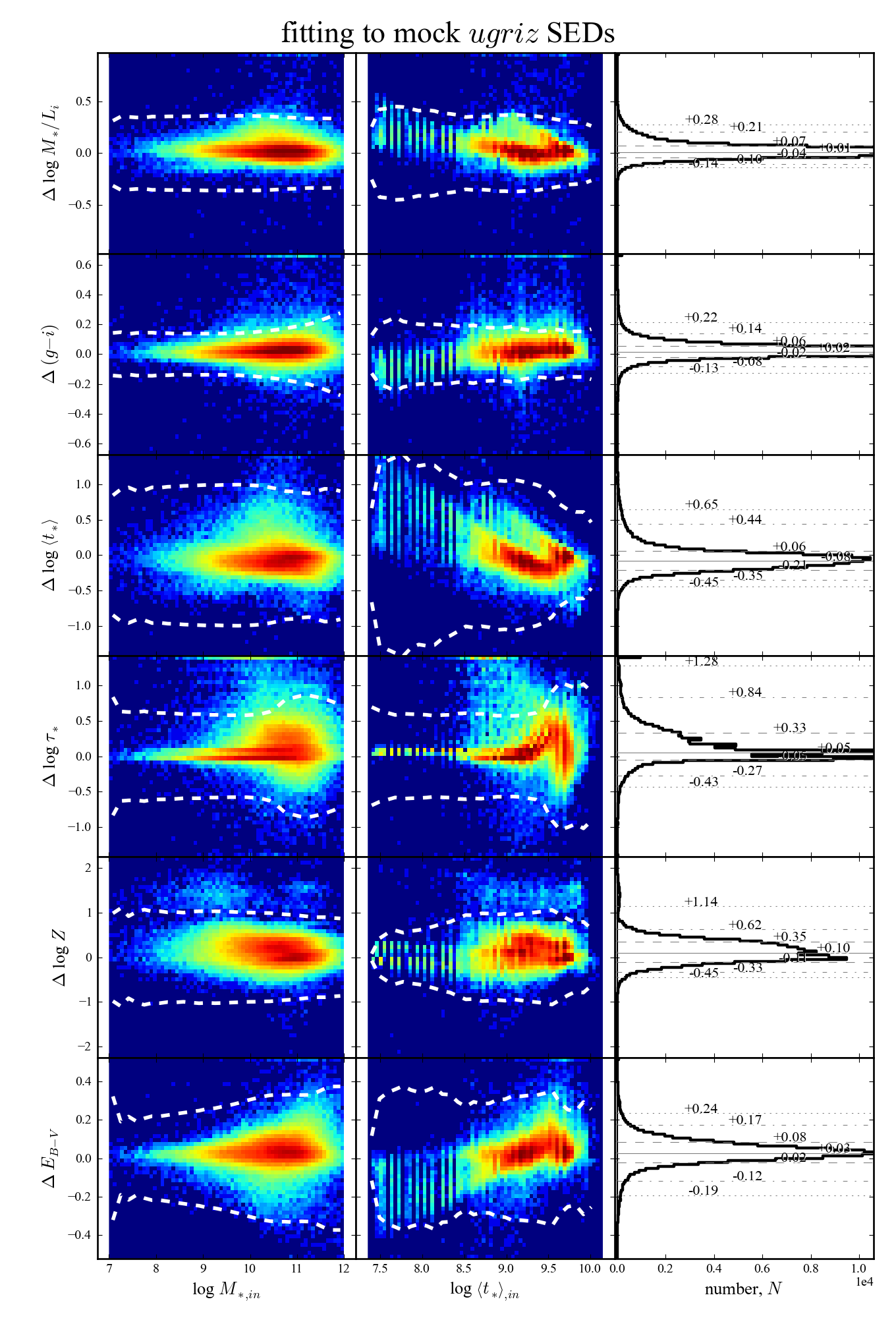}
\includegraphics[width=8cm]{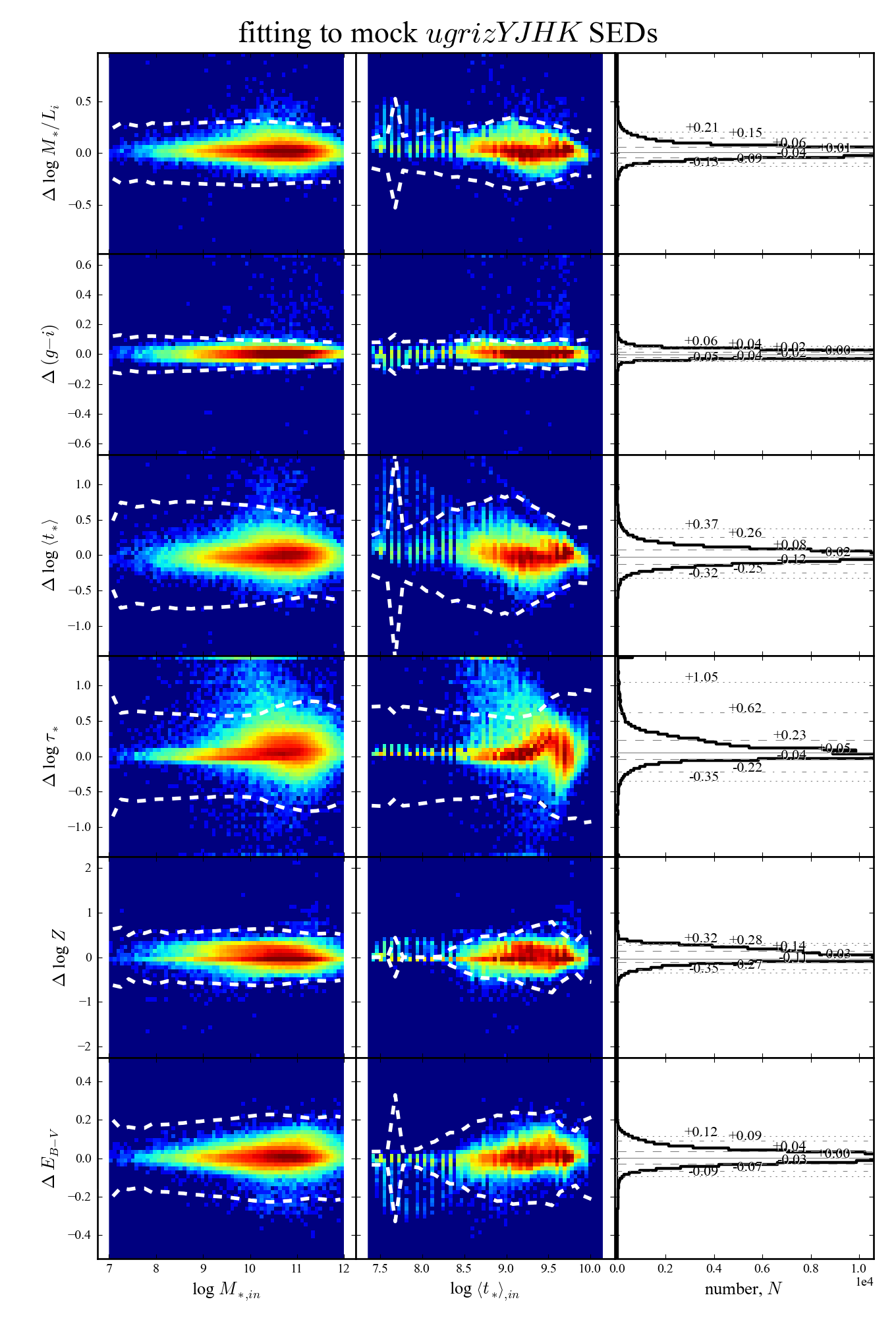} \caption{Stellar population parameter
recovery for mock galaxy photometry.--- These plots are based on the mock
galaxy photometry described in Appendix \ref{ch:mocks}; each panel shows
the difference between the `known' parameter of a mock galaxy and that
inferred from a fit to optical--only (left panels) or the optical--plus--NIR
(right panels) photometry. In all cases, the `$\Delta$'s on the $y$ axes
should be understood as `recovered--minus--input'; the quantities on the $x$
axes relate to the `known' value. As in Figures \ref{fig:optnir} and
\ref{fig:bc03m05}, the histograms show the distribution in the `$\Delta$'s,
with the percentile equivalents of the $\pm 0$/1/2/3$\sigma$ points as marked.
In both cases, we are able to recover the SP parameters of the mock galaxies
with little to no systematic bias. This is particularly true for $M_*/L_i$:
the reliability of the optical--plus--NIR--derived estimates (median error
$\sim 0.05$ dex) is not significantly better than that based on only the
optical (median error $\sim 0.06$ dex). \label{fig:reliable}} \end{figure*}

\subsubsection{Implications for SED fitting---the subtle role of priors}

Given the above, what are we to make of the (very slight) residuals in the
$ugriz$ fits? Since we are fitting to the same $ugriz$ photometry in both of
the above experiments, the additional information provided by the NIR
photometry must exclude some of those models that are consistent with the
optical data on its own. In other words, the models allowed by the 5-band fits
span a broader range of SP parameter values than those allowed by the 9-band
fits; the set of SPL templates allowed by the 5-band fits must be a superset
of those allowed by the 9-band fits. (We will look at precisely how the SP
parameter estimates change with the inclusion of the NIR data in a moment; for
now, let us keep the discussion general.) The implication of this, as is well
known, is that an optical SED simply does not encode sufficient information to
fully constrain a galaxy's SP parameters.

Naturally, the models are distinguished by their SED shapes. From the fact
that the $ugriz$ fits tend to over-predict the `true' NIR photometry of our
mock galaxies, we know that the optical-only fits are consistent with a range
of SPL models, and that these models are on average redder than the `real'
solution. Now, the fiducial parameter estimate is derived from marginalising
over the PDF \'a la Equation \ref{eq:margin}. Again, for the optical-only
fits, this includes a disproportionally large number of models with the
`wrong' SED shape; specifically, models that are substantially too red in the
NIR. But these models will also have (very) slightly different optical SED
shapes. Hence the very slight offsets seen in the optical bands when the NIR
data are excluded.

Consider what would happen if we were to significantly change the form of our
assumed priors in such a way as to make these redder fits less likely---for
example, by making higher dust extinctions or metallicities less likely than
lower ones. Reducing the prior probability of these models directly reduces
their contribution to the integral in Equation \ref{eq:margin}, which defines
the Bayesian `most likely' parameter value. This implies that if we were to
use more realistic priors, we might be able to do a substantially better job
of predicting the NIR photometry based on the optical SEDs. (Parenthetically,
this may be why \texttt{kcorrect} is so successful at predicting NIR
photometry from optical colours.)

In this context, it is significant that the $ugriz$ residuals seen when
fitting only to the optical bands disappear when the NIR data are included.
This shows that our SPS algorithm is in fact able to near-perfectly match
galaxies' SEDs given sufficient information, where we have also now shown that
`sufficient information' means both optical and NIR photometry. Furthermore,
this is possible even despite the fact that the assumed priors are very
different from the real distribution intrinsic to the data. These experiments
thus suggest that the GAMA optical--plus--NIR dataset can, in principle, be
used to constrain galaxies' SP parameters to the extent that the quality of
the fits is not sensitive to the assumed priors. Said another way (and more
accurately), so long as NIR data are available, the way that the photometry
uncertainties map onto SP parameter space means the allowed range of SP
parameters is small enough that the assumption that the priors are {\em
locally} flat is a good one.

\begin{figure*} \centering
	\includegraphics[width=8cm]{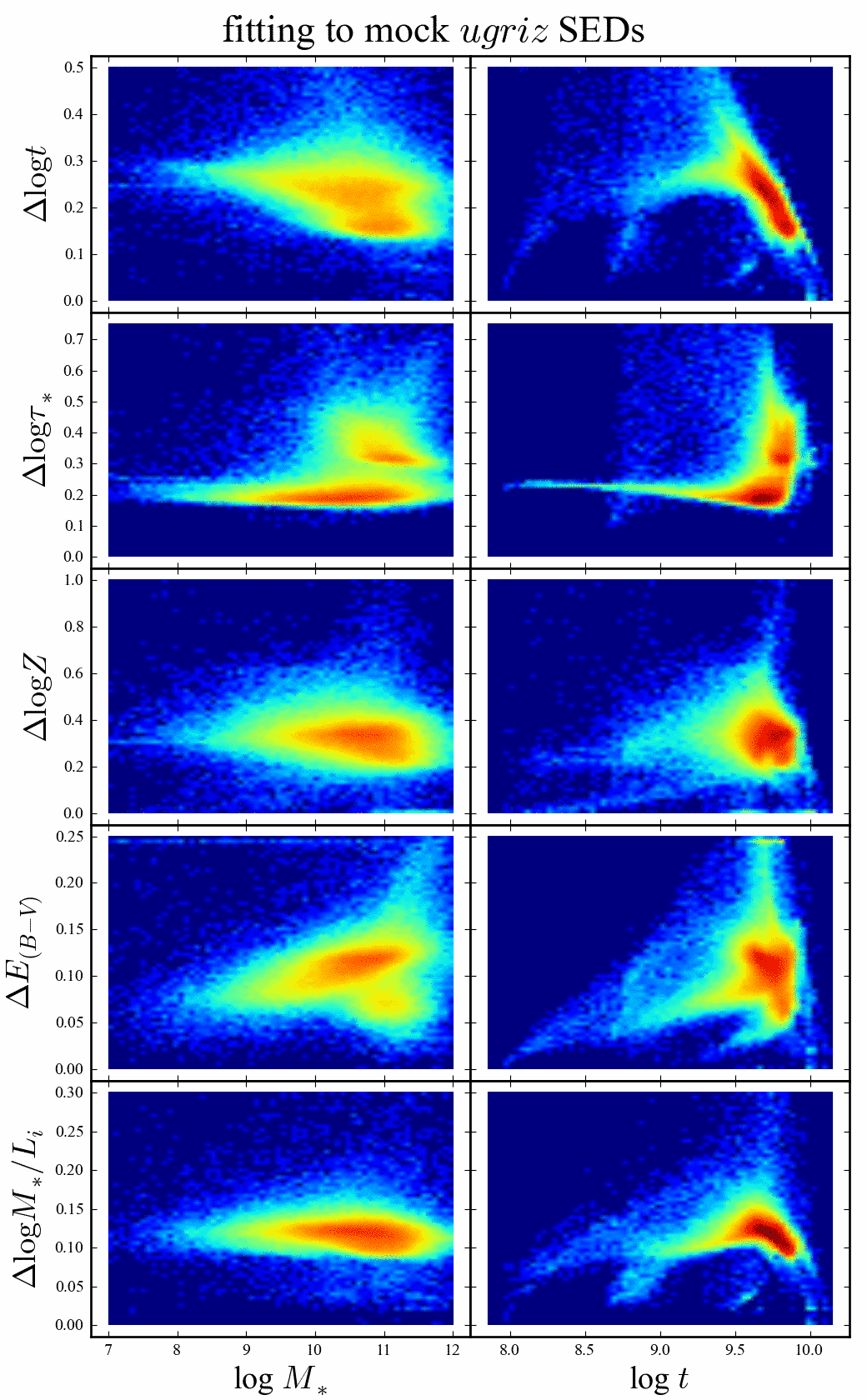}
	\includegraphics[width=8cm]{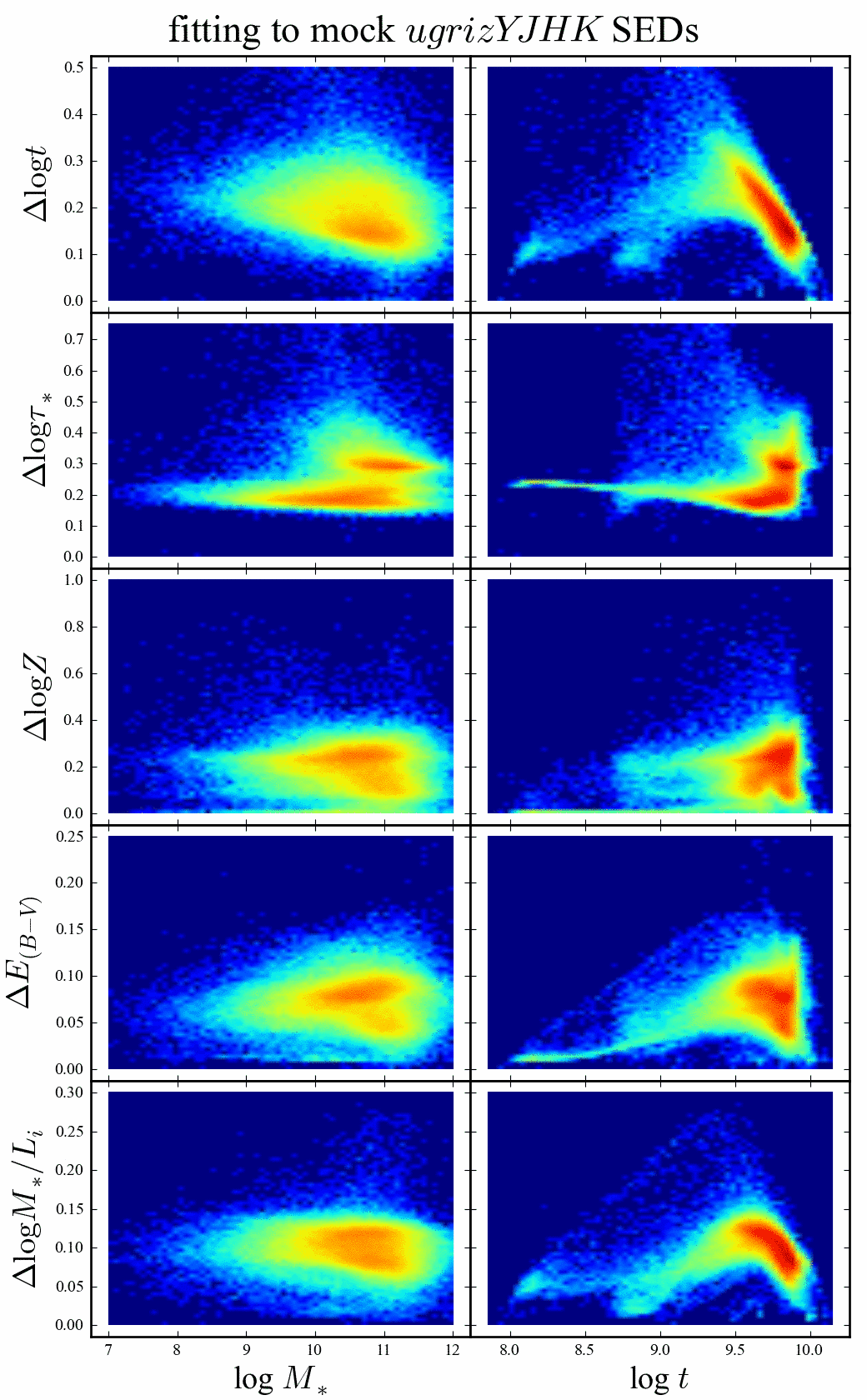}
\caption{Precision of stellar population parameter estimates for mock galaxies.--- In analogy to Figure \ref{fig:errors}, these plots show the distributions of the formal uncertainties in the inferred SP parameter values for the mock galaxies based either on the optical-only (left panels), or the optical--plus--NIR (right panels) SED fits. If the NIR data are included, the formal uncertainties in the recovered values of all of $t$, $\tau$, $Z$, and $\dust$ are considerably smaller than if they are excluded. However, the formal uncertainties in $M_*/L$ are virtually unchanged. By breaking the age--metallicity--dust degeneracies, NIR data provides a better estimate of the ancillary SP parameters, but this has little to no bearing on the precision with which $M_*/L$ can be constrained. \label{fig:robust}}
\end{figure*}

As we have already pointed out in \secref{ch:nirsummary}, there is an
important corollary to the idea that we are, in principle, able to
near-perfectly match the `observed' SEDs in our mock catalogue. Once NIR data
are included, the quality of the fits is no longer limited by the amount of
information that is encoded in the data, but instead by how closely the
stellar populations that comprise the SPL represent those found in the wild.
This means that, to the extent that more complex stellar
populations---including more complicated SFHs, a mix of stellar metallicities,
and patchy dust geometries---change the shape of a galaxies SEDs, these
effects must be adequately folded into the construction of the SPL. In other
words, precisely {\em because} NIR data provides the additional SP parameter
information not found in the optical, robust and reliable fits to
optical--plus--NIR SEDs require more sophisticated SPLs (see also
\secref{ch:future}.)

\subsection{Parameter recovery} \label{ch:recovery}

Whereas in the previous section we have focussed on how well our SPS fitting
algorithm is able to describe or reproduce the SED shapes of mock galaxies, we
now turn to the question of how well galaxies' SP parameters can be
constrained from their broadband SED shapes.

\subsubsection{Reliability}

In \textbf{Figure \ref{fig:reliable}}, we show how accurately we are able to
recover the SP parameters associated with the mock galaxies based either on
their optical-only (left panels) or optical--to--NIR (right panels)
photometry, and in the face of realistic observational uncertainties. In all
cases, the `$\Delta$s' on the $y$ axis should be understood as being the
`output--minus--input' parameter value, plotted as a function of the `known',
input value from the mock catalogue. As in Figures \ref{fig:optnir} and
\ref{fig:bc03m05}, the colour-scale shows the logarithmic data density, with
the percentile equivalents of the $\pm 0/1/2/3 \sigma$ points of the
distributions given with the histograms at right.

The first---and, in the context of our main argument, the most crucial---point
to be made from these plots is that we are able to recover the $M_*/L_i$s of
the mock galaxies with no discernible bias based on the optical SEDs alone.
Further, the $M_*/L_i$ determinations derived from fits including NIR
photometry are not all that much more reliable than those based only on the
optical data: on the one hand, the 1 $\sigma$ `errors' are $^{+0.07}_{-0.04}$
dex; on the other, they are $^{+0.06}_{-0.04}$ dex.

In line with the results of the previous section, where we have shown that our
priors tend to over-weight models with redder SED shapes, the optical-only
fits imply slightly too-red $(g-i)$ colours. Empirically, the error is
$0.015^{+0.057}_{-0.020}$ mag; using the slope of the empirical
$(g-i)$--$M_*/L_i$ relation, this translates to an error of
$0.010^{+0.040}_{-0.014}$ dex in $M_*/L_i$.  

For the other SP parameters, as for $M_*/L_i$, the inclusion of NIR data does
not appear to be crucial to obtaining reliable parameter estimates. The median
offset between the known and the recovered SP parameter values for the
optical--plus--NIR fits is not clearly less than for the optical--only fits.
That said, the inclusion of NIR data clearly does reduce the `random' error in
the derived SP parameters---that is, the robustness---particularly in the case
of $Z$ and $\dust$, as well as $\tau$ for those `passive' galaxies with
$t/\tau \gg 1$.

\subsubsection{Robustness}

The final question to be considered here is how precisely galaxies' SP
parameters can be constrained based on SED fits with or without NIR
photometry. We address this question with reference to \textbf{Figure
\ref{fig:robust}}, which shows the distribution of the formal uncertainties in
SP parameter estimates derived from the mock photometric catalogues, based on
fits to optical--only (left panels) or optical--plus--NIR (right panels) SEDs.

Looking at the global distribution of uncertainties for all galaxies in the
mock catalogues, the greatest effect of the NIR is to reduce the uncertainties
on $Z$ and on $\lwage$. The median value of $\Delta \log Z$ goes from 0.35 dex
to 0.25 dex with the inclusion of the NIR, while the median value of $\Delta
\log \lwage$ goes from 0.26 dex to 0.19 dex. By comparison, the improvement in
$\Delta \log \dust$ is relatively minor: the median value goes from 0.10 dex
to 0.07 mag. It is interesting to compare this improvement in $\Delta \log
\lwage$ to that in $\Delta \log t$, which goes from 0.21 to 0.18 dex, and in
$\Delta \log \tau$, which remains nearly unchanged at 0.21 dex. This suggests
that while NIR photometry helps to break degeneracies between $\lwage$ and $Z$
(and to a lesser extent $\dust$), and so helps provide a better constraint on
instantaneous mean stellar age, it does not provide much additional
information concerning the precise SFH.

But again, the NIR data does not lead to a substantial improvement in the
accuracy with which $M_*/L_i$ can be determined: the median value of $\Delta
\log M_*/L_i$ goes from 0.11(4) dex ($\approx 30$ \%) to 0.09(8) dex ($\approx
25 \%$). The NIR encodes virtually no additional information concerning a
galaxy's stellar mass that cannot be found in the optical.

This fact has one important implication for future stellar mass catalogues. In
the previous section, we found that we were able to recover $M_*/L_i$ for the
mock galaxies with an empirical $1\sigma$ `error' on the order of $\pm 0.05$
dex, both with and without the inclusion of NIR data; \ie, more precisely than
might be expected from the formal uncertainties of $\pm 0.10$ dex. The reason
for this is that, in generating the mock photometry, we have added random
photometric errors commensurate with the random photometric errors; in the
fitting, on the other hand, we include an error `floor' of 0.05 mag. This
error is intended to account for potential differential systematic errors
between the different photometric bands. The implication is thus that the
accuracy of our stellar mass determinations is not limited by
signal--to--noise (\ie, the random observational uncertainties in the
photometry in each band), but instead by the relative accuracy of the
photometry in the different bands with respect to one another (\ie,
differential systematic errors between the different bands). This means that
the extent to which the considerably deeper VST and VISTA photometry will
improve our ability to constrain galaxies' stellar masses will depend
crucially on how well we are able to control systematic photometric errors in
the different bands, including the accuracy of the basic photometric
calibrations.

\section{Comparisons between GAMA and SDSS} \label{ch:cfsdss}

In this Appendix, as a means of validating our stellar mass estimates, we
compare them to the latest generation of stellar mass estimates from the
MPA-JHU catalogue for SDSS DR7\footnote{available via
http://www.mpa-garching.mpg.de/SDSS/}. The motivation for this comparison
stems from the fact that the MPA-JHU mass-to-light ratios have been well
tested; they thus provide a useful set of benchmark measurements. They are in
excellent agreement with other frequently used MPA-JHU mass determinations;
\eg , the \citet{Kauffmann2003a} DR4 catalogue.\footnote{The median offset is
$-0.01$ dex, with a scatter on the order of 0.1 dex; see
http://www.mpa-garching.mpg.de/SDSS/DR7/mass\_comp.html} That is, the
(SED-derived) DR7 mass estimates are wholly consistent with values derived
from spectra. Further, \citet{Taylor2010b} have compared the DR7 MPA-JHU
stellar masses to dynamical mass estimates, derived using the \Sersic -fit
structural parameters of \citet{Guo2009}. Based on the consistency between
these stellar mass estimates and dynamical mass estimates, \citet{Taylor2010b}
have argued that any differential biases in the stellar-to-dynamical mass
ratio as a function of stellar population parameters may be as low as
$\lesssim 0.12$ dex ($\sim 40$ \%).\looseness-2

There are two facets to this comparison: differences in SDSS and GAMA
photometry from which the mass estimates are derived, and differences in the
algorithms used to actually derive the mass estimates. We compare the GAMA and
SDSS photometry in \secref{ch:sdssphot}. After describing the key differences
between the SDSS and GAMA algorithms in \secref{ch:sdssmasses}, we will then
look at our ability to reproduce the SDSS stellar mass and stellar
mass-to-light ratio values using first SDSS \model\ the and then the GAMA
\auto\ photometry in \secref{ch:automasses} and \secref{ch:modelmasses},
respectively. In this way, we will hope to identify whether and how these
differences affect the derived values for $M_*/L$ and $M_*$. Before we begin,
let us again stress that the rationale behind this comparison is that the SDSS
mass-to-light ratios have been well tested; our main concern is thus our
ability to reproduce the SDSS values for the galaxies that are common to both
SDSS and GAMA.

\subsection{Comparing the GAMA and SDSS photometry \label{ch:sdssphot}}

The basic SDSS catalogue contains two different photometric measures in each of
the $ugriz$ bands. Following the recommendation of \citet{Stoughton2002}, it
is standard practice to use \texttt{model} photometry to construct multi-band
SEDs. This photometry comes from fitting either an exponential or a de
Vaucouleurs profile to the observed light distribution. The choice of profile
shape and structural parameters (\ie, effective radius, ellipticity, and
position angle) are based on the $r$-band image. For the $ugiz$-bands these
parameters are then held fixed during fits, so that only overall normalisation
is allowed to vary; this is then the \texttt{model} flux. The MPA-JHU mass
estimates are based on the \model\ SEDs taken from the basic SDSS catalog.

The second photometric measure is the \texttt{petro} magnitude, which is based
on the idea of \citet{Petrosian}. This flux is measured within a flexible
circular aperture, the size of which is based on the observed (radial) surface
brightness profile. Again following the recommendations of
\citet{Stoughton2002}, it is standard practice to use the \texttt{petro}
photometry as a measure of total flux. Accordingly, when we consider $\log
M_*$ below, we will scale the MPA-JHU mass estimates by $-0.4 (
r_{\texttt{petro}}-r_{\texttt{model}} )$ to obtain `total' mass estimates.
This is directly analogous to our use of SEDs based on matched aperture
\texttt{auto} photometry, scaled to match \texttt{sersic} total magnitudes.

\subsubsection{$r$-band Magnitudes} \label{ch:missedflux}

\begin{figure} \includegraphics[width=8.4cm]{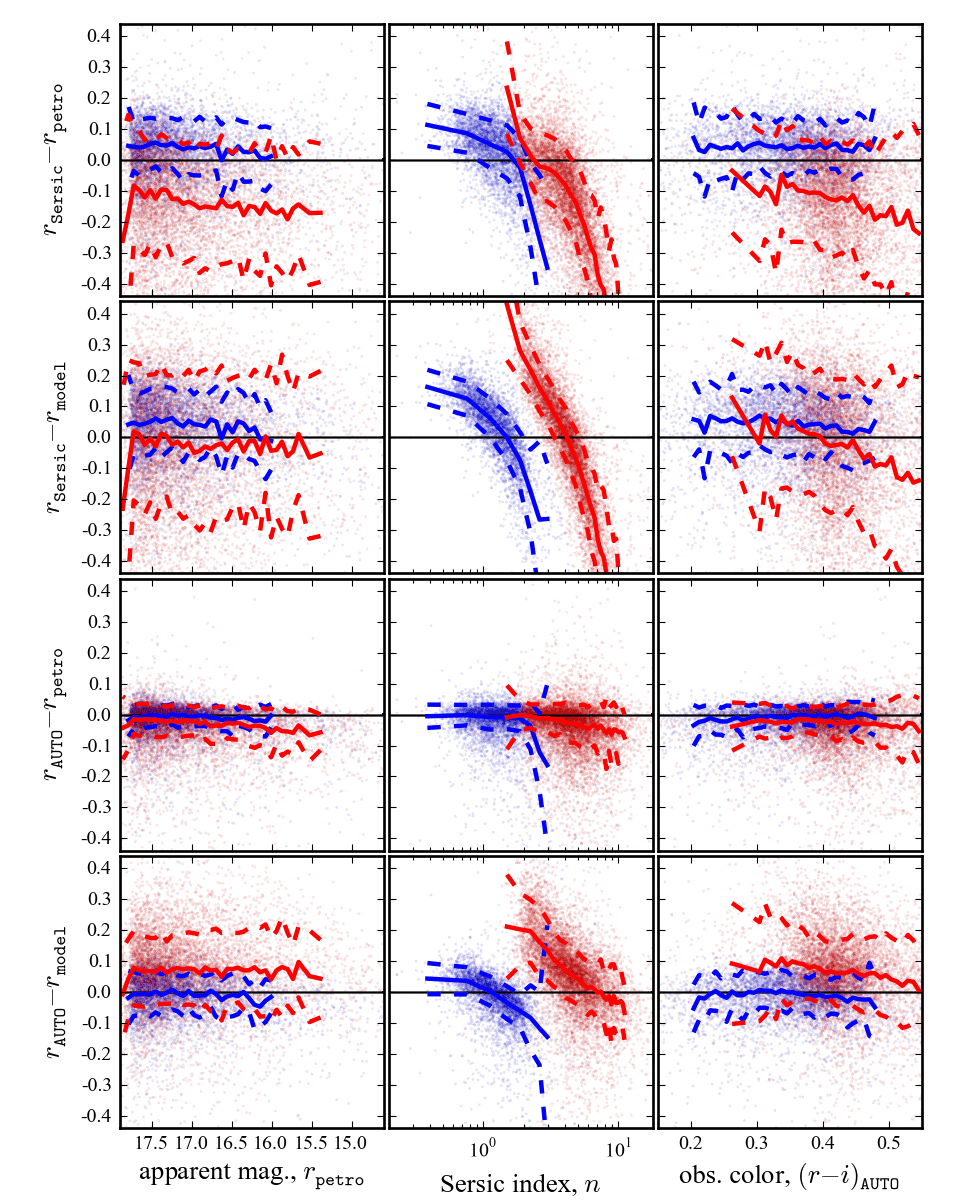}
\caption{Comparison between GAMA and SDSS $r$-band magnitude measurements.---
Each panel of this Figure shows the difference between a GAMA and an SDSS
measure of $r$-band magnitude as a function of (left to right) apparent
magnitude, GAMA-derived \Sersic\ index, $n$, or observed colour. Within each
panel, we make the distinction between those objects that are fit using an
exponential profile (blue) or a de Vaucouleurs profile (red) for the SDSS
\model\ photometry. For GAMA, we use \auto\ magnitudes to construct
multicolour SEDs, and the $r$-band \sersictt\ magnitude to measure total flux;
it is standard SDSS practice to use \model\ magnitudes for SEDs, and the
\petro\ magnitude as a measure of total flux. As expected, the \petro\
magnitude misses an increasingly large fraction of total flux for galaxies
with higher values of $n$. It seems that the GAMA \sersictt\ magnitude may
miss up to $\lesssim 10$ \% of flux for $n \lesssim 2$ galaxies. The SDSS
\model\ magnitudes (which assume either $n=1$ or $n=4$) have strong
$n$-dependent systematics: where the assumed value of $n$ in the \model\
underestimates the `true' value of $n$, the \model\ flux overestimates the
total magnitude by up to $\sim 0.3$ mag. Between $n \sim 2.5$ and $n \sim 8$,
the size of the differential effect for de Vaucouleurs-like galaxies is
greater than a factor of 2. \label{fig:mags}} \end{figure}

In \textbf{Figure \ref{fig:mags}}, we compare the different GAMA and SDSS
photometric measures of $r$-band flux. This comparison is based on the $\sim
12000$ SDSS-targeted galaxies that appear in the GAMA catalogue. In this
Figure, we distinguish between those galaxies whose SDSS \model\ photometry
is based on an exponential (blue points) or a de Vaucouleurs (red points)
profile.\looseness-2

In each case, it is clear that the relation between different photometric
measures depends most strongly on profile shape (parameterised by the GAMA
$r$-band derived \Sersic\ index, $n$). For $n \gtrsim 2$, the difference
between the GAMA \Sersic -fit and SDSS \petro\ fluxes is more or less as
expected: the fraction of missed flux increases rapidly from $\approx 0$ for
$n \sim 2$, to $\approx$ 0.1 mag for $n \approx 4$ galaxies, to $\approx 0.5$
mag for $n \approx 8$, and so on.

The most striking feature of Figure \ref{fig:mags} is the large $n$-dependent
differences between the SDSS \model\ and GAMA \sersictt\ and \auto\
photometry. The crucial assumption behind the SDSS \model\ photometry is that
galaxies can be well described by a \Sersic\ profile with either $n=1$ or
$n=4$. For those galaxies with a GAMA-derived $n \approx 1$, the results in
Figure \ref{fig:mags} suggest that the GAMA \sersictt\ photometry may miss
$\lesssim 10$ \% ($\lesssim 0.04$ dex) of the flux for low $n$ galaxies; for
galaxies with $n \approx 4$, there is excellent agreement between the GAMA
\sersictt\ and SDSS \model\ photometry.

Away from these points, however, the \model\ photometry has large systematic biases: for both exponential- and de Vaucouleurs--like galaxies, where the \model\ \Sersic\ index ($n=1$ or $n=4$) is higher than the `true' value, the \model\ flux significantly over-estimates the `true' total flux.  For exponential-like galaxies, the size of the differential effect is nearly 0.3 mag between $n \approx 0.5$ and $n \approx 2$; for de Vaucouleurs-like galaxies, the effect is greater than 0.7 mag (a factor of 2!) between $n \approx 2$ and $n \approx 8$. This is thus a major, if not the largest, source of error in the SDSS mass estimates.

\begin{figure}
\includegraphics[width=8.4cm]{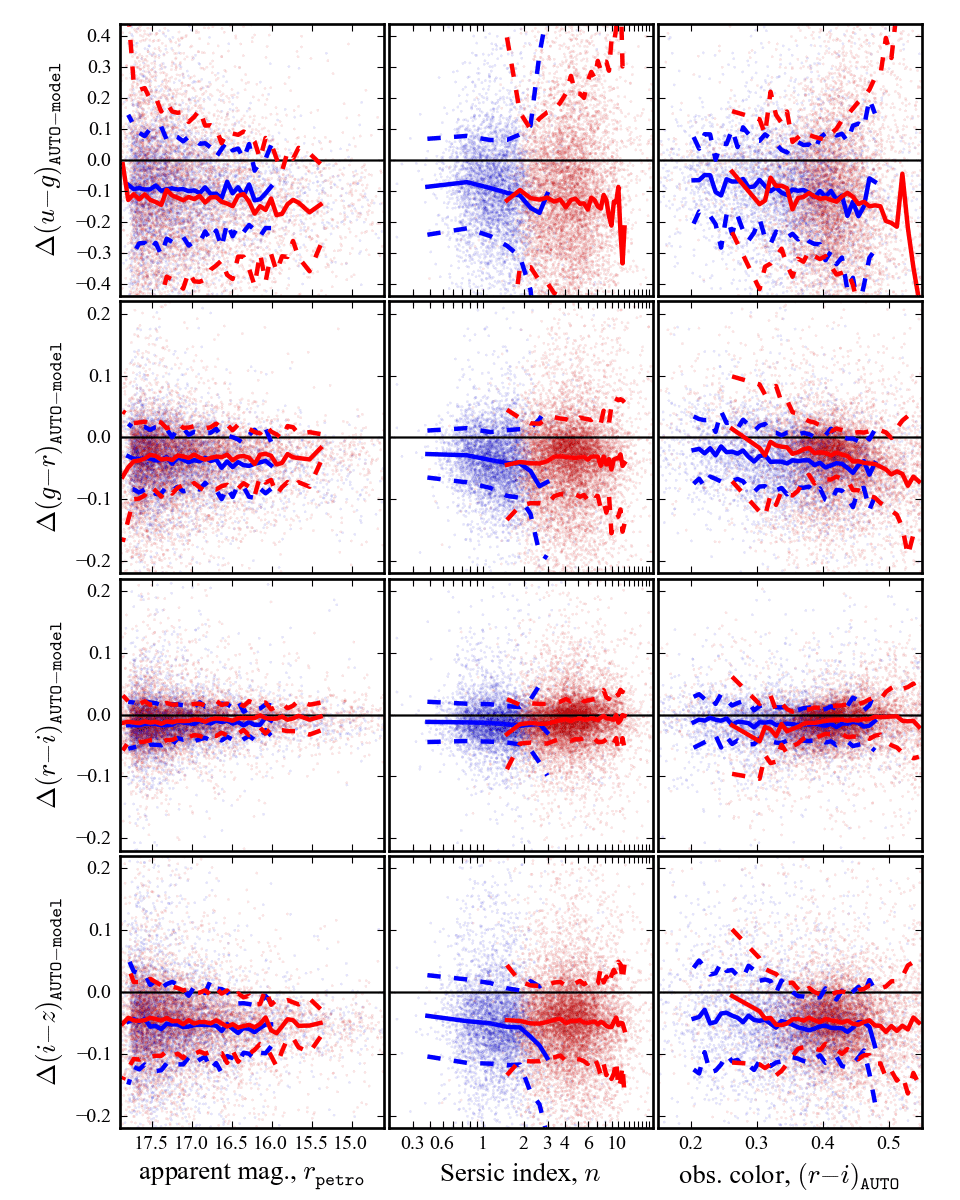}
\caption{Comparison between GAMA \texttt{AUTO} and SDSS \texttt{model}
photometry.--- Each row shows the difference in the measured, observers'
frame colours of galaxies as reported in the GAMA and SDSS catalogues; the
`$\Delta$' should be understood as meaning GAMA--minus--SDSS. From left to
right, the panels show the systematic differences in observed colour as a
function of apparent magnitude, \Sersic\ index, and $(r-i)$ colour. As in
Figure \ref{fig:mags}, we make the distinction between those objects that
are fit using an exponential profile (blue) or a de Vaucouleurs profile
(red) for the SDSS \model\ photometry. The points show the data themselves;
the lines show the binned biweight mean and scatter. The SDSS \model\ SEDs
are systematically redder than the GAMA \auto\ ones: the cumulative
difference in $(u-z)$ is 0.2 mag. Note that we find no such systematic differences between the GAMA \auto\ and SDSS \petro\ colours. This suggests that the SDSS data may be better analysed using \petro\ rather than \model\ SEDs.
\label{fig:seds}} \end{figure}

\subsubsection{$ugriz$ SEDs} \label{ch:modelseds}

In \textbf{Figure \ref{fig:seds}}, we show a comparison between
galaxies' optical colours as reported in the GAMA and SDSS catalogues.
Although the current GAMA optical photometry is derived from the SDSS imaging
data, there are systematic differences between the galaxy colours---as
measured using the GAMA \texttt{AUTO} and SDSS \texttt{model}
photometry---that are used as the basic inputs to the stellar mass estimation
calculation. 

In comparison to the GAMA \texttt{AUTO} photometry, the SDSS \texttt{model}
SEDs are systematically redder across all bands. Quantitatively, the observed
`GAMA \auto --minus--SDSS \model ' offsets are $\Delta(u-g)=-0.10$ mag,
$\Delta(g-r)=-0.03$ mag, $\Delta(r-i) = -0.01$ mag, and $\Delta(i-z)=-0.05$
mag; the cumulative offset between $u$ and $z$ is thus $-0.2$ mag. These
offsets are not a strong function of apparent brightness. Particularly for the
bluer bands, they may depend weakly on \Sersic\ index. Further, looking at the
right-hand panels of this Figure, there is the hint that the offsets vary
systematically with observed colour: this immediately suggests that colour
gradients may play a role in one or the other of these measurements.

Note that we find no such systematic offsets between the GAMA \auto\ and SDSS
\petro\ colours. That is, whatever the cause of the discrepancies seen in
Figure \ref{fig:seds}, it is specific to the SDSS \model\ photometry. We also
note that the fact that the \model\ photometry is so sensitive to $n$ implies
that \model -derived SEDs may be badly biased by colour gradients: a small
change in \Sersic\ index across different bands will produce a relatively
large differential bias in the inferred fluxes. Taken together, these two
points suggest that that when using SDSS photometry, despite the fact that the
\petro\ photometry is not PSF-matched, it may provide a better basis for
constructing multi-colour SEDs than \model\ photometry.

\subsection{Differences between the MPA-JHU and GAMA mass estimation
algorithms \label{ch:sdssmasses}}

Unlike previous MPA-JHU catalogues \citep[\eg][] {Kauffmann2003a,
Brinchmann2004, Gallazzi2005} that were based on the SDSS spectroscopy, the
DR7 MPA-JHU stellar mass estimates are based on fits to the $ugriz$
photometry. The first difference between the MPA-JHU and GAMA algorithms is
that for the MPA-JHU mass estimates, the observed photometry has been
corrected for contributions from emission lines (which are not included in the
\citet{BC03} models), under the assumption that the global emission line
contribution is the same as in the spectroscopic fiber aperture.

Like the one described here, the MPA-JHU SPL is based on the \citet{BC03} SSP
models, and assuming a \citet{Chabrier} IMF. Whereas we use a \citet{Calzetti}
dust law, however, the SDSS SPL spectra use the \citet{CharlotFall} curve
to account for dust obscuration. At least in terms of the values of $M_*/L$,
as we shall show, this difference is not important.

The biggest structural difference between the GAMA and SDSS stellar mass
calculations is that, whereas we have constructed our SPL by sampling a
semi-regular grid in ($\age,~ \tau,~ Z,~ \dust$) parameter space, the MPA-JHU
masses are based on a library made up of large number of Monte Carlo
realizations of different star formation histories. The priors in the MPA-JHU
algorithm are applied in the Monte Carlo sampling of the allowed parameter
space \citep[see also][]{Gallazzi2005}. Specifically, the model ages are
randomly sampled from a uniform distributions in both formation time (over the
range $1.5 < t_\mathrm{form}/[\mathrm{Gyr}] < 13.5$) and in the exponential
decay rate ($0 < \gamma/[\mathrm{Gyr}^{-1}] < 1$; here, $\gamma$ can be
thought of as $1/\tau$). The models also include a number of secondary bursts
of star formation. The burst probabilities are normalised such that 10 \% of
galaxies experience a burst in the last 2 Gyr, with the burst times uniformly
distributed between $t_\mathrm{form}$ and and the time of observation.
Individual bursts are treated as constant star formation rate events lasting
for $10^{7.5}$---$10^{8.5}$ yr. The strength of each burst is parameterised by
the mass relative to the `underlying' population, which is logarithmically
distributed between $0.03 < f_{M_*,\mathrm{burst}} < 0.4$. Finally, the
assumed metallicity prior is logarithmic for super-solar metallicites, with
lower metallicites downweighted through an assumed prior distribution of the
form $(\log Z)^{1/3}$ for $0.02 < Z < 0.2$. The prior distribution of dust
extinctions is derived from the SDSS H$\alpha$/H$\beta$ ratios (Jarle
Brinchmann, private communication; 24/09/2009). In terms of their SPLs, the
major differences between the MPA-JHU and GAMA calculations are thus the
inclusion of bursts, and the different form of the metallicity distribution
prior.

The decision to randomly sample parameter space, rather than to use a
(semi-)regular grid has two consequences. First, it makes it possible to
accommodate bursts in the SPL (as described above); this would not be
computationally practical to include into a SPL grid like ours, since it would
expand the parameter space by (at least) an additional three dimensions.
Second, the nominal SDSS parameter values given in the MPA-JHU catalogues are
the median of the posterior probability distribution; \ie, the 50 \%
confidence upper/lower limits, rather than the `most likely' value from
explicit marginalisation over the PDF. That said, at least for the GAMA mass
estimates, we find that the (probability weighted) median and mean values of
$M_*/L$ are in extremely good agreement. This implies, albeit weakly, that the
posterior probability distributions for $M_*/L$ are roughly symmetric about
the mean/median value.

\begin{figure*} \includegraphics[width=17.8cm]{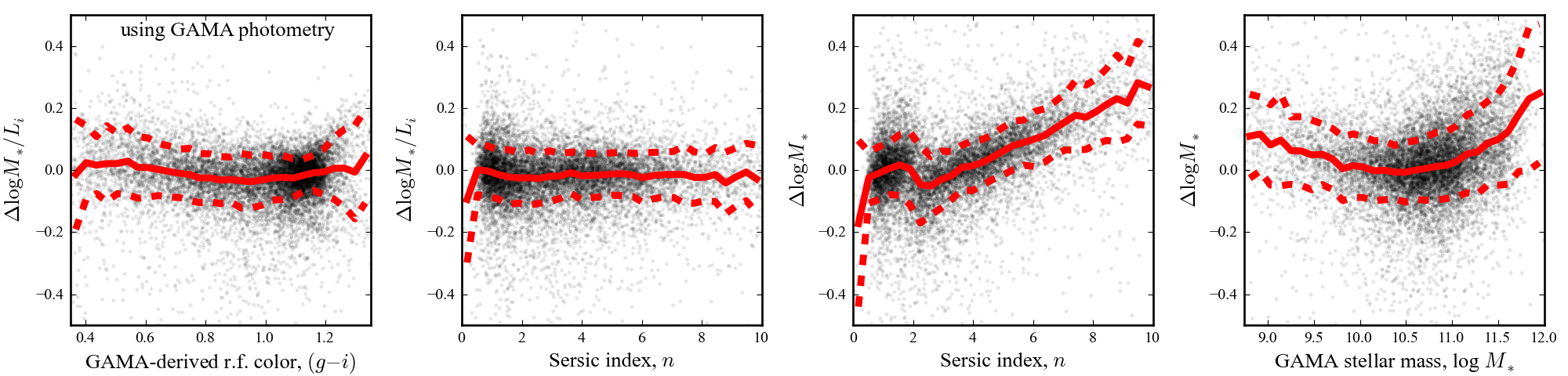}
\includegraphics[width=17.8cm]{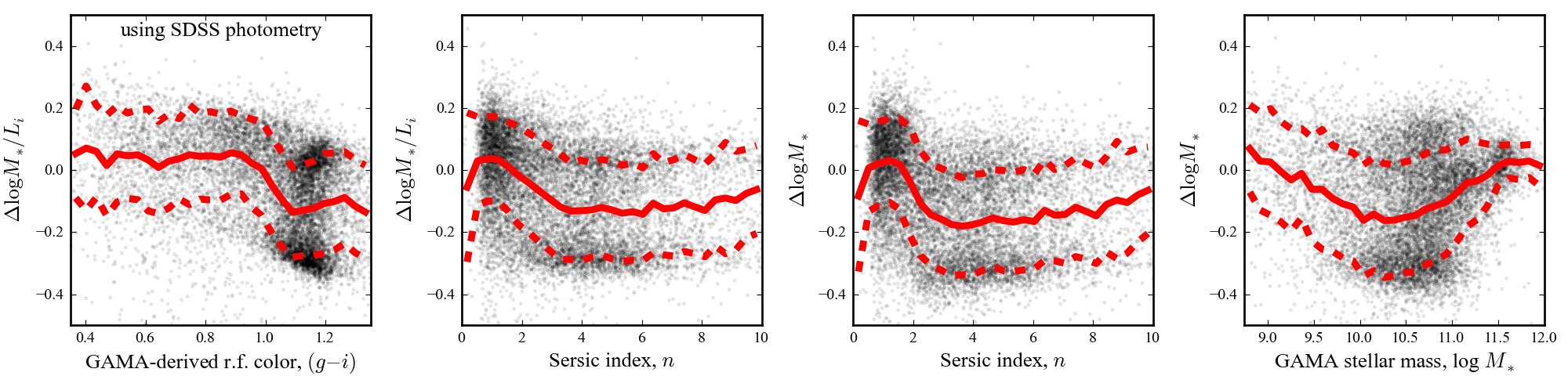}
\caption{Stellar mass-to-light ratios and stellar masses inferred from SDSS
\texttt{model} and GAMA \texttt{AUTO} photometry.--- In the lower panels, the `GAMA-derived' values are based on SEDs constructed from the SDSS \model\ photometry; in the upper panels, the `GAMA' values are derived from the GAMA \auto\ SEDs.  In each case, the `$\Delta$'s should be understood as GAMA-minus-SDSS.  Using the SDSS \model\ photometry, we do not do a particularly good job at reproducing the (well-tested) SDSS-derived values of $M_*/L$; using the GAMA \auto\ photometry, the agreement is very good (the reasons for this are discussed at greater length in \secref{ch:modelseds} and \secref{ch:modelmasses}).  While the fiducial GAMA $M_*/L$s agree very well with those from SDSS, the GAMA masses are systematically larger than the SDSS values.  This can only be explained by missed flux in the SDSS \model\ photometry.  For the highest values of $n$ and $M_*$, GAMA finds $\gtrsim 0.15$ dex more light/mass than SDSS (see also \secref{ch:missedflux} and Figure \ref{fig:mags}). \label{fig:cfsdss}}
\end{figure*}

\subsection{Comparison between the MPA-JHU- and GAMA-derived mass estimates I.\ Using GAMA \auto\ SEDs \label{ch:automasses}}

How well are we able to reproduce the MPA-JHU values for $M_*/L$ and $M_*$? We
address this question in \textbf{Figure \ref{fig:cfsdss}}. In the left-hand
panels of this Figure, we compare the stellar masses that we derive based on
SDSS photometry to those given in the MPA-JHU catalogue: these panels thus
probe differences in the GAMA and SDSS algorithms applied to the same data. In
the upper panels of this Figure, we compare our fiducial stellar mass
estimates based on GAMA photometry. It is thus these panels that most interest
us, inasmuch as these panels show a direct comparison between the well-tested
MPA-JHU values and our own.

Looking at the upper panels of Figure \ref{fig:cfsdss}, the agreement between
our fiducial mass estimates and the MPA-JHU values is very good: the random
scatter between the two values of $M_*/L$ is small, and there are no obvious
systematics. More quantitatively, our \texttt{AUTO}-derived $M_*/L$s agree
with the MPA-JHU values with a biweight mean and scatter in $\Delta M_*/L$ of
$-0.01$ and $0.07$ dex, respectively. The offsets in $M_*/L$ as a function of
restframe colour are at the level of a few percent ($\lesssim 0.02$ dex). That
said, there are large differences in the total inferred $M_*$ as a function of
$n$. Given that we can faithfully reproduce the $M_*/L$s, this discrepancy can
only be explained by differences in the total $L$s. These results thus suggest
that missed flux is a significant problem in the MPA-JHU SDSS masses.

\subsection{Comparison between the MPA-JHU- and GAMA-derived mass estimates II.\ Using SDSS \model\ SEDs \label{ch:modelmasses}}

While we have now shown very good agreement between the MPA-JHU- and
GAMA-derived $M_*/L$s for SDSS galaxies, the comparison presented in the
previous Section mixes the effects of differences in both the input photometry
and the mechanics of the stellar mass estimation algorithms. The next obvious
question is how well the two algorithms agree when applied to the same data.

Looking at the lower panels of Figure \ref{fig:cfsdss}, it is clear that we do
not do a particularly good job of reproducing the MPA-JHU masses when using
the SDSS \model\ photometry to construct galaxy SEDs. We see mild systematic
differences between the GAMA- and SDSS-derived values of $M_*/L$ as a function
of both colour and structure, and the random scatter between the two estimates
is not small: $\sim 0.15$ dex. Further, there is a distinct population of de
Vaucouleurs-like galaxies ($n \sim 2.5$--6) galaxies with rest frame $(g-i)$
colours of $\sim 1.1$ whose GAMA-derived $M_*/L$s are lower by $\sim 0.25$
dex.

How can it be that we do a better job at reproducing the MPA-JHU $M_*/L$s when
using the GAMA photometry than we do using the SDSS \model\ photometry? At
least part of the answer is thus directly tied to differences between the SDSS
\model\ and GAMA \auto\ photometry. We have made similar comparisons using
SDSS \petro\ photometry. Perhaps unsurprisingly, given the close agreement
between the \petro\ and \auto\ colours, we do not find any strong systematic
differences between the GAMA- and SDSS-derived values in this case. The median
value of $\Delta M_*/L_i$ is $-0.02$ dex; the RMS $\Delta M_*/L_i$ is 0.11
dex. There are no obvious trends in $\Delta M_*/L_i$ with apparent magnitude,
\Sersic\ index, or inferred rest-frame colour. That is, the problem appears to
be specific to the SDSS \model\ photometry.

We have seen that the SDSS \model\ SEDs are systematically redder than those
constructed using GAMA \auto\ magnitudes. The effect of these differences can
be understood by looking at Figure \ref{fig:ugri}. Using the \model\ SEDs, the
`problem' objects (that is, those galaxies where there are large differences
in the GAMA- and SDSS- derived stellar mass estimates) prefer templates with
young ages ($\lwage \lesssim 3$ Gyr), high metallicity ($Z \approx 0.05$; the
highest value allowed in our BC03 library) and moderately heavy dust
extinction ($\dust \sim 0.2$). When using the GAMA \texttt{AUTO} photometry,
these objects come out to be considerably older ($\lwage \sim 6$ Gyr), lower
metallicity ($Z \sim 0.01$) and less dusty ($\dust \sim 0.05$---0.10).

Looking carefully at the $1.0 < (g-i) < 1.2$ region of Figure
\ref{fig:ugri}, one can see that immediately above the broad strip defined
by the older, low-SSFR models (colour-coded red in the upper-left panel),
there is a narrower strip of models with ages $\lwage \sim 1$ Gyr
(colour-coded green). In the lower-lefthand panel of Figure \ref{fig:ugri},
these models can also be seen to have $Z = 0.05$ (colour-coded red). Note
that this is precisely the regime where we see the largest differences in
the GAMA- and MPA-JHU-derived $M_*/L$s.  

This explains the differences between our stellar mass estimates based on
the \auto\ and \model\ photometry: the 0.13 mag offset in $(u-r)$ between
the \model\ and \auto\ photometry pushes these galaxies up towards the upper
edge of the region of colour space spanned by the models. The redder $(u-r)$
\model\ colours thus open up a qualitatively different, young,
high-metallicity stellar population solution for what would otherwise be
old, lower-metallicity galaxies. 

But how is it that the SDSS stellar mass estimates, which are based on the
`wrong' \model\ SEDs still get the `right' value for $M*/L$? We speculate that
the answer may lie in the different metallicity and/or dust priors used by the
MPA-JHU team. The priors most strongly affect galaxies with low metallicities
and dust extinction (Jarle Brinchmann, private communication; 24/09/2009);
these are precisely the kinds of galaxies where we see the greatest
discrepancies between the GAMA- and MPA-JHU-derived values of $M_*/L$. That
is, it would seem that the `problem' dustier, younger, and high-metallicity
solutions preferred by the \model\ photometry are down-weighted by the
inclusion of a dust prior in the MPA-JHU algorithm. Here, too, the SDSS
decision to use the median, rather than the mean, of the PDF will help to
reduce any susceptibility to a `bimodality' in the PDF, and so reduce the
likelihood of choosing these `problem' solutions. In our case, using the GAMA
\auto\ photometry, the inclusion of such a prior is unnecessary.

Again, our primary motivation for performing this comparison is to test our
ability to reproduce the well tested MPA-JHU values for $M_*/L$. Given that we
have demonstrated our ability to do so using our own $ugriz$ photometry, and
the fact that without access to the MPA-JHU algorithm we are unable to perform
any more detailed tests or comparisons, we have not investigated this issue
any further.

\subsection{Summary}

In this Appendix, we have compared the GAMA photometry and stellar mass
estimates to those from SDSS. Our primary motivation for this comparison is
that is has been argued, based on their consistency with dynamical mass
estimates, the SDSS stellar mass estimates have no strong systematic
differential biases for galaxies with different stellar populations
\citep{Taylor2010b}. When using the GAMA photometry, we find excellent
agreement between our fiducial estimates of $M_*/L$s and those from SDSS, with
no strong differential biases as a function of mass, colour, or structure.
This argues against there being any strong biases in the GAMA $M_*/L$
estimates. We will investigate this further through comparison between the
GAMA stellar and dynamical mass estimates in a separate work.

We have also shown that there are significant differences between the GAMA and
SDSS estimates of total flux, which come from \sersictt\ and \model\
photometry, respectively. These differences are a strong function of $n$: for
$n=1$ and $n=4$ galaxies, where the SDSS \model\ assumes the `right' value of
$n$, we find excellent agreement between the two surveys' photometry. Away
from these points, however, the \model\ photometry is strongly biased. For de
Vaucouleurs-like galaxies, the size of the differential bias in the \model\
photometry is as large as a factor of 2. This will have a significant impact
on a number of stellar mass-centric measurements like the mass function or the
size--mass relationship. 

We have shown that there are significant systematic differences between the
GAMA- and SDSS-derived colours, which are derived from \auto\ and \model\
photometry, respectively. The SDSS \model\ photometry is systematically
redder, with a net offset of $\Delta (u-z) = 0.2$ mag. We suggest that it may
be better to use \petro , rather than \model\ photometry when analysing SDSS
data: we find no such differentials between the GAMA \auto\ and SDSS \petro\
photometry . When we apply our SPS pipeline to the \model\ photometry, our
stellar mass estimates no longer agree well with those from SDSS. We suggest
that these differences may be explained by the different priors used in the
SDSS pipeline, which act to downweight young, moderately dusty, and high
metallicity SPS fits. In the case of SDSS, their priors would seem to
effectively circumvent the potential biases in $M_*/L$ that these photometric
biases might produce. In our case, we are able to reproduce the well-tested
SDSS values with no need for such priors.

\end{document}